\documentclass[hyperref]{acmtrans2e}

\usepackage[all]{xy}
\usepackage{amssymb,alltt}

\newcommand{\OUT}[1]{\ignorespaces}

\newcommand{\TXF}{{\cal T}({\cal F},{\cal X})}

\newcommand{\TFXA}{{\cal T}({\cal F},{\cal X}_A)}
\newcommand{\TFXN}{{\cal T}({\cal F},{\cal X \cup \cal N})}
\newcommand{\TFXXA}{{\cal T}({\cal F},{\cal X} \cup {\cal X}_A)}

\newcommand{\TF}{{\cal T}({\cal F})}
\newcommand{\XA}{{\cal X}_A}

\newcommand{\R}{{\cal R}}
\newcommand{\U}{{\cal U}}
\newcommand{\OC}{{\cal O}}

\newcommand{\OS}{{\overline{\cal O}}}
\newcommand{\F}{{\cal F}} 
\newcommand{\X}{{\cal X}}

\newcommand{\ra}{\rightarrow}

\newcommand{\da}{\mathord{\downarrow}}

\newcommand{\surred}{\leadsto}

\newcommand{\con}{\da  =}

\newcommand{\red}{\twoheadrightarrow^*}
\newcommand{\sur}{/}

\newcommand{\Var}{{\cal V}{\it ar}}

\newcommand{\Def}{{\cal D}{\it ef}}

\newcommand{\gtree}[2]{#2} 
\newcommand{\comment}[1]{}

\newcommand{\rname}[1]{\mbox{$\bf #1$}} 

\newlength{\Labelwidth}

\newcommand{\cinfr}[4]
   {\makebox[\Labelwidth][l]{\bf #1:}\ 
     \(\displaystyle {#2} \over \displaystyle {#3}\) \\[3mm]
         \(\displaystyle ~~{\rm if}\;{#4}\)}
\newcommand{\wcinfr}[5]
   {\makebox[\Labelwidth][l]{\bf #1:}\ 
     \(\displaystyle {#2} \over \displaystyle {#3}\) \\[3mm]
         \(\displaystyle {#4}\) \\[1mm]
         \(\displaystyle {#5}\)}

\newcommand{\dubblewcinfr}[6]
   {\makebox[\Labelwidth][l]{\bf #1:}\ 
     \(\displaystyle {#2} \over \displaystyle {#3}\) \\[3mm]
         \(\displaystyle {#4}\) \\[1mm]
         \(\displaystyle {#5}\) \\[1mm]
         \(\displaystyle {#6}\)}

\newcommand{\fivecinfr}[8]
   {\makebox[\Labelwidth][l]{\bf #1:}\ 
     \(\displaystyle {#2} \over \displaystyle {#3}\) \\[3mm]
         \(\displaystyle {#4}\) \\[1mm]
         \(\displaystyle {#5}\) \\[1mm]
         \(\displaystyle {#6}\) \\[1mm]
         \(\displaystyle {#7}\) \\[1mm]
         \(\displaystyle {#8}\)}

\newcommand{\sixcinfr}[9]
   {\makebox[\Labelwidth][l]{\bf #1:}\ 
     \(\displaystyle {#2} \over \displaystyle {#3}\) \\[3mm]
         \(\displaystyle {#4}\) \\[1mm]
         \(\displaystyle {#5}\) \\[1mm]
         \(\displaystyle {#6}\) \\[1mm]
         \(\displaystyle {#7}\) \\[1mm]
         \(\displaystyle {#8}\) \\[1mm]
         \(\displaystyle {#9}\)}

\newtheorem{lemma}{Lemma}[section]
\newtheorem{sublemma}{Lemma}[subsection]
\newtheorem{proposition}{Proposition}[section]
\newtheorem{subproposition}{Proposition}[subsection]

\newtheorem{subtheorem}{Theorem}[subsection]
\newdef{definition}[theorem]{Definition}
\newdef{subdefinition}[subtheorem]{Definition}
\newdef{example}[theorem]{Example}
\newdef{subexample}[subtheorem]{Example}

\newcommand{\BibTeX}{{\rm B\kern-.05em{\sc i\kern-.025em b}\kern-.08em
    T\kern-.1667em\lower.7ex\hbox{E}\kern-.125emX}}

\firstfoot{ACM Transactions on Computational Logic,
  Vol.\ X, No.\ X, Date XXXX, Pages xxx-xxx.}

\runningfoot{ACM Transactions on Computational Logic,
  Vol.\ X, No.\ X, Date XXXX.}

\title{Termination of rewriting strategies:
a generic approach}
\author{ISABELLE GNAEDIG\\Loria-INRIA
 \and
HELENE KIRCHNER\\Loria-CNRS}
\begin{abstract}
We propose a generic termination proof method for rewriting under strategies, 
based on an explicit induction on the termination property.
Rewriting trees on ground terms are modeled by proof trees, generated by 
alternatively applying narrowing and abstracting steps.
The induction principle is applied through the abstraction mechanism, 
where terms are replaced by variables representing any 
of their normal forms.
The induction ordering is not given a priori, but defined with ordering constraints, 
incrementally set during the proof. Abstraction constraints can be used to control
the narrowing mechanism, well known to easily diverge.
The generic method is then instantiated for the innermost, outermost and 
local strategies. 
\end{abstract}





  





\begin{document}

\setcounter{page}{1}

\begin{bottomstuff}
Author's address: Isabelle Gnaedig, H\'el\`ene
Kirchner,
LORIA, 615, rue du Jardin Botanique, BP 101, F-54602 Villers-l\`es Nancy
Cedex ,
Fax: + 33 3 83 27 83 19 \\
{\tt e-mail:~Isabelle.Gnaedig@loria.fr,Helene.Kirchner@loria.fr}
\permission
\copyright\ 2005 ACM ... \$00.75
\end{bottomstuff}
\maketitle

\bibliographystyle{acmtrans}

\section{Introducing the problem}

Rewriting techniques are now widely used in automated deduction,
especially to handle equality, as well as in programming, in
functional, logical or rule-based languages.
Termination of rewriting is a crucial property, important in itself
to guarantee a result in a finite number of steps, but it is also
required to decide properties like confluence and sufficient
completeness, or to allow proofs by
consistency. Existing methods for proving termination of 
rewrite systems essentially tackle the 
termination problem  on free term algebras for rewriting without strategies.

Most are based on
syntactic or semantic noetherian orderings containing the rewriting
relation induced by the
rewrite system~\cite{Plaisted-well-founded-78,Lankford-79,Kamin-Levy,DershowitzTCS-1982,Cherifa-Lescanne-SCP87,Dershowitz-Hoot-1995,BorFerRubio-SPO-CADE-2000}.
Other methods consist in transforming the termination problem of a rewrite system
into the decreasingness problem of another rewrite system or of pairs of terms, 
then handled with
techniques of the previous category.  Examples are 
semantic labelling \cite{Zantema-1995}, and the dependency pair
method~\cite{Arts-Giesl-TCS-2000,Giesl-Thi-all-improvingDP-2003}.  For most approaches, finding an
appropriate ordering is the key problem, that often  comes down
to solving a set of ordering constraints.  

In the context of proof environments for rule-based programming
languages, such as ASF+SDF \cite{Klint-ACM93},
Maude~\cite{Maude-RwLg1996}, Cafe\-OBJ~\cite{FutatsugiN-IEEE97},
ELAN~\cite{BorovanskyKKMR-WRLA98}, or TOM~\cite{MoreauRV-2003}, 
where a program is a rewrite system
and the evaluation of a query consists in rewriting a ground
expression, more specific termination proof tools are required, to
allow termination proofs on ground terms, and
under specific reduction strategies.  There
are still few results in this domain.  To our knowledge, methods have
only been given on the free term algebra with the innermost
strategy~\cite{Arts-Giesl-inner-tech-96,Giesl-Mid-inn-context-sens-2003} 
and for the context-sensitive
rewriting~\cite{LucasIC02}, which involves particular kinds of local
strategies~\cite{Luc01}. In previous works, we already have obtained
termination results on ground terms for the innermost
strategy~\cite{GKF-in-out-2001,FGK-PPDP-2002}, for general local
strategies on the operators~\cite{FGK-local-strat-ENTCS-2001}, and for
the outermost strategy~\cite{FGK-WRLA-2002}.

In this paper, we propose a generic proof principle, based on an
explicit induction mechanism on the termination property, which is
a generalization of our three previous results. 
We then show how it can be instantiated to give an effective 
termination proof algorithm for the innermost strategy, the outermost
strategy, and local strategies on operators.
This generalizing work allowed not only to propose a generic version
of our proof method, but also to considerably simplify the technical
features of the algorithms initially designed for the different strategies.

The three above strategies have been chosen for their
relevance to programming languages.
The most widely used innermost strategy consists in rewriting always
at the lowest possible positions. It is often used as a built-in
mechanism in evaluation of rule-based or functional languages.  In
addition, for non-overlapping or locally confluent overlay systems
\cite{Gramlich-FI96}, or systems satisfying critical peak conditions
\cite{Gramlich-RTA96}, innermost termination is equivalent to standard
termination (i.e. termination for standard rewriting, which consists
in rewriting without any strategy).  As proved in
\cite{Rao-innermost-TCS-2000}, termination of rewriting is equivalent
for the leftmost innermost and the innermost strategies.

The outermost strategy for evaluating expressions in the context of
programming is essentially used when one knows that computations can
be non-terminating. The intuition suggests that rewriting a term at
the highest possible position gives more chance than with another
strategy to lead to an irreducible form.  Indeed, outermost rewriting
may succeed when innermost fails, as illustrated by the expression
$\mathit{second}(\mathit{dec}(1), 0)$, with the rewrite rules 
$\mathit{second}(x,y) \ra y$  and
$\mathit{dec}(x) \ra \mathit{dec}(x-1)$ on integers.  Innermost rewriting fails to
terminate, because it first evaluates $\mathit{dec}(1)$ into $\mathit{dec}(0)$,
$\mathit{dec}(-1)$, and so on.  Outermost rewriting, however, gives $0$ in one
rewriting step.  Moreover, outermost derivations are often shorter~:
in our example, to reduce $\mathit{second}(u,v)$, one does not need to reduce
$u$, which can lead to infinite computations or, at least, to a
useless evaluation.  This advantage makes the outermost strategy an
interesting strategy for rule-based languages, by allowing the
interpreters to be more efficient, as well as for theorem proving, by
allowing the rewriting-based proofs to be shorter.

Outermost computations are of interest in particular for functional
languages,
where interpreters or compilers generally involve a strategy for call
by name. Often, lazy evaluation is used instead: operators are
labelled in terms as lazy or eager, and the strategy consists in
reducing the eager subterms only when their reduction allows a
reduction step higher in the term~\cite{NG01a}. However, lazy
evaluation may diverge while the outermost computation terminates,
which gives an additional motivation for studying outermost
termination.  For instance, let us consider the evaluation of the
expression $\mathit{inf}(0)$ with the following two rules~: 
$\mathit{cons}(x,\mathit{cons}(y,z))
\ra \mathit{big}, \; \; \mathit{inf}(x) \ra \mathit{cons}(x, \mathit{inf}(s(x)))$.  
If $\mathit{inf}$ is
labelled as eager, $\mathit{inf}(0)$ is reduced to 
$\mathit{cons}(0, \mathit{inf}(s(0)))$, and
then, since application of the first rule fails, the sub-expression
$\mathit{inf}(s(0))$ has to be evaluated before considering the whole
expression, which leads to an infinite evaluation.  Evaluated in an
outermost manner, $\mathit{inf}(0)$ is also reduced to 
$\mathit{cons}(0, \mathit{inf}(s(0)))$,
but then $\mathit{inf}(s(0))$ is reduced to $\mathit{cons}(s(0), \mathit{inf}(s(s(0))))$, 
and the
whole expression is reduced to $\mathit{big}$. Lazy termination of functional
languages has already been studied (see for
example~\cite{Panitz-Schmidt-Schauss-1997}), but to our knowledge, 
except our previously cited work, no
termination proof method exists for specifically proving outermost
termination of rewriting.

Local strategies on operators are used
in particular to force the evaluation of expressions 
to terminate. A famous example is the evaluation of a recursive function 
defined with an $\mathit{if\_then\_else}$ expression, for which evaluating the 
first argument in priority may allow to avoid divergence.

This kind of strategy is allowed by languages such that 
OBJ3, Cafe\-OBJ or Maude, and studied
in~\cite{Eker98} and~\cite{Nakamura-Ogata-2000}. 
It is defined
in the following way: 
to any operator $f$ is attached an  ordered list
of integers, giving the positions of the subterms to be
evaluated in a given term, whose top operator is $f$. For example, the 
rewrite system

{\footnotesize
\[
\begin{array}{ll}
f(i(x))         & \ra   if\_then\_else(zero(x),g(x),f(h(x)))\\
zero(0)         & \ra   true\\
zero(s(x))      & \ra   false\\
if\_then\_else(true,x,y)        & \ra  x\\
if\_then\_else(false,x,y)       & \ra  y\\
h(0)            & \ra i(0)\\
h(x)            & \ra s(i(x))
\end{array}
\]
}
\noindent
using the conditional expression,
does not terminate for the standard rewriting relation, but does with
the following strategy:
$\mathit{LS}(\mathit{ite}) = [1 ; 0]$, $\mathit{LS}(f) = \mathit{LS}(\mathit{zero}) =$ 
$\mathit{LS}(h) = [1 ; 0]$ and 
$\mathit{LS}(g) = \mathit{LS}(i) = [1]$, where $\mathit{ite}$ denotes 
$\mathit{if\_then\_else}$ for short.

Local strategies have to be compared with context-sensitive rewriting,
where rewriting is also allowed at some specified positions only in
the terms: as local strategies specify in addition an ordering on these
rewriting positions, they are more specific. 

The termination problem for these various strategies is always
different: in~\cite{FGK-out-interne-2002}, the interested reader can
find examples showing that termination for one of these strategies
does not imply termination for any other of them. A better knowledge
of these differences would be interesting, and could help to choose the good
one when programming in these languages.


Despite of these distinct behaviours, the termination proofs we
propose rely on
a generic principle and a few common concepts, that are emphasized in
this paper.
Our approach
is based on an explicit induction mechanism on the termination
property.  The main idea is to proceed by induction on the ground term
algebra with a noetherian ordering
$\succ$, assuming that for any $t'$ such that $t \succ t'$, 
$t'$
terminates, i.e. there is no infinite derivation
chain starting from $t'$.
The general proof principle relies on the simple idea that
for establishing  termination of a ground term $t$, it
is enough to suppose that subterms of $t$ are smaller than $t$ for
this ordering, and that rewriting the context only leads to 
terminating chains. Iterating this process until a non-reducible context
is obtained establishes  termination
of $t$. 

Unlike classical induction
proofs, where the ordering is given, we do not need to define it {\em
a priori}.  We only have to check its existence by ensuring
satisfiability of ordering constraints incrementally set along the
termination proof.
Thanks to the power of induction, the generated constraints are often simpler 
to solve than for other approaches, and even, in many cases, do not need any 
constraint solving algorithm.

Directly using the termination notion on terms has also been proposed
in~\cite{Goubault-Larrecq-well-found-2001}, 
but for inductively proving well-founded\-ness of
binary relations, among which path orderings. 

In order to explain the basic idea of this work, let us consider 
the classical example, due to Toyama, of a rewrite system
that does not terminate,  but terminates with the innermost strategy:

{\footnotesize
\[
\begin{array}{ll}
f(0,1,x)        & \ra   f(x,x,x) \\
g(x,y)        & \ra   x\\
g(x,y)      & \ra   y
\end{array}
\]
}
\noindent

Let us prove by induction on the set $\TF$ of ground terms 
built on ${\cal F}=\{0,1,f,g\}$
with a noetherian 
ordering $\succ$, that any term
$t$ innermost terminates (i.e. there is no infinite innermost rewriting
chain starting from $t$). The terms of
$\mathit{\TF}$ are $0$, $1$, or terms of the form 
$f(t_1,t_2,t_3)$, or
$g(t_1,t_2)$, with $t_1,t_2,t_3 \; \in \TF$. The terms 
$0$ and $1$
are obviously terminating.

Let us now prove that $f(t_1,t_2,t_3)$ is innermost terminating. 
First, $f(t_1,t_2,t_3) \succ$ $t_1, t_2, t_3$ for any term ordering with
the subterm property (i.e. any term is greater than any of its subterms). 
Then, by induction
hypothesis, assume that $t_1, t_2$ and $t_3$ innermost terminate. Let 
$t_1\da, t_2\da, t_3\da$ be respectively any of their normal forms.
The problem is then reduced to innermost termination of all 
$f(t_1\da, t_2\da, t_3\da)$.
If $t_1\da = 0 \; ,\; t_2\da = 1$, then $f(0,1, t_3\da)$ 
only rewrites at the top position into $f(t_3\da, t_3\da,$ $t_3\da)$, 
which is in normal form.
Else $f(t_1\da, t_2\da, t_3\da)$ is already in normal form.

Let us finally prove that $g(t_1,t_2)$ is innermost terminating. 
First, $g(t_1,t_2) \succ t_1,$ $t_2$. Then, by induction
hypothesis, assume that $t_1$ and  $t_2$ innermost terminate.
Let $t_1\da, t_2\da$ be respectively any of their normal forms.
It is then sufficient to prove that $g(t_1\da, t_2\da)$ is innermost 
terminating. 
The term $g(t_1\da, t_2\da)$ rewrites either into $t_1\da$ or into 
$t_2\da$ at the
top position, with both $t_1\da$ and $t_2\da$ in normal form.
Remark that for $\succ$ in this proof, any ordering having the subterm
property is convenient.
Our goal is to provide a procedure implementing such a reasoning.

The paper is organized as follows: in Section~\ref{sec:background},
the background is presented.  Section~\ref{sec:inductive} introduces
the inductive proof principle of our approach.
Section~\ref{sec:abstraction-narrowing-constraints} gives the basic
concepts of our inductive proof mechanism based on abstraction and
narrowing, and the involved constraints.
Section~\ref{sec:generic-algoritm} presents the generic termination
proof procedure that is further applied to different rewriting
strategies.  In Section \ref{sec:innermost}, the mechanism is
instantiated for the case of innermost termination.  In Section
\ref{sec:outermost}, the procedure is applied to outermost
termination.  Finally, in section \ref{sec:local-strat}, the same
method is adapted to the case of local strategies.

\section{The background}
\label{sec:background}

We assume that the reader is familiar with the basic definitions and notations
of term rewriting given for instance in~\cite{DershowitzJ-1989}. 
$\TXF$ is the set of terms built from a given finite set $\F$
of function symbols $f$ having arity $n \in \mathbb N$ (denoted $f:n$), and a set $\X$ of 
variables denoted $x, y \ldots $.
$\TF$ is the set of ground terms (without variables). The
terms reduced to a symbol of arity $0$ are called {\em constants}.
Positions in a term are represented as sequences of integers.
The empty sequence $\epsilon$ denotes the top position. 
The symbol at the top position of a term $t$ is written $top(t)$.
Let $p$ and $p'$ be two positions. The position $p$ is said to be (a strict)
prefix of $p'$ (and $p'$ suffix of $p$) if $p'=p \lambda$, where
$\lambda$ is a non-empty sequence of integers.
Given a term $t$, $Var(t)$ is the set of variables of $t$,
$\OC(t)$ is the set of positions in $t$, inductively
defined as follows:
$\OC(t) = 
\{\epsilon\} \; if  \; t \in \X, \; 
\OC(t) = \{\epsilon\} \cup \{i.p \mid 1 \leq i \leq n \mbox{ and } p \in \OC(t_i)\}
\; if  \; t = f(t_1, \ldots ,t_n)$.
This set is partitioned into $\OS(t) = \{p \in \OC(t) \mid t|_p \not \in \X\}$
and $\OC_{\cal V}(t) = \{p \in \OC(t) \mid t|_p \in \X \}$ where the notation
$t|_p$ stands for the subterm of $t$ at position $p$.
If $p \in \OC(t)$, then $t[t']_p$ denotes the term obtained from $t$
by replacing the subterm at position $p$ by the term $t'$.

A substitution is an assignment from $\X$ to $\TXF$,
written $\sigma = (x \mapsto t) \ldots (y \mapsto u)$.
It uniquely extends to an endomorphism of $\TXF$. 
The result of applying 
$\sigma$ to a term $t \in \TXF$ is written $\sigma(t)$ or $\sigma t$.
The domain of $\sigma$, denoted $Dom(\sigma)$ is the finite subset of
$\X$ such 
that $\sigma x \neq x$. The range of $\sigma$, denoted $Ran(\sigma)$, is
defined by $Ran(\sigma) = \bigcup_{x \in Dom(\sigma)} Var(\sigma x)$.
We have in addition $Dom(\sigma) \cap Ran(\sigma) = \emptyset$.
A ground substitution or instantiation is an assignment from $\X$ to
$\TF$.
$Id$ denotes the identity substitution. The composition of substitutions
$\sigma_1$ followed by $\sigma_2$ is denoted $\sigma_2 \sigma_1$.
Given a subset ${\cal X}_1$ of $\cal X$, we write 
${\sigma}_{{\cal X}_1}$ for the {\em restriction} of
$\sigma$ 
to the variables of ${\cal X}_1$, i.e. the substitution such that
$Dom({\sigma}_{{\cal X}_1}) \subseteq {\cal X}_1$ and
$\forall x \in Dom({\sigma}_{{\cal X}_1}):{\sigma}_{{\cal X}_1} x = \sigma x.$

Given a set $\R$ of rewrite rules (a set of pairs of terms of $\TXF$, denoted 
$l \ra r$, such that $Var(r) \subseteq Var(l)$) or rewrite system 
on $\TXF$, a function symbol in $\F$
is called a {\em constructor} iff it does not occur in $\R$ at the top position
of a left-hand side of rule, and is called a {\em defined function symbol}
otherwise. 
The set of defined function symbols of $\F$ for $\R$ is denoted
by $\Def_R$ ($\R$ is omitted when there is no ambiguity).

The  rewriting relation induced by $\R$ is denoted by $\ra^{\R}$ ($\ra$ if 
there is no ambiguity on $\R$), and defined by $s \ra t$ iff there exists a 
substitution $\sigma$ and a position $p$ in $s$ such that $s|_p = \sigma l$ 
for some rule $l \ra r$ of $\R$, and $t=s[\sigma r]_p$. 
This is written $ s \ra^{p, l \ra r, \sigma}_{\cal R} t$ 
where either $p$ either $l \ra r$ either $\sigma$ or ${\cal R}$ may be omitted;
$s|_p$ is called a redex.
The reflexive transitive closure of the rewriting
relation induced by $\R$ is denoted by 
$\stackrel{*}{\ra}_{\R}$.
If $t \stackrel{*}{\ra} t'$ and $t'$ cannot be rewritten anymore, then $t'$ is 
called a normal form  of $t$ and denoted by $t\da$. Remark that given $t$, 
$t\da$ may be not unique.

Let $\cal R$ be a rewrite system on $\TXF$. A term $t$ is {\em narrowed} into $t'$, at
the non-variable position
$p$, using the rewrite rule $l\ra r$ of $\R$ and the substitution
$\sigma$, when $\sigma$ is a most general unifier of $t|_p$ and $l$,
and $t'=\sigma(t[r]_p)$.
This is denoted 
$t\surred_{R}^{p,l\ra r,\sigma}t'$ where either $p$, either $l\ra r$ or $\sigma$
may be omitted.
It is always assumed that there is no variable in common
between the rule and the term, i.e. that $Var(l) \cap Var(t) = \emptyset$.

An ordering $\succ$ on $\TXF$ is said to be noetherian  iff there is no
infinite decreasing chain for this ordering. It is 
$\F$-stable iff for any pair of terms $t, t'$ of $\TXF$, for any
context $f( \ldots  \;  \ldots )$, $t \succ t'$ implies $f( \ldots t \ldots
)$ $ \succ f( \ldots t' \ldots )$. It 
has the subterm property iff for any $t$ of $\TXF$, $f( \ldots t \ldots ) \succ
t$. 
Observe that, for $\F$ and $\X$ finite,  if $\succ$ is $\F$-stable and has the subterm 
property, then it is noetherian~\cite{Kruskal60}. If, in addition, 
$\succ$ is stable under substitution (for any 
substitution $\sigma$, any pair of terms $t, t' \in \TXF, \; t \succ t'$
implies $\sigma t \succ \sigma t'$), then it is called a simplification
ordering.
Let $t$ be a term of  $\TF$; let us recall that $t$  terminates  
if and only if any  rewriting
derivation (or derivation chain) starting from $t$ is finite.

Rewriting strategies are in general aimed at reducing the derivation
tree (for standard rewriting) of terms. The following definition
expresses that rewriting a term with a strategy $S$ can only give a
term that would be obtained with the standard rewriting relation.

\begin{definition}[\thetheorem\ (rewriting strategy)]
Let $\R$ a rewrite system on $\TXF$.
A {\em  rewriting strategy} $S$ for $\R$ is
a mapping $S : \TXF \ra \TXF$ such that for every $t \in \TXF$, 
$S(t)=t'$ (we write $t \ra^S t'$) where $t'$ is such that $t
\ra_{\R} t'$.
\end{definition}





\begin{definition}[\thetheorem\ (innermost/outermost strategy)]
Let $\R$ a rewrite system on $\TXF$. The innermost (resp. outermost) strategy is a
rewriting strategy written $\mathit{S=Innermost}$ \ (resp. $\mathit{S=Outermost}$) 
such that 
for any term $t \in \TXF$, if $t \ra^S t'$, the rewriting position
$p$ in $t$ is such that
there is no suffix (resp. prefix) position $p'$ of $p$ such
that $t$ rewrites at position $p'$. 
\end{definition}

Rewriting strategies may be more complex to define. This is the case
for local strategies on operators, used in the OBJ-like languages.
We use here the notion of local strategy as expressed in \cite{OBJ3}
and studied in \cite{Eker98}. 

\begin{definition}[\thetheorem\ (LS-strategy)]
\label{def:local-strat}
An {\em LS-strategy}
is given by a function $LS$
from ${\cal F}$ to the set of lists of integers ${\cal L}(\Bbb{N})$,
that induces a rewriting strategy as follows. 

Given a LS-strategy such that $LS(f) = [p_1, \ldots ,p_k]$,
$p_i \in [0..arity(f)]$ for all $i \in [1..k]$,
for some symbol $f \in {\cal F}$,
normalizing a term $t =f(t_1, \ldots ,t_m) \in \TXF$ with
respect to $LS(f) = [p_1, \ldots,$ $p_k]$, consists in normalizing all subterms of $t$ at
positions $p_1, \ldots ,p_k$ successively, according to the strategy.
If there exists $i \in [1..k]$ such that $p_1, \ldots ,p_{i-1} \neq 0$
and $p_i=0$ ($0$ is the top position), then 

\begin{itemize}
\item if the current term $t'$ obtained
after normalizing $t|_{p_1},$ $ \ldots ,t|_{p_{i-1}}$ is reducible at the top
position into a term $g(u_1,$ $ \ldots ,u_n)$, then $g(u_1, \ldots ,u_n)$ is
normalized with respect to $LS(g)$ and the rest of the strategy
$[p_{i+1}, \ldots ,$ $p_k]$ is ignored,

\item if $t'$ is not reducible at the top
position, then $t'$ is normalized with respect to 
$p_{i+1}, \ldots ,p_k$.
\end{itemize}

\end{definition}

Let $t$ be a term of  $\TF$; we say that $t$  terminates  (w.r.t. to
the strategy $S$) if and only if every  rewriting
derivation (or derivation chain) (w.r.t. to the strategy $S$) 
starting from $t$ is finite. 
Given a term $t$, we call normal form (w.r.t. to the strategy $S$) 
or S-normal form of $t$, denoted $t\da$, any irreducible term, if it
exists, such that $t \stackrel{*}{\ra}^{S} t\da$.

\section{The inductive proof process}
\label{sec:inductive}

\subsection{Lifting rewriting trees into  proof trees}

For proving that a term $t$ of $\TF$ terminates (for the considered
strategy), we proceed by induction on $\TF$ with a noetherian ordering
$\succ$, assuming that for any $t'$ such that $t \succ t'$, $t'$
terminates.  To warrant non emptyness of $\TF$, we assume that $\F$
contains at least a constructor constant.

The main intuition is to observe the rewriting derivation tree (for
the considered strategy) starting from a ground term $t \in \TF$ which is any
instance of a term $g(x_1, \ldots ,x_m)$, for some defined function
symbol $g \in \Def$, and variables $x_1, \ldots ,x_m$.  Proving
termination on ground terms amounts proving that all rewriting
derivation trees have only finite branches, using the same induction
ordering $\succ$ for all trees.

Each rewriting derivation tree is simulated, using a lifting mechanism,
by a proof tree, developed from $g(x_1,\ldots,x_m)$ on $\TXF$, 
for every $g \in \Def$, by
alternatively using two main operations, namely narrowing and
abstraction, adapted to the considered rewriting strategy.
More precisely, narrowing schematizes all rewriting
possibilities of terms. The abstraction process simulates the
normalization of subterms in the derivations, according to the
strategy.  It consists in replacing these subterms by special
variables, denoting one of their normal forms, without computing
them. This abstraction step is performed on subterms that can be
assumed terminating by induction hypothesis.

The schematization of ground rewriting derivation trees is achieved
through constraints. The nodes of the developed proof trees are
composed of a current term of $\TXF$, and a
set of ground substitutions represented by a constraint progressively
built along the successive abstraction and narrowing steps. 
Each node in a proof tree schematizes  a set of ground terms:
the ground instances of the current term, that are solutions of the constraint. 

The constraint is in fact composed of two kinds of formulas: ordering
constraints, set to warrant the validity of the inductive steps, and
abstraction constraints combined to narrowing substitutions, which
effectively define the relevant sets of ground terms.  The latter are
actually useful for controlling the narrowing process, well known to
easily diverge.

The termination proof procedures given in this paper are described by
deduction rules applied with a special control $\mathit{Strat{-}Rules(S)}$,
depending on the studied rewriting strategy $S$.
To prove termination of $\R$
on any term $t \in \TF$ w.r.t. the strategy $S$, we consider a so-called
reference term $t_{\mathit{ref}} = g(x_1, \ldots , x_m)$ for each
defined symbol $g \in \Def$, and empty
sets $\top$ of constraints.  Applying the deduction rules according to the
strategy $\mathit{Strat{-}Rules(S)}$ to the initial state $(\{g(x_1, \ldots ,
x_m)\}, \top, \top)$ builds a proof tree, whose nodes are the states
produced by the inference rules. Branching is produced by the
different possible narrowing steps.

Termination is established when the procedure terminates because the
deduction rules do not apply anymore and all terminal states of all proof trees
have an empty set of terms.

\subsection{A generic mechanism for strategies}
\label{subsec:generic}

As said previously, we consider any term  of $\TF$ as a ground
instance of 
a term $t$ of $\TXF$ occurring in a proof tree issued from
a reference term $t_{\mathit{ref}}$.
Using the termination induction hypothesis on $\TF$ naturally
leads us to simulate the rewriting relation by two mechanisms:

\begin{itemize}
\item
first, some subterms $t_j$ 
of the current term $t$  of the proof tree are supposed
to have  only terminating ground instances, by induction hypothesis, if $\theta
t_{\mathit{ref}} \succ \theta t_j$ for the induction ordering $\succ$
and for every $\theta$ solution of the constraint associated to $t$.
They are replaced in $t$
by {\em  abstraction variables} $X_j$ representing respectively one
of their  normal forms $t_j \da$.
Reasoning by induction allows us to
only suppose the existence of the $t_j \da$ 
{\em without  explicitly computing them};

\item
second, narrowing (w.r.t. to the strategy $S$) the resulting term
$u=t[X_j]_{j\in \{i_1,\ldots,i_p\}}$ (where
$i_1,\ldots,i_p$ are the positions of the abstracted subterm $t_j$ in
$t$) into terms $v$, according to the
possible instances of the $X_j$.
This corresponds to  rewriting (w.r.t. to the strategy $S$) 
the possible ground instances of $u$ (characterized
by the constraint associated to $u$) in all possible ways.

In general, the narrowing step of $u$ is not unique.  We
obviously have to consider all terms $v$ such that $\theta u$
rewrites into $\theta v$, which
corresponds to considering all narrowing steps from $u$.

\end{itemize}

Then the  termination problem of the ground instances of $t$ is reduced
to the  termination problem of the ground instances of $v$.  If 
$\theta t_{\mathit{ref}} \succ \theta v$ for every ground
substitution $\theta$ solution of the constraint associated to $v$,
by induction hypothesis, $\theta v$ is supposed to be  terminating. 
Else, the process is iterated on $v$,  until getting
a term $t'$ such that either $\theta t_{\mathit{ref}} \succ \theta t'$, or 
$\theta t'$ is irreducible.  

We introduce in the next section the necessary concepts  to formalize and automate  
this technique.

\section{Abstraction, narrowing, and the involved constraints}
\label{sec:abstraction-narrowing-constraints}

\subsection{Ordering constraints}

The induction ordering is constrained
along the proof by imposing constraints between terms that must be
comparable, each time the induction hypothesis is used in the
abstraction mechanism.
As we are working with a lifting mechanism on the proof trees
with terms of $\TXF$, we directly work with an ordering $\succ_{\cal
P}$ on $\TXF$ 
such that $t\succ _{\cal P} u$ implies $\theta t \succ \theta u$,
for every $\theta$ solution of the constraint associated to $u$.

So inequalities of
the form $t > u_1, \ldots ,u_m$ are accumulated, which are called
{\em ordering constraints}. Any ordering $\succ_{\cal P}$ on $\TXF$
satisfying them and which is stable under substitution fulfills the
previous requirements on ground terms. The ordering $\succ_{\cal P}$,
defined on $\TXF$,
can then be seen as an extension of the induction ordering $\succ$,
defined on $\TF$. For convenience, the ordering $\succ_{\cal P}$
will also be written $\succ$.

It is important to remark that, for establishing the inductive
termination proof, it is sufficient to decide whether
there exists such an ordering.

\begin{subdefinition}[\thesubtheorem\ (ordering constraint)]
\label{def:ordering-constraint}
{\em An ordering constraint} is a pair of terms of $\TXF$ noted $(t>t')$.
It is said to be {\em satisfiable} if there exists an ordering
$\succ$, such that for every instantiation $\theta$ whose domain
contains $\Var(t) \cup \Var(t')$, we  have
$\theta t \succ \theta t'$. We say that $\succ$ satisfies $(t>t')$.

A conjunction $C$ of ordering constraints is 
satisfiable if  there exists an ordering  
satisfying all conjuncts.
The empty conjunction, always satisfied,  is denoted by $\top$.
\end{subdefinition}

Satisfiability of a constraint conjunction $C$ of this form is undecidable. 
But a sufficient condition for an ordering $\succ$ to satisfy $C$
is that $\succ$ is stable under substitution 
and $t \succ t'$ for any constraint $t > t'$ of $C$.

\subsection{Abstraction}
\label{subsec:generic-abstraction}

To abstract a term $t$ at positions $i_1,\ldots,i_p$, 
where
the $t|_j$ are supposed to have a normal
form $t|_j \da$, we  replace the $t|_j$ by abstraction variables $X_j$
representing respectively one of their possible  normal forms.
Let us define these special variables more formally.

\begin{subdefinition}
\label{def:abstraction-var}
Let $\XA$ be a set of variables disjoint from ${\cal X}$.
Symbols of  $\XA$ are called {\em abstraction variables}.
Substitutions and instantiations are extended to $\TFXXA$
in the following way:
let $X \in \XA$; for any substitution $\sigma$ 
(resp. instantiation $\theta$)
such that $X \in Dom(\sigma)$, $\sigma X$ (resp. $\theta X$)
 is in S-normal form.
\end{subdefinition}


\begin{subdefinition}[\thesubtheorem\  (term abstraction)]
\label{def:abstraction}
The term $t[t|_j]_{j \in \{i_1, \ldots , i_p\}}$ is said to be 
{\em abstracted} into the term $u$ (called {\em abstraction} of $t$) at
positions $\{i_1, \ldots, i_p\}$ iff 
$u =$ $t[X_j]_{j \in \{i_1, \ldots, i_p\}}$, where the $X_j, j \in \{i_1,
\ldots, i_p\}$ are fresh distinct  abstraction variables.
\end{subdefinition}

Termination on $\TF$ is proved by reasoning on terms with
abstraction variables, i.e. on terms of $\TFXXA$.  Ordering constraints
are extended to pairs of terms of $\TFXXA$. When subterms $t|_j$
are abstracted by $X_j$, we state constraints on abstraction
variables, called {\em abstraction constraints} to express that their
instances can only be normal forms of the corresponding instances
of $t|_j$. Initially, they are of the form $t \con X$ where $t \in
\TFXXA$, and $X \in \XA$, but we will see later how they are
combined with the substitutions used for the narrowing process.

\subsection{Narrowing}
\label{sub:narrowing}

After abstraction of the current term $t$ into $t[X_j]_{j \in \{i_1, \ldots , i_p\}}$, 
we check whether the possible ground instances of $t[X_j]_{j \in \{i_1, \ldots , i_p\}}$ 
are reducible, according to the possible values of the instances of
the $X_j$. This is achieved by narrowing $t[X_j]_{j \in \{i_1, \ldots , i_p\}}$.

The narrowing relation depends on the considered strategy $S$ and
the usual definition needs to be refined. The first idea is to use
innermost (resp. outermost) narrowing. Then, if a position $p$ in a term
$t$ is a narrowing position, a suffix (resp.  prefix) position of $p$ 
cannot be a narrowing position too.
However, if we consider ground instances of $t$, we can have
rewriting positions $p$ for some instances, and $p'$ for some other
instances, such that $p'$ is a suffix (resp. a prefix) of $p$.
So, when narrowing at some position $p$, the 
set of relevant ground instances of $t$ is defined by excluding the
ground instances  that would be narrowable at some suffix (resp. prefix)
position of $p$, that we call $S$-better position:
a position $S$-better than a position $p$ in $t$ is a suffix
position of $p$ if $S$ is the innermost strategy, a prefix position of
$p$ if $S$ is the outermost strategy. 
Note that local strategies are not of the same nature, and there is  no
$S$-better position in this case.

%

Moreover, to preserve the fact that a narrowing step of $t$ schematizes a 
rewriting step of  possible ground instances of $t$, 
we have to be sure that an innermost (resp. outermost) narrowing redex in $t$ 
corresponds to the same rewriting redex in a ground instance of $t$.
This is  the case only if, in the rewriting chain of the ground instance of $t$,
there is no rewriting redex at a suffix position of variable of $t$ anymore.
So before each narrowing step, we schematize the
longest rewriting chain of any ground instance of $t$, whose redexes
occur in the variable part of the instantiation, by a linear variable renaming.
Linearity is crucial to express that, in the previous rewriting chain, 
ground instances of the same variables can be reduced in different ways.
For the innermost strategy, abstraction of variables performs 
this schematization. For the outermost strategy, a reduction renaming 
will be introduced.
For local strategies however, this variable renaming is not relevant. 

The $S$-narrowing steps applying to a given term $t$ are computed in 
the following way. 
After applying the variable renaming to $t$, 
we look at every position $p$ of $t$ such that $t|_p$ 
unifies with the left-hand side of a rule using a substitution
$\sigma$. The position $p$ is a $S$-narrowing position of $t$,
iff there is no $S$-better position $p'$ 
of $t$ such that $\sigma t|_{p'}$ unifies with a left-hand side of
rule. Then we look for every  $S$-better position $p'$ than $p$ in $t$ such that 
$\sigma t|_{p'}$ narrows with some substitution $\sigma'$ and 
some rule $l' \ra r'$, and we set a constraint to exclude these 
substitutions.
So the substitutions used to narrow a
term have in general to satisfy a set
of disequalities coming from the negation of previous substitutions. 
To formalize this point, we need the following notations
and definitions.

In the following, we identify a substitution 
$\sigma = (x_1 \mapsto t_1) \ldots (x_n \mapsto t_n)$ on $\TFXXA$
with the finite set 
of solved equations $(x_1 = t_1) \wedge \ldots \wedge (x_n = t_n)$,
also denoted by the equality formula $\bigwedge_i (x_i=t_i)$,
with $x_i \in {\cal X} \cup \XA$, $t_i \in \TFXXA$,  where $=$ is
the syntactic equality. Similarly, we call {\em negation
$\overline{\sigma}$} of the substitution $\sigma$ the formula 
$\bigvee_i (x_i \neq t_i)$.

\begin{subdefinition}
[\thesubtheorem\ (constrained substitution)]
A {\em constrained substitution} $\sigma$ is a formula $\sigma_0
\wedge \bigwedge_j \bigvee_{i_j} (x_{i_j} \neq t_{i_j})$, where
$\sigma_0$ is a substitution.
\end{subdefinition}

\begin{subdefinition}[\thesubtheorem\ ($S$-narrowing)]
\label{def:narrowing}
A term $t \in \TFXXA$ {\em $S$-narrows} into a term $t' \in \TFXXA$ at
the non-variable position $p$ of $t$, using the rule $l \ra r \in \R$ with
the constrained substitution 
$\sigma = \sigma_0 \wedge \bigwedge_{j \in [1..k]} \overline{\sigma_j}$, 
which is written $t \leadsto^{S}_{p, l \ra r, \sigma} t'$ iff
\begin{center}
        $ \sigma_0(l) = \sigma_0(t|_p)$ and $ t' = \sigma_0(t[r]_p)$ 
\end{center}
\noindent
where $\sigma_0$ is the most general unifier of $t|_p$ and $l$ and 
$\sigma_j, j \in [1..k]$ are
all most general unifiers of $\sigma_0 t|_{p'}$ and a left-hand side
$l'$ of a rule of $\R$, for all position $p'$  which are  $S$-better
positions than $p$ in $t$.
\end{subdefinition}

It is always assumed that there is no variable in common
between the rule and the term, i.e. that $Var(l) \cap Var(t) =
\emptyset$.
This requirement of disjoint variables is easily fulfilled by an
appropriate renaming of variables in the rules when narrowing is
performed.
The most general unifier $\sigma_0$ used in
the above definition can be taken such that its
range only contains fresh variables.
Since we are interested in the
narrowing substitution applied to the current term $t$, but not in its
definition on the variables of the left-hand side of the rule, the
narrowing substitutions can be restricted to the variables of
the narrowed term $t$.

The following lifting lemma, generalized from~\cite{MiddeldorpH-AAECC94}, 
ensures the correspondence between the narrowing relation, used during the
proof, and the rewriting relation.

\begin{sublemma}[\thesubtheorem\ ($S${-}lifting Lemma)]
\label{lemma:surred-S}
Let $\R$ be a rewrite system.
Let $s \in \TXF$, $\alpha$ a ground
substitution  such that $\alpha s$ is $S${-}reducible 
at a non variable position $p$ of $s$, and ${\cal Y} \subseteq {\cal X}$
a set of variables 
such that $Var(s) \cup Dom(\alpha) \subseteq {\cal Y}$.
If $\alpha s \ra^{S}_{p, l \ra r} t'$, then there exist a term 
$s' \in \TXF$ and substitutions
$\beta, \sigma = \sigma_0 \wedge \bigwedge_{j \in [1..k]} \overline{\sigma_j}$ 
such that:

\[
  \begin{array}{ll}

        1.~s \surred^{S}_{p, l \ra r, \sigma} s', \\
        2.~\beta s' = t', \\
        3.~\beta \sigma_0 = \alpha [{\cal Y}]\\
        4.~\beta {\rm ~satisfies~} \bigwedge_{j \in [1..k]} \overline{\sigma_j}

  \end{array}
\]
where $\sigma_0$ is the most general unifier of $s|_p$ and $l$ and 
$\sigma_j, j \in [1..k]$ are
all most general unifiers of $\sigma_0 s|_{p'}$ and a left-hand side
$l'$ of a rule of $\R$, for all position $p'$  which are  $S$-better
positions than $p$ in $s$.

\end{sublemma}

\subsection{Cumulating constraints}
\label{subsec:cumulating-constraints}

Abstraction constraints have to be combined with the narrowing
constrained substitutions to characterize the ground terms schematized
by the proof trees.
A narrowing step effectively corresponds to a 
rewriting step of ground instances of $u$
if the narrowing constrained substitution $\sigma$ 
is {\em compatible} with the abstraction constraint formula A associated to
$u$ (i.e. $A \wedge \sigma$ is satisfiable). Else, the narrowing step 
is meaningless.
So the narrowing constraint attached to the  narrowing step is added 
to $A$.
Hence the introduction of abstraction constraint formulas.



\begin{subdefinition}[\thesubtheorem\ (abstraction constraint formula)]
An {\em abstraction constraint formula} (ACF in short) is a formula 
$\bigwedge_i (t_i \con t'_i) \wedge \bigwedge_j (x_j = t_j) \wedge 
\bigwedge_k \bigvee_{l_k} (u_{l_k} \neq v_{l_k})$, 
where $t_i, t'_i, t_j, u_{l_k}, v_{l_k} \in \TFXXA$, $x_j
\in \X \cup \XA$.
\end{subdefinition}

\begin{subdefinition}[\thesubtheorem\ (satisfiability of an ACF)]
\label{def:satACF}
An {\em abstraction constraint formula} 
$\bigwedge_i (t_i \con t'_i) \wedge \bigwedge_j (x_j = t_j) \wedge 
\bigwedge_k \bigvee_{l_k} (u_{l_k} \neq v_{l_k})$, 
is {\em satisfiable} iff there exists at least one instantiation $\theta$ such
that $\bigwedge_i (\theta t_i \con \theta t'_i) \wedge 
\bigwedge_j (\theta x_j = \theta t_j) \wedge 
\bigwedge_k \bigvee_{l_k} (\theta u_{l_k} \neq \theta v_{l_k})$. 
The instantiation
$\theta$ is then said to satisfy the ACF $A$ and is called solution of
$A$.
\end{subdefinition}

Integrating a constrained substitution $\sigma = \sigma_0 
\wedge \bigwedge_i \bigvee_{j_i} (x_{j_i} \neq t_{j_i})$ to an ACF $A$
is done by adding the formula defining $\sigma$ to $A$, thus giving the formula 
$A \wedge \sigma$.
For a better readability on examples,
we can propagate $\sigma$ into $A$ (by applying $\sigma_0$ to $A$), 
thus getting instantiated abstraction constraints of the form $t_i \con t'_i$
from initial abstraction constraints of the form $t_i \con X_i$.

An ACF $A$ is attached to each term $u$ in the proof trees;
its solutions characterize the interesting ground instances of this
term, i.e. the $\theta u$  such that $\theta$ is a solution of $A$.
When $A$ has no solution, the current node of the proof tree 
represents no ground term. Such nodes are then irrelevant for the 
termination proof.
Detecting and suppressing them during a narrowing step allows to control
the narrowing mechanism.
So we have the choice between 
generating only the relevant nodes of the
proof tree, by testing satisfiability of $A$ at each step, or stopping the
proof on a branch on an irrelevant node, by testing unsatisfiability
of $A$. These are both facets of the same question,  but in practice,
they are handled in different ways.

Checking satisfiability of $A$ is in general undecidable.
The disequality part of an ACF is a particular instance of a
disunification problem (a quantifier free equational formula), whose satisfiability
has been addressed in~\cite{Comon-rob91}, that provides rules
to transform any disunification problem into a solved form.  
Testing satisfiability of the equational part of an ACF is
undecidable in general, but sufficient conditions can be given, 
relying on a characterization of normal forms.

Unsatisfiability of $A$ is also undecidable in general, but simple sufficient conditions
can be used, very often applicable in practice. They rely
on reducibility, unifiability, narrowing and constructor tests.

According to Definition~\ref{def:satACF}, 
an ACF
$\bigwedge_i (t_i \con t'_i) \wedge \bigwedge_j (x_j = t_j) \wedge 
\bigwedge_k \bigvee_{l_k} (u_{l_k} \neq v_{l_k})$ 
is unsatisfiable if for instance, one of its conjunct $t_i \con t'_i$ is
unsatisfiable, i.e. is such that $\theta t'_i$ is not a normal form of
$\theta t_i$ for any ground substitution $\theta$.
Hence, we get four automatable conditions for unsatisfiability of 
an abstraction constraint $t \con t'$:
\begin{description}

\item[Case 1:]
$t \con t'$, with $t'$ reducible.
Indeed, in this case, any ground instance of $t'$ is reducible, and hence
cannot be a normal form.

\item[Case 2:]
$t \con t' \wedge \ldots \wedge t' \con t''$, with $t'$ and $t''$ not unifiable.
Indeed, any ground substitution $\theta$ satisfying the above conjunction is
such that (1)~$\theta t \con \theta t'$ and (2)~$\theta t' \con \theta t''$.
In particular, (1) implies that $\theta t'$ is in normal form and hence
(2) imposes $\theta t' = \theta t''$, which is impossible if $t'$ and $t''$
are not unifiable.

\item[Case 3:]
$t \con t'$ where $top(t)$ is a constructor, and $top(t) \neq top(t')$.
Indeed, if the top symbol of $t$ is a constructor $s$, then any normal form
of any ground instance of $t$ is of the form $s(u)$, where $u$ is a 
ground term in normal form.
The above constraint is therefore unsatisfiable if the top symbol of $t'$ 
is $g$, for some
$g \neq s$.

\item[Case 4:]
$t \con t'$ with $t,t' \in \TFXA$ not unifiable and
$\bigwedge_{t \surred^S v} v \con t'$ unsatisfiable.
This criterion is of interest if unsatisfiability of each conjunct $v \con t'$
can be shown with one of the four criteria we present here.

\end{description}

So both satisfiability and unsatisfiability checks need to use
sufficient conditions.
But in the first case, the proof process stops with failure as soon as satisfiability
of $A$ cannot be proved. In the second one, it can go on, until $A$ is
proved to be unsatisfiable, or until other stopping conditions are fulfilled.

Let us now come back to ordering constraints.  If we check satisfiability
of $A$ at each step, we only generate states in the proof trees, that
represent non empty sets of ground terms.  So in fact, the ordering
constraints of $C$ have not to be satisfied for every ground
instance, but only for those instances that are solution of $A$, hence
the following definition, that can be used instead of
Definition~\ref{def:ordering-constraint},
when constraints of this  definition cannot be proved satisfiable,
and solutions of $A$ can easily be characterized.

\begin{subdefinition}[\thesubtheorem\ (constraint problem)]
\label{def:constraint-problem}
Let $A$ be an abstraction constraint formula 
and $C$ a conjunction of ordering constraints.
The constraint problem $C \sur A$ is satisfied by an ordering $\succ$
iff for every instantiation $\theta$ 
satisfying $A$,  then $\theta t \succ \theta t'$ for 
every conjunct $t>t'$ of $C$.
$C \sur A$ is satisfiable iff there exists an
ordering $\succ$ as above.
\end{subdefinition}

Note that $C \sur A$ may be satisfiable even if $A$ is not.

\subsection{Relaxing the induction hypothesis}
\label{subsec:relaxing}

\newcommand{\APos}{{\cal AP}{\it os}}

It is important to point out the flexibility of the proof method
that allows the combination with auxiliary termination proofs using 
different techniques: when the induction hypothesis cannot be applied
on a term $u$, i.e. when it is not possible to decide whether the
ordering constraints are satisfiable, it is often possible to
prove termination 
(for the considered strategy) of any ground instance of $u$ by another way. In
the following we use a predicate $\mathit{TERMIN(S,u)}$ that is true iff
every ground instance of $u$ terminates for the considered strategy
$S$.

In particular, $\mathit{TERMIN(S,u)}$ is true when every instance of $u$ is in
normal form. This is the case when $u$ is not narrowable, and all
variables of $u$ are in $\XA$.  Indeed, by Lemma~\ref{lemma:surred-S}
and Definition~\ref{def:abstraction-var}, every instance of $u$ is in
normal form.  This includes the cases where $u$ itself is an
abstraction variable, and where $u$ is a non narrowable ground term.

Every instance of a narrowable $u$ whose variables are all in $\XA$,
and whose narrowing substitutions are not compatible with $A$, is also in
normal form. As said in Section~\ref{subsec:cumulating-constraints},
these narrowing possibilities do not represent any reduction step for
the ground instances of $u$, which are then in normal form.


Otherwise, in many cases, for proving that $\mathit{TERMIN(S,u)}$ is true, the
notion of usable rules~\cite{Arts-Giesl-inner-tech-96} is relevant.
Given a rewrite system $\R$ on $\TXF$ and a term $t \in \TFXXA$, the {\em usable
rules} of $t$ are a subset of $\R$, which is a computable superset of the 
rewrite rules that are likely to be used in any rewriting chain 
(for the standard strategy) starting from any ground instance of $t$, 
until its ground normal forms are reached, if they exist.

Proving termination of any ground instance of $u$ then comes
down to proving termination of its usable rules, which is in general
much easier than proving termination of the whole rewrite system $\R$.  In
general, we try to find a simplification ordering $\succ_N$ that
orients these rules.  Thus any ground instance $\alpha t$ is bound to
terminate for the standard rewriting relation, and then for the
rewriting strategy $S$.  Indeed, if $\alpha t \ra t_1 \ra t_2 \ra
\ldots $, then, thanks to the previous hypotheses, $\alpha t \succ_N
t_1 \succ_N t_2 \succ_N \ldots $ and, since the ordering $\succ_N$ is
noetherian, the rewriting chain cannot be infinite. As a particular
case, when a simplification ordering can be found to orient the whole rewrite system, 
it also orients the usable rules of any term, and our inductive approach
can also conclude to termination.
  If an appropriate
simplification ordering cannot be found, termination of the usable
rules may also be proved with our inductive process itself. The fact
that the induction ordering used for usable rules is independent of
the main induction ordering, makes the proof very flexible.
Complete results on usable rules for the innermost strategy are given in 
Section~\ref{subsec:relaxing-innermost}. For the outermost and local 
strategies, this is developed in~\cite{FGK-WRLA-2002} and \cite{FGK-local-strat-ENTCS-2001}.

\section{The termination proof procedure}
\label{sec:generic-algoritm}

\subsection{Strategy-independent proof steps}
\label{sec:proof-steps}

We are now ready to describe the different steps of the proof
mechanism presented in Section~\ref{sec:inductive}. 

The proof steps generate proof trees in transforming  3-tuples $(T,A,C)$ where 
\begin{itemize}
\item
$T$ is a set of terms of $\TFXXA$, containing the current term $u$ whose termination
has to be proved. $T$ is either a singleton or the empty set.
For local strategies, the term is enriched by
the list of positions where $u$ has to be evaluated, $LS(top(u))$.
This is denoted by $u^{LS(top(u))}$.

\item
$A$ is a conjunction of abstraction constraints. At each abstraction
step, constraints of the form 
$u \con X, u \in \TFXXA, X \in {\XA}$  are stated for each subterm term $t$ 
abstracted into a new abstraction variable $X$.
At each narrowing step with narrowing substitution $\sigma$, $A$ is replaced by 
$A \wedge \sigma$.

\item
$C$ is a conjunction of ordering constraints stated by the
abstraction steps.
\end{itemize}

Starting from initial states
$(T=\{t_{\mathit{ref}}=g(x_1,\ldots,x_m)\},A=\top,C=\top)$, where $g
\in \Def$,
the proof process consists in iterating 
the following generic steps:

\begin{itemize}
\item  
The first step abstracts the current term $t$ at
given positions $i_{1}, \ldots,$ $i_{p}$.
If the conjunction of ordering constraints
$\bigwedge_j t_{\mathit{ref}}>t|_j$ is satisfiable for some $j
\in \{i_{1}, \ldots ,i_{p}\}$, we
suppose,  by induction, the existence of  irreducible 
forms for the $t|_j$. 
We must have $\mathit{TERMIN(S,t|_j)}$ for the other $t|_j$.
Then, $t|_{i_1}, \ldots ,t|_{i_p}$ are
abstracted into abstraction variables $X_{i_{1}}, \ldots ,$
$X_{i_{p}}$.
The abstraction constraints $t|_{i_1} \con X_{i_{1}}, \ldots ,t|_{i_p} \con
X_{i_{p}}$ are added to the ACF $A$.
We call that step the \emph{abstract} step.

\item The second step narrows the resulting term $u$
in one step with all 
possible rewrite rules of the rewrite system $\R$, and all possible
substitutions $\sigma$, into terms $v$, according to Definition~\ref{def:narrowing}. 
This step is a branching step,
creating as many states as narrowing possibilities.
The  substitution $\sigma$ is added to $A$.
This is the \emph{narrow} step.

\item We then have a \emph{stop} step halting the proof process on the
current branch of the proof tree, when $A$ is detected to be unsatisfiable,
or when the  ground instances of the current term
can be stated terminating for the considered strategy. This happens 
when the whole
current term $u$ can be abstracted, i.e. when the induction hypothesis applies
on it, or when we have $\mathit{TERMIN(S,u)}$.

\end{itemize}

The satisfiability and unsatisfiability tests of $A$ are integrated in
the previously presented steps.  If testing unsatisfiability of $A$ is
chosen, the unsatisfiability test is integrated in the \emph{stop} step. If
testing the satisfiability of $A$ is chosen, the test is made at each
attempt of an abstraction or a narrowing step, which are then
effectively performed only if $A$ can be proved
satisfiable. Otherwise, the proof cannot go on anymore and stops with
failure.

As we will see later, for a given rewriting strategy $S$,
these generic proof steps are instantiated by more
precise mechanisms, depending on $S$,
and  taking advantage of its specificity.
We will define these specific instances by inference rules.

\subsection{Discussion on abstraction and narrowing positions}
\label{sec:discussion}

There are different ways to simulate the rewriting relation on ground
terms, using abstraction and narrowing. 

For example, the abstraction positions can be chosen so that the
abstraction mechanism captures the greatest possible number of
rewriting steps. For that, we abstract the greatest subterms in the term, that
are the immediate subterms of the term. Then, if a narrowing step
follows, the abstracted term has to be narrowed in all possible ways
at the top position only.  This strategy may yield a deadlock  if some
of the direct subterms cannot be abstracted.  We can instead abstract
all greatest possible subterms of $t=f(t_1, \ldots , t_n)$.  More
concretely, we try to abstract $t_1, \ldots, t_n$ and, for each $t_i =
g(t'_1,\ldots,t'_p)$ that cannot be abstracted, we try to abstract
$t'_1,\ldots,t'_p$, and so on.  In the worst case, we are driven to
abstract leaves of the term, which are either variables, that do not
need to be abstracted if they are abstraction variables, or constants.

On the contrary, we can choose in priority the smallest
possible subterms $u_i$, that are constants or variables. The ordering
constraints $t > u_i$ needed to apply the induction hypothesis, and
then to abstract the term, are easier to satisfy than in the previous
case since the $u_i$ are smaller.  
 
Between these two cases, there are a finite
but possibly big number of ways to choose the positions where terms
are abstracted.  
Anyway it is not useful to abstract the subterms, whose
ground instances are in normal form. Identifying these subterms is
made in the same way that for the study of $\mathit{TERMIN(S,u)}$ (see
Section~\ref{subsec:relaxing}).

From the point of view of the narrowing step following the abstraction, 
there is no general optimal abstracting strategy either:
the greater the term to be narrowed, 
the greater is the possible number of narrowing positions.
On another side, more general the term to be narrowed, 
greater is the possible number of narrowing substitutions for a given redex.

\subsection{How to combine the proof steps}

The previous proof steps, applied to every
reference term $t_{\mathit{ref}}=g(x_1,$ $\ldots ,x_m)$, where $x_1, \ldots ,x_m \in
{\cal X}$ and $g \in \Def$,
can be combined in the same way whatever $\mathit{S \in
\{Innermost, Outermost, Local{-}Strat\}}$:

\begin{center}
$\mathit{Strat{-}Rules(S) = repeat^*(try(abstract), \; try(narrow), \; try(stop))}$.
\end{center}
\noindent

$\mathit{"repeat^*}(T_1, \ldots, T_n)"$ repeats the strategies of the set  
$\{T_1, \ldots, T_n\}$ until it is not  possible anymore.
The operator $"try"$ is a generic operator that can be instantiated, 
following $S$, by
$\mathit{try{-}skip(T)}$, expressing that the strategy or rule $T$
is tried, and skipped when it cannot be applied, or by $\mathit{try{-}stop(T)}$, 
stopping the strategy if $T$ cannot be applied.

\subsection{The termination theorem}
For each strategy $\mathit{S \in \{Innermost, Outermost, Local{-}Strat\}}$,
we write $\mathit{SUCCESS(g,}$ $\mathit{\succ)}$ if the application of $\mathit{Strat{-}Rules(S)}$
on $(\{g(x_1, \ldots,$ $ x_m)\}, \top,\top)$  
gives a finite  proof tree, whose sets $C$ of ordering constraints
are satisfied by a same ordering $\succ$, and
whose leaves
are either states of the form $(\emptyset,A,C)$  or states
whose set of constraints $A$ is unsatisfiable.

\begin{subtheorem}
\label{theo:termin}
Let $R$ be a rewrite system on a set $\F$ of symbols containing at least a
constructor constant.  If there exists an $\F$-stable ordering $\succ$
having the subterm property, such that for each symbol $g \in \Def$,
we have $\mathit{SUCCESS(g,\succ)}$, then every term of $\TF$ terminates with
respect to the strategy $S$.

\end{subtheorem}


We are now ready to instantiate this generic proof process, according
to the different rewriting strategies.

\section{The innermost case}
\label{sec:innermost}


\subsection{Abstraction and narrowing}

When rewriting a ground instance of the current term according to the innermost 
principle, the ground instances of variables in the current term have to be
normalized before a redex appears higher in the term. So the variable
renaming performed before narrowing corresponds here to abstracting variables
in the current term.
Then, here, narrowing has only to be performed on terms of $\TFXA$. 

Moreover for the most general unifiers $\sigma$ produced during the proof process,
all variables of $Ran(\sigma)$ are abstraction variables. 
Indeed, by Definition~\ref{def:abstraction-var}, 
if $X \in Dom(\sigma)$, $\sigma X$ is in normal form, 
as well as $\theta X$ for any instantiation $\theta$.
By definition of the innermost strategy,  this requires that
variables of $\sigma X$ can only be instantiated by terms in normal
form, i.e. variables of $\sigma X$ are abstraction variables. 

Then, since before the first narrowing step, all variables are renamed into 
variables of $\XA$, and the narrowing steps only introduce variables of  $\XA$,
it is superfluous to rename the variables of the current term after the 
first narrowing step.

\subsection{Relaxing the induction hypothesis}
\label{subsec:relaxing-innermost}

To establish $\mathit{TERMIN(Innermost,u)}$, a simple narrowing test of $u$ can
first be tried. Except for the initial state, the variables of $u$ are
in $\XA$. So if $u$ is not narrowable, or if $u$ is narrowable with a 
substitution $\sigma$ that is not compatible with $A$,
then every ground instance of
$u$ is in innermost normal form.  Else, we compute the usable
rules. 

When $t$ is a  variable of $\X$, the usable
rules of $t$ are $\R$ itself.
When $t \in \XA,$ the set of  usable rules of $t$ is
empty, since the only possible instances of such a variable are ground terms 
in normal form.

\begin{subdefinition}[\thesubtheorem\ Usable rules]
Let $\R$ be a rewrite system on a set $\F$ of symbols.
Let $Rls(f) = \{l \ra r \in \R \mid root(l)~= f\}$.
For any $t \in \TFXXA$, the set of usable rules of $t$, denoted $\U(t)$,
is defined by:

\begin{itemize}
        \item $\U(t) = \R$ if $t \in \X$,
        \item $\U(t) = \emptyset$ if $t \in \XA$,
        \item $\U(f(u_1, \ldots ,u_n)) = 
Rls(f) \cup \bigcup_{i=1}^n \U(u_i) \cup \bigcup_{l \ra r \in Rls(f)} \U(r)$.
\end{itemize}
\end{subdefinition}

\begin{sublemma}\label{lemma:U-correction}
Let $\R$ be a rewrite system on a set $\F$ of symbols and $t \in \TFXXA$.
Whatever $\alpha t$ ground instance of $t$ and
$\alpha t \ra_{p_1,l_1 \ra r_1} t_1 \ra_{p_2,l_2 \ra r_2} t_2
\ra  \ldots  \ra_{p_n,l_n \ra r_n} t_n$ rewrite chain starting from $\alpha t$,
then $l_i \ra r_i \in \U(t), \;  \forall i \in [1..n]$.
\end{sublemma}

A sufficient criterion for ensuring standard
termination (and then innermost termination) of
{\ any ground instance of} a term $t$ can be given.

\begin{subproposition}\label{prop:TC}
Let $\R$ be a rewrite system on a set $\F$ of symbols, and $t$ a term of $\TFXN$.
If there exists a simplification ordering $\succ$ such that
$\forall l \ra r \in \U(t): l \succ r$, then any ground instance of $t$
is terminating. 
\end{subproposition}

\subsection{The innermost termination proof procedure}

The inference rules \rname{Abstract}, \rname{Narrow} and  \rname{Stop}
instantiate respectively the proof steps
\emph{abstract}, \emph{narrow}, and \emph{stop} defined in 
Section~\ref{sec:proof-steps}.
They are given in Table~\ref{innermost-inference}. 
Their application conditions depend on whether
satisfiability of $A$ or unsatisfiability of $A$ is checked. These conditions are
specified in Tables~\ref{conditions-satisfiability-A} and
\ref{conditions-unsatisfiability-A} respectively.

\begin{table*}
\centering
\caption{Inference rules for the innermost strategy} 
\label{innermost-inference}
\framebox[\textwidth]{
\begin{minipage}{\textwidth}
\setlength{\parindent}{0pt}
\setlength{\parskip}{2ex}


\wcinfr{~Abstract}
{ \{t\},~~A,~~C }
{ \{u\},~A \wedge  t|_{i_1} \con X_{i_1} \ldots  \wedge t|_{i_p} \con X_{i_p},
~C \wedge  H_C(t|_{i_1}) \ldots  \wedge H_C(t|_{i_p})}
{{\rm ~where~} t {\rm ~is~abstracted~into~} 
u {\rm~at~positions~} i_1, \ldots ,i_p \neq \epsilon}
{\rm ~if~ \mathit{COND{-}ABSTRACT}}

\cinfr{~Narrow}
{ \{t\},~~A,~~C }
{ \{u\},~~A \wedge \sigma, ~~ C}
{t \surred_{R}^{Inn,\sigma} u {\rm ~and~} {\rm \mathit{COND{-}NARROW}}}

\cinfr{~Stop}
{ \{t\},~~A,~~C }
{ \emptyset,~~A \wedge  H_A(t),~~C \wedge H_C(t)}
{\mathit{COND{-}STOP}}

${\rm ~and~}  H_A(t) = 
\left\{
\begin{array}{ll}
\mathit{\top}           & {\rm ~if~any~ground~instance~of~} t \\
& {\rm ~is~in~normal~form}\\
t \con X     & {\rm ~otherwise.}
\end{array}
\right.
$
\\
$ ~~~~~~~~H_C(t) = 
\left\{
\begin{array}{ll}
\mathit{\top}           & {\rm ~if~} \mathit{TERMIN(Innermost,t)}\\
t_{\mathit{ref}}>t      & {\rm ~otherwise.}
\end{array}
\right.
$

\end{minipage}}
\end{table*}

%



As said above, the ground terms whose termination is studied are
defined by the solutions of $A$.
When satisfiability of $A$ is checked at each inference step, the nodes of the proof
tree exactly  model the ground terms generated during the rewriting
derivations. Satisfiability of $A$, although undecidable in general, 
can be proved by exhibiting a ground substitution satisfying the
constraints of $A$.

When satisfiability of $A$ is not checked, nodes are generated in the
proof tree, that can represent empty sets of ground terms, so the
generated proof trees can have branches that do not represent any derivation on the
ground terms. The unsatisfiability test  of $A$ is only used  to stop
the development of meaningless branches as soon as possible, with
the sufficient conditions presented in Section~\ref{subsec:cumulating-constraints}.

\begin{table*}
\centering
\caption{Conditions for inference rules dealing with satisfiability of
$A$} 
\label{conditions-satisfiability-A}
\framebox[\textwidth]{
\begin{minipage}{\textwidth}
\setlength{\parindent}{0pt}
\setlength{\parskip}{2ex}

${\rm \mathit{~COND{-}ABSTRACT:~~}}
{ (A \wedge  t|_{i_1} \con X_{i_1} \ldots  \wedge t|_{i_p} \con X_{i_p})}\\
 {\rm ~and~} (C \wedge  H_C(t|_{i_1}) \ldots  \wedge H_C(t|_{i_p})) \;
{\rm  ~are~satisfiable}$ 

${\rm \mathit{~COND{-}NARROW:~~}}
{A \wedge \sigma} {\rm { ~is~satisfiable}}$

${\rm \mathit{~COND{-}STOP:~~}}
{ (A \wedge  H_A(t))} {\rm ~and~} (C \wedge H_C(t)) 
{\rm ~are~satisfiable}$

\end{minipage}}
\end{table*}

\begin{table*}
\centering
\caption{Conditions for inference rules dealing with unsatisfiability of $A$} 
\label{conditions-unsatisfiability-A}
\framebox[\textwidth]{
\begin{minipage}{\textwidth}
\setlength{\parindent}{0pt}
\setlength{\parskip}{2ex}

${\rm \mathit{~COND{-}ABSTRACT:~~}}
C \wedge  H_C(t|_{i_1}) \ldots  \wedge H_C(|t_{i_p}) {\rm ~is~satisfiable}$ 

${\rm \mathit{~COND{-}NARROW:~~}}
true $ 

${\rm \mathit{~COND{-}STOP:~~}}
(C \wedge H_C(t))  {\rm ~is~satisfiable~or~} {A} {\rm  ~is~unsatisfiable}. $


\end{minipage}}
\end{table*}

Once instantiated, the generic strategy $\mathit{Strat{-}Rules(S)}$ simply becomes:

\begin{center}
$\mathit{repeat*(try{-}skip(\rname{Abstract}); try{-}stop(\rname{Narrow});
try{-}skip(\rname{Stop}))}$
\end{center}
with conditions of Table~\ref{conditions-satisfiability-A}, and
\begin{center}
$\mathit{repeat*(try{-}skip(\rname{Abstract}); try{-}skip(\rname{Narrow});
try{-}skip(\rname{Stop}))}$
\end{center}
with conditions of Table~\ref{conditions-unsatisfiability-A}.
Note that \rname{Narrow} with conditions of
Table~\ref{conditions-satisfiability-A} is the only rule stopping the
proof procedure when it cannot be applied: in this case, when $A \wedge \sigma$
is satisfiable, the narrowing step can be applied, while, if
satisfiability of $A \wedge \sigma$ cannot be proved, the procedure must
stop.

The procedure can  diverge, with  infinite alternate applications
of \rname{Abstract} and \rname{Narrow}. With conditions of 
Table~\ref{conditions-satisfiability-A}, it can stop on \rname{Narrow} with
at least in a branch of the proof tree, a state of the form $(\{t\} \neq
\emptyset,A,C)$. In both cases, nothing can be said on termination.
Termination is proved when, for all proof trees,  the procedure stops with an 
application of  \rname{Stop} on each branch, generating only final
states of the form $(\emptyset,A,C)$.\\

According to the strategy $\mathit{Strat{-}Rules(Innermost)}$, testing satisfiability of $A$ in conditions of
Table~\ref{conditions-satisfiability-A}
can be optimized on the basis of the following remarks. 
In the first application of \rname{Abstract} for each initial state,
$(A \wedge  t|_{i_1} \con X_{i_1} \ldots  \wedge t|_{i_p} \con X_{i_p}) =
(\top \wedge  x_1 \con X_1 \ldots  \wedge x_m \con X_m)$, which is
always satisfiable, since the signature admits at least a constructor constant.
Moreover, the following possible current application of  \rname{Abstract}
comes after an application of \rname{Narrow}, for which it has been
checked that $A \wedge \sigma {\rm {\bf ~is~satisfiable}}$. So 
$(A \wedge \sigma \wedge  t|_{i_1} \con X_{i_1} \ldots  \wedge t|_{i_p} \con
X_{i_p})$ is also satisfiable since $X_{i_1}, \ldots, X_{i_p}$ are
fresh variables, not used in $A \wedge \sigma$.
So it is useless to verify satisfiability of
$(A \wedge  t|_{i_1} \con X_{i_1} \ldots  \wedge t|_{i_p} \con X_{i_p})$
in $\mathit{COND{-}ABSTRACT}$.

In a similar way, as \rname{Stop} is applied with a current
abstraction constraint formula $A$, which is satisfiable,
$A \wedge  t \con X$ is also satisfiable since $X$ is a
fresh variable, not used in $A$.
So it is also useless to verify that $A \wedge  t \con X$
is satisfiable in $\mathit{COND{-}STOP}$.

This leads to the conditions expressed in
Table~\ref{conditions-satisfiability-A-optimized}, simplifying those
of Table~\ref{conditions-satisfiability-A}.

\begin{table*}
\centering
\caption{Conditions for inference rules dealing with satisfiability of
$A$} 
\label{conditions-satisfiability-A-optimized}
\framebox[\textwidth]{
\begin{minipage}{\textwidth}
\setlength{\parindent}{0pt}
\setlength{\parskip}{2ex}

${\rm \mathit{~COND{-}ABSTRACT:~~}}
(C \wedge  H_C(t|_{i_1}) \ldots  \wedge H_C(t|_{i_p})) \;
{\rm  ~is~satisfiable}$ 

${\rm \mathit{~COND{-}NARROW:~~}}
A \wedge \sigma {\rm ~is~satisfiable}$

${\rm \mathit{~COND{-}STOP:~~}}
(C \wedge H_C(t)) 
{\rm ~is~satisfiable}$

\end{minipage}}
\end{table*}

\subsection{Examples}

For a better readability, when a constrained substitution $\sigma$ is 
added to the ACF $A$, we propagate the new constraint $\sigma$ into $A$ 
in applying the substitution part $\sigma_0$ of $\sigma$ to $A$.

\begin{subexample}

Let \(R\) be the previous example of Toyama.
We prove that $R$ is innermost terminating on $\TF$, where $\F =\{f\!:\!3, g\!:\!2, 0\!:\!0, 1\!:\!0\}$.

\[
  \begin{array}{ll}
    f(0,1,x) \ra f(x,x,x)\\
    g(x,y) \ra x \\
    g(x,y) \ra y\\
  \end{array}
\]

The defined symbols of $\F$ are here $f$ and $g$.
Applying the rules on $f(x_1,x_2,x_3)$, we get:

\footnotesize{
$$
\xymatrix{
\txt{
$t_{ref} = f(x_1, x_2, x_3)$\\
$A = \top$\\
$C = \top$
} \ar[d]|-{\txt{\footnotesize \rname{Abstract}}}\\
\txt{
$f(X_1,X_2,X_3)$\\
$A = (x_1 \con X_1 \wedge x_2 \con X_2 \wedge x_3 \con X_3)$\\
$C = (f(x_1,x_2,x_3) > x_1, x_2, x_3)$
}
\ar[d]^-{\txt{\footnotesize \rname{Narrow}}}_-{\sigma = (X_1 = 0 \wedge X_2 = 1)}\\
\txt{
$f(X_3,X_3,X_3)$\\
$A = (x_1 \con X_1 \wedge x_2 \con X_2 \wedge x_3 \con X_3)$\\
$C = (f(x_1,x_2,x_3) > x_1, x_2, x_3)$
}
\ar[d]|-{\txt{\footnotesize \rname{Stop}}}\\
\txt{
$\emptyset$\\
$A = (x_1 \con X_1 \wedge x_2 \con X_2 \wedge x_3 \con X_3)$\\
$C = (f(x_1,x_2,x_3) > x_1, x_2, x_3)$
}
}
$$ 
}
\normalsize

\rname{Abstract} applies since $f(x_1,x_2,x_3) > x_1,x_2,x_3$ 
is satisfiable by any simplification ordering.

If we are using the conditions for inference rules dealing with satisfiability of $A$
given in Table~\ref{conditions-satisfiability-A-optimized}, we have to justify the \rname{Narrow} application.
Here, \rname{Narrow} applies because $A \wedge \sigma= (x_1 \con 0 \wedge x_2 \con 1 \wedge x_3 \con X_3)$, where 
$\sigma = (X_1=0 \wedge X_2=1)$,
is satisfiable by any ground instantiation $\theta$ such that 
$\theta x_1 = 0$, $\theta x_2 = 1$ and $\theta x_3 = \theta X_3 = 0$.

Then \rname{Stop} applies because $f(X_3,X_3,X_3)$ is a non narrowable term whose all variables
are abstraction variables, and hence we have 
$\mathit{TER\-MIN(Innermost, f(X_3,X_3,}$ $\mathit{X_3))}$.

Considering now $g(x_1,x_2)$, we get:

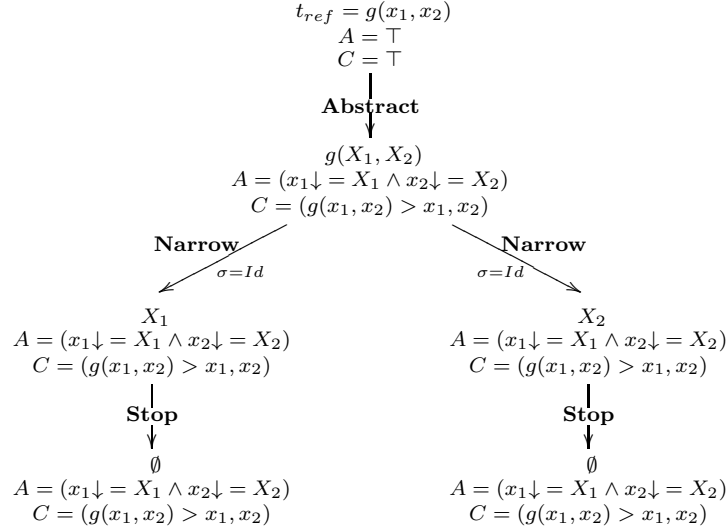
\footnotesize{
$$
\xymatrix{
& \txt{
$t_{ref} = g(x_1,x_2)$\\
$A = \top$\\
$C = \top$
}
\ar[d]|-{\txt{\footnotesize \rname{Abstract}}} 
&\\
& \txt{
$g(X_1, X_2)$\\
$A = (x_1 \con X_1 \wedge x_2 \con X_2)$\\
$C = (g(x_1,x_2) > x_1,x_2)$
} 
\ar[dl]_-{\txt{\footnotesize \rname{Narrow}}}^-{\sigma = Id}
\ar[dr]^-{\txt{\footnotesize \rname{Narrow}}}_-{\sigma = Id} & \\
\txt{}
&& \txt{}
}
$$
\vspace{-4mm}
$$
\xymatrix{
\txt{
$X_1$\\
$A = (x_1 \con X_1 \wedge x_2 \con X_2)$\\
$C = (g(x_1,x_2) > x_1,x_2)$
}
\ar[d]|-{\txt{\footnotesize \rname{Stop}}}
&&
\txt{
$X_2$\\
$A = (x_1 \con X_1 \wedge x_2 \con X_2)$\\
$C = (g(x_1,x_2) > x_1,x_2)$
}
\ar[d]|-{\txt{\footnotesize \rname{Stop}}}\\
\txt{
$\emptyset$\\
$A = (x_1 \con X_1 \wedge x_2 \con X_2)$\\
$C = (g(x_1,x_2) > x_1,x_2)$
}
&&
\txt{
$\emptyset$\\
$A = (x_1 \con X_1 \wedge x_2 \con X_2)$\\
$C = (g(x_1,x_2) > x_1,x_2)$
}
}
$$
}
\normalsize 


\rname{Abstract} applies since $g(x_1,x_2) > x_1,x_2$ 
is satisfiable by any simplification ordering.

Again, we have to justify the \rname{Narrow} application.
Here, \rname{Narrow} applies because $A \wedge \sigma = 
(x_1 \con X_1 \wedge x_2 \con X_2)$, where 
$\sigma = Id$,
is satisfiable by any ground instantiation $\theta$ such that 
$\theta x_1 = \theta X_1 = 0$ and $\theta x_2 = \theta X_2 = 0$.

Then \rname{Stop} applies on both branches because $X_1$ and $X_2$ 
are abstraction variables,
hence we trivially have $\mathit{TERMIN(Innermost, X_1)}$ and 
$\mathit{TERMIN(Innermost,}$ $\mathit{X_2)}$.

\end{subexample}


\begin{subexample}
Let us now give an example illustrating how the usable rules can be helpful and why detecting
unsatisfiability of $A$ can be important.
Let us consider the following system $\R$:
\[
\begin{array}{lll}
plus(x,0) & \ra x & (1)\\
plus(x,s(y)) & \ra s(plus(x,y)) & (2)\\
f(0,s(0),x) & \ra f(x,plus(x,x),x) & (3)\\
g(x,y) & \ra x & (4)\\
g(x,y) & \ra y & (5)
\end{array}
\]
Let us first remark that $\R$ is not terminating, as illustrated by
the following cycle, where successive redexes are underlined:\\

\footnotesize{
$
\begin{array}{ll}
\underline{f(0,s(0),g(0,s(0)))}
& \ra^{(3)} f(\underline{g(0,s(0))}, plus(g(0,s(0)),g(0,s(0))), g(0,s(0)))\\
& \ra^{(4)} f(0, plus(\underline{g(0,s(0))},g(0,s(0))), g(0,s(0)))\\
& \ra^{(5)} f(0, plus(s(0),\underline{g(0,s(0))}), g(0,s(0)))\\
& \ra^{(4)} f(0, \underline{plus(s(0),0)}, g(0,s(0)))\\
& \ra^{(1)} \underline{f(0, s(0), g(0,s(0)))}\\
& \ra^{(3)} \ldots
\end{array}
$
}
\normalsize

Let us prove the innermost termination of $\R$ on $\TF$,
where $\F = \{0\!:\!0, s\!:\!1, plus\!:\!2, g\!:\!2, f\!:\!3\}$.
The defined symbols of $\F$ are $f, plus$ and $g$.

Let us apply the inference rules checking unsatisfiability of $A$,
whose conditions are given in Table~\ref{conditions-unsatisfiability-A}.
Applying the rules on $f(x_1,x_2,x_3)$, we get:

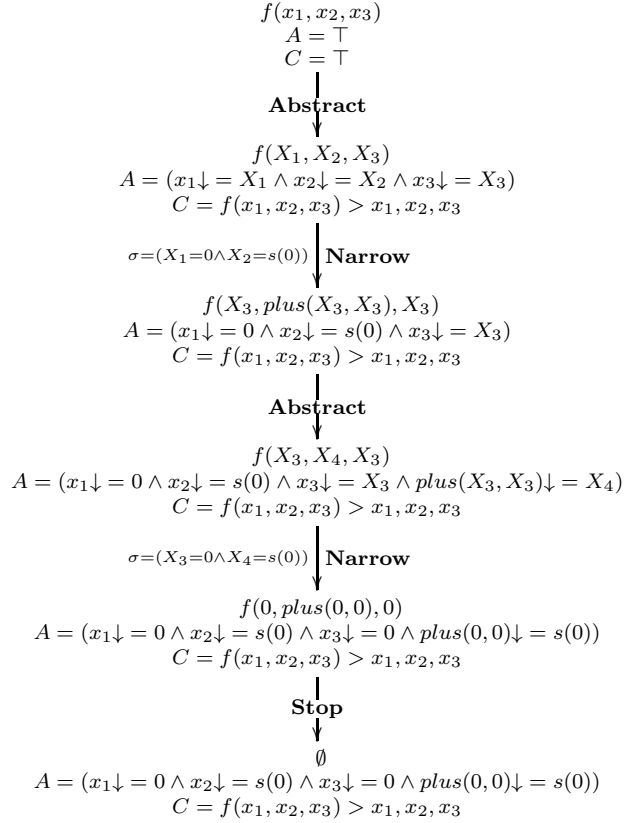
\footnotesize{
$$
\xymatrix{
\txt{
$f(x_1,x_2,x_3)$\\
$A = \top$\\
$C = \top$
}
\ar[d]|-{\txt{\footnotesize \rname{Abstract}}}
\\
\txt{
$f(X_1,X_2,X_3)$\\
$A = (x_1 \con X_1 \wedge x_2 \con X_2 \wedge x_3 \con X_3)$\\
$C = f(x_1,x_2,x_3) > x_1,x_2,x_3$
}
\ar[d]^-{\txt{\footnotesize \rname{Narrow}}}_-{\sigma = (X_1=0 \wedge X_2 = s(0))}
\\
\txt{
$f(X_3,plus(X_3,X_3),X_3)$\\
$A = (x_1 \con 0 \wedge x_2 \con s(0) \wedge x_3 \con X_3)$\\
$C = f(x_1,x_2,x_3) > x_1,x_2,x_3$
}
\ar[d]|-{\txt{\footnotesize \rname{Abstract}}}
\\
\txt{
$f(X_3,X_4,X_3)$\\
$A = (x_1 \con 0 \wedge x_2 \con s(0) \wedge x_3 \con X_3 \wedge plus(X_3,X_3) \con X_4)$\\
$C = f(x_1,x_2,x_3) > x_1,x_2,x_3$
}
\ar[d]^-{\txt{\footnotesize \rname{Narrow}}}_-{\sigma = (X_3=0 \wedge X_4 = s(0))}
\\
\txt{
$f(0,plus(0,0),0)$\\
$A = (x_1 \con 0 \wedge x_2 \con s(0) \wedge x_3 \con 0 \wedge plus(0,0) \con s(0))$\\
$C = f(x_1,x_2,x_3) > x_1,x_2,x_3$
}
\ar[d]|-{\txt{\footnotesize \rname{Stop}}}
\\
\txt{
$\emptyset$\\
$A = (x_1 \con 0 \wedge x_2 \con s(0) \wedge x_3 \con 0 \wedge plus(0,0) \con s(0))$\\
$C = f(x_1,x_2,x_3) > x_1,x_2,x_3$
}
}
$$
}
\normalsize

The first \rname{Abstract} applies since $f(x_1,x_2,x_3) > x_1,x_2,x_3$ 
is satisfiable by any simplification ordering.

Since we are using the inference rules checking unsatisfiability of $A$
given in Table~\ref{conditions-unsatisfiability-A}, we do not have to justify
the \rname{Narrow} applications.

The second \rname{Abstract} applies by using the $\mathit{TERMIN}$ predicate.
Indeed, the usable rules of $plus(X_3,X_3)$ consist of the system
$\{plus(x,0)$ $ \ra x, plus(x,s(y)) \ra s(plus(x,y))\}$, that can be
proved terminating with any precedence based ordering,
independent of the induction ordering, with the
precedence $plus \succ_{\F} s$, which ensures the property
$\mathit{TERMIN(Innermost, plus(X_3,X_3))}$.
Without abstraction here, the process would have generated a branch 
containing an infinite number of \rname{Narrow} applications.

Finally, \rname{Stop} applies because the constraint $A$ becomes unsatisfiable.
Indeed, it contains the abstraction constraint $plus(0,0) \con s(0)$, which is
not true since the unique normal form of $plus(0,0)$ is $0$.
Note that if we would have chosen to apply the inference rules checking  
satisfiability of $A$, whose conditions are given in 
Table~\ref{conditions-satisfiability-A-optimized}, 
then the last narrowing step would not have applied, and would have been 
replaced by a \rname{Stop} application.

Considering now $g(x_1,x_2)$, we get:

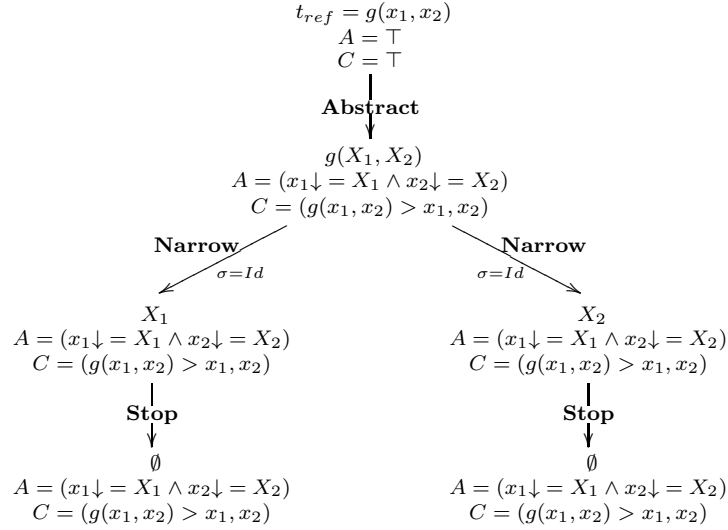
\footnotesize{
$$
\xymatrix{
& \txt{
$t_{ref} = g(x_1,x_2)$\\
$A = \top$\\
$C = \top$
}
\ar[d]|-{\txt{\footnotesize \rname{Abstract}}} 
&\\
& \txt{
$g(X_1, X_2)$\\
$A = (x_1 \con X_1 \wedge x_2 \con X_2)$\\
$C = (g(x_1,x_2) > x_1,x_2)$
} 
\ar[dl]_-{\txt{\footnotesize \rname{Narrow}}}^-{\sigma = Id}
\ar[dr]^-{\txt{\footnotesize \rname{Narrow}}}_-{\sigma = Id} & \\
\txt{}
&& \txt{}
}
$$ 
\vspace{-4mm}
$$
\xymatrix{
\txt{
$X_1$\\
$A = (x_1 \con X_1 \wedge x_2 \con X_2)$\\
$C = (g(x_1,x_2) > x_1,x_2)$
}
\ar[d]|-{\txt{\footnotesize \rname{Stop}}}
&&
\txt{
$X_2$\\
$A = (x_1 \con X_1 \wedge x_2 \con X_2)$\\
$C = (g(x_1,x_2) > x_1,x_2)$
}
\ar[d]|-{\txt{\footnotesize \rname{Stop}}}\\
\txt{
$\emptyset$\\
$A = (x_1 \con X_1 \wedge x_2 \con X_2)$\\
$C = (g(x_1,x_2) > x_1,x_2)$
}
&&
\txt{
$\emptyset$\\
$A = (x_1 \con X_1 \wedge x_2 \con X_2)$\\
$C = (g(x_1,x_2) > x_1,x_2)$
}
}
$$
}
\normalsize

\rname{Abstract} applies since $g(x_1,x_2) > x_1,x_2$ 
is satisfiable by the previous precedence based ordering.
\rname{Stop} applies on both branches because $X_1$ and $X_2$ 
are abstraction variables,
so we trivially have $\mathit{TERMIN(Innermost, X_1)}$ and 
$\mathit{TERMIN(Innermost,}$ $\mathit{X_2)}$.

Let us finally apply the inference rules of Table~\ref{conditions-unsatisfiability-A}
on $plus(x_1,x_2)$:

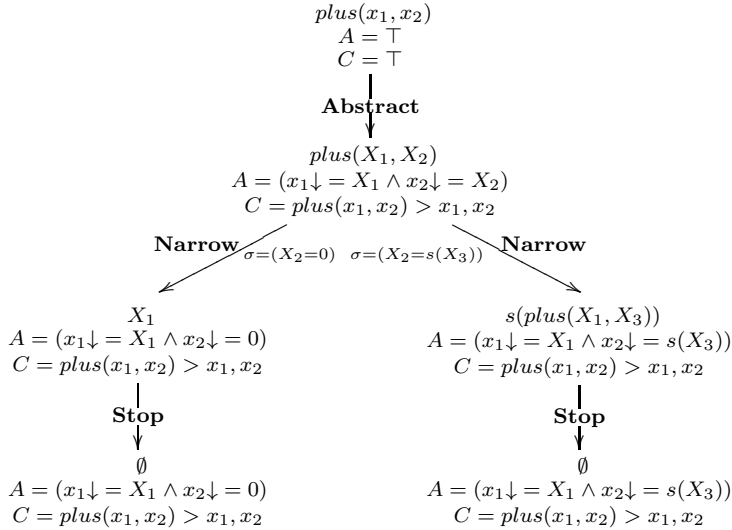
\footnotesize{
$$
\xymatrix{
&
\txt{
$plus(x_1,x_2)$\\
$A = \top$\\
$C = \top$
}
\ar[d]|-{\txt{\footnotesize \rname{Abstract}}}
&
\\
&
\txt{
$plus(X_1,X_2)$\\
$A = (x_1 \con X_1 \wedge x_2 \con X_2)$\\
$C = plus(x_1,x_2) > x_1,x_2$
}
\ar[dl]_-{\txt{\footnotesize \rname{Narrow}}}^{\sigma = (X_2=0)}
\ar[dr]^-{\txt{\footnotesize \rname{Narrow}}}_{\sigma = (X_2=s(X_3))~~~~}
&
\\
\txt{}
&& \txt{}
}
$$
\vspace{-4mm}
$$
\xymatrix{
\txt{
$X_1$\\
$A = (x_1 \con X_1 \wedge x_2 \con 0)$\\
$C = plus(x_1,x_2) > x_1,x_2$
}
\ar[d]|-{\txt{\footnotesize \rname{Stop}}}
&&
\txt{
$s(plus(X_1,X_3))$\\
$A = (x_1 \con X_1 \wedge x_2 \con s(X_3))$\\
$C = plus(x_1,x_2) > x_1,x_2$
}
\ar[d]|-{\txt{\footnotesize \rname{Stop}}}
\\
\txt{
$\emptyset$\\
$A = (x_1 \con X_1 \wedge x_2 \con 0)$\\
$C = plus(x_1,x_2) > x_1,x_2$
}
&&
\txt{
$\emptyset$\\
$A = (x_1 \con X_1 \wedge x_2 \con s(X_3))$\\
$C = plus(x_1,x_2) > x_1,x_2$
}
}
$$
}
\normalsize

\rname{Abstract} applies since $g(x_1,x_2) > x_1,x_2$ 
is satisfiable by the previous precedence based ordering.
\rname{Stop} applies on the left branch because $X_1$ is an abstraction variable,
hence we trivially have $\mathit{TERMIN(Innermost, X_1)}$.
\rname{Stop} applies on the right branch by using the $\mathit{TER\-MIN}$ predicate.
Indeed, the usable rules of $s(plus(X_1,X_3))$ consist of the previous 
terminating system
$\{plus$ $(x,0) \ra x, plus(x,s(y)) \ra s(plus(x,y))\}$.

\end{subexample}

\section{The outermost case}
\label{sec:outermost}


\subsection{Abstraction}
\label{subsec:abstraction}


According to the outermost strategy, abstraction can be performed on
subterms $t_i$  only if
during their normalization, the $t_i$ do not introduce outermost
redexes higher in the term $t$.
More formally, the induction hypothesis is applied to the subterms
$t|_{p_1}, \ldots ,t|_{p_n}$ of the current term $t$, provided 
$\alpha t_{\mathit{ref}} \succ \alpha t|_{p_1}, \ldots ,\alpha t|_{p_n}$ 
for every ground substitution $\alpha$,
for the induction ordering $\succ$
and provided $u = t[y_1]_{p_1}$ $ \ldots [y_n]_{p_n}$ is not narrowable 
at prefix positions of $p_1,\ldots, p_n$, for 
the outermost narrowing relation defined below. 

As already mentioned in Section~\ref{subsec:relaxing}, if in addition, 
the variables of 
$u$ are all in $\XA$, and $u$ is not narrowable, then every ground
instance of the term $u$ outermost terminates.

\subsection{The narrowing mechanism}

Outermost narrowing is defined by Definition~\ref{def:narrowing},
where a $S$-better position is a prefix position.
In order to support intuition, let us consider for instance the
system $\{f(g(a))\ra a, f(f(x)) \ra b, g(x) \ra f(g(x))\}$.
With the standard narrowing relation used at the outermost position, 
$f(g(x_1))$ only narrows into $a$ with the first rule and the
substitution $\sigma=(x_1=a)$. 
With the outermost narrowing relation,
$f(g(x_1))$  narrows into $a$ with the first rule and $\sigma=
(x_1=a)$, and into $f(f(g(x_2)))$ with the third rule and  
the constrained substitution
$\sigma=(x_1=x_2 \wedge x_2 \neq a)$. 

The variables of the narrowed terms are in $\X \cup \XA$: 
as we will see, renaming variables 
of $\X$ still gives variables of $\X$, and abstraction,
replacing subterms by variables of $\XA$, may not cover all 
variables of $\X$ in the abstracted term.

In the outermost termination proof, 
the variable renaming performed before the narrowing step
has a crucial meaning for the schematization of outermost derivations.
This renaming, applied on the current term $g(x_1, \ldots ,x_m)$,
replaces the variable occurrences $x_1,\ldots,x_m$ by new and all 
different variables $x'_1,\ldots,$ $x'_m$, defined as follows.
Given any ground instance $\alpha g(x_1, \ldots ,x_m)$ of
$g(x_1, \ldots ,$ $x_m)$, the  $x'_1,\ldots,x'_m$ represent
the first reduced form of $\alpha x_1,\ldots, \alpha x_m$ 
generating an outermost  reduction higher 
in the  term (here, at the top),
in any outermost rewriting chain starting from 
$\alpha g(x_1, \ldots ,x_m)$. 
This replacement is memorized in a reduction formula before applying  
a step of outermost narrowing to $g(x'_1,$ $\ldots ,x'_m)$.
The abstraction variables are not renamed: since their ground instances
are in normal form, they are not concerned by the rewriting chain schematized by
the variable renaming.
 
Formally, the definition of the variable replacement
performed before a narrowing step is the following.

\begin{subdefinition}
\label{def:renaming-red}
Let $t \in \TXF$ be a term whose variable occurrences from left to
right in $t$ are $x_1,\ldots,x_m$.
The reduction renaming of $t$, noted $\rho = (x_1 \red x'_1)...(x_m \red x'_m)$,
consists in replacing the $x_i$ by new and all different variables $x'_i$ in
$t$, giving a term $t^\rho$. This is denoted by the so-called reduction formula
\begin{center}
$ R(t)= t \red t^\rho$.
\end{center}
\end{subdefinition}

Notice that the reduction renaming linearizes the term. For instance, 
the two occurrences of $x$ in $g(x,x)$ are respectively renamed into 
$x'_1$ and $x'_2$, and $g(x,x) \red g(x'_1, x'_2)$.

\begin{subdefinition}
\label{sat-red-formula}
Let $t \in \TXF$ be a term whose variable occurrences from left to
right are $x_1,\ldots,x_m$,
at positions $p_1,\ldots,p_m$ respectively.
A ground substitution $\theta$ satisfies the reduction formula
$ R(t)=  t \red t^\rho$, where $\rho = (x_1 \red x'_1)...(x_m \red x'_m)$,
iff there exists an outermost rewriting chain 
$\theta t \ra^{* out}_{p \not \in \OS(t)} \theta t^\rho 
\ra^{out}_{p \in \OS(t)} u$, i.e. such that:
\begin{itemize}
\item either $t [\theta x'_1]_{p_1} \ldots [\theta x'_m]_{p_m}$ 
is the first reduced form of 
$\theta t = t [\theta x_1]_{p_1}$ $ \ldots [\theta x_m]_{p_m}$
on this chain 
having an outermost rewriting position
at a non variable position of $t$, if this position exists, 

\item or
$\theta x'_1 = (\theta x_1\da), \ldots, \theta x'_m = (\theta x_m\da)$
if there is no such position.
\end{itemize}

\end{subdefinition}

Before going on, a few remarks on this definition can be made.
In the second case of satisfiability, $t[\theta x_1 \da]_1 \ldots
[\theta x_m \da]_m$ is in  normal form.
In any case, $ R(t)$ is always satisfiable~:
it is sufficient to take a ground substitution $\theta$ such that
$t[\theta x_1]_{p_1} \ldots [\theta x_m]_{p_m}$ has an outermost
rewriting position at a non variable position of $t$,
and then to extend its domain $\{x_1, \ldots, x_m\}$ to
$\{x_1, \ldots, x_m, x'_1, \ldots, x'_m\}$ by choosing
for each $i \in \{1,...,m\}$, $\theta x'_i = \theta x_i$.
If such a substitution does not exist, then every ground instance
of $t$ has no outermost rewriting position at a non variable
position of $t$, and it is sufficient to take a ground substitution
$\theta$ such that
$\theta x_1 = \ldots = \theta x_m = \theta x'_1 = \ldots = \theta x'_m =
u$, with $u$ any ground term in normal form.

However, there may exist several instantiations solution of such constraints.
Let us consider for instance the rewrite system 
$R= \{f(a) \ra f(c), b \ra a\}$ and the reduction formula
$R(f(x)) =  f(x) \red f(x')$. The substitution
$\theta_1(x) = \theta_1(x') = a$ and  
$\theta_2(x) = b, \theta_2(x') = a$ are two distinct solutions.
With the substitution $\theta_2$, $f(a)$ is the first reduced form 
of $f(b)$  having an outermost rewriting position at a non variable
position of $f(x)$ (here at top).


Notice also that if $t$ is outermost reducible at position $p$, 
variables of $t$ whose position is a suffix of $p$ are not affected 
by the reduction renaming.

Indeed, if $t$ is reducible at position $p$, a ground instance $\alpha t$ of $t$
cannot be outermost reduced in the instance of $x$, whose positions are suffix 
of $p$.
So $x'$, representing the first reduced form of $\alpha x$
in any outermost rewriting chain starting from 
$\alpha t$, such that the reduction is performed higher 
in the current term, is equal to $x$.




To illustrate this, let us consider the system $\{g(x) \ra x, f(x,x) \ra x\}$
(the right-hand sides of the rules are not important here).
Then, since $f(x,g(y))$ outermost rewrites at the position of $g$,
the variable $y$ does not need to be renamed.
So $R(f(x,g(y))) = (f(x,g(y)) \red f(x',g(y)))$.



Because of the previously defined renaming process, the formula $A$ for 
cumulating constraints has to be completed in the following way.

\begin{subdefinition}
A {\em renaming-abstraction constraint formula} (RACF for short) is a 
formula \\
$\bigwedge_m u_m \red u_m^\rho
\bigwedge_i (t_i \con t'_i) \wedge \bigwedge_j (x_j = t_j) \wedge 
\bigwedge_k \bigvee_{l_k} (u_{l_k} \neq v_{l_k})$, 
where
$u_m, u_m^\rho, t_i, t'_i, t_j,$ $ u_{l_k}, v_{l_k} \in \TFXXA$, $x_j
\in \X \cup \XA$.
The empty formula is denoted  $\top$.
\end{subdefinition}

\begin{subdefinition}
A  renaming-abstraction constraint formula \\
$\bigwedge_m u_m \red u_m^\rho
\bigwedge_i (t_i \con t'_i) \wedge \bigwedge_j (x_j = t_j) \wedge 
\bigwedge_k \bigvee_{l_k} (u_{l_k} \neq v_{l_k})$
is said to be {\em satisfiable} iff there
exists at least one instantiation $\theta$ such that 
$\bigwedge_i (\theta t_i \con \theta t'_i)
 \wedge \bigwedge_j (\theta x_j = \theta t_j) 
\wedge \bigwedge_k \bigvee_{l_k} (\theta u_{l_k} \neq \theta v_{l_k})$
and $\theta$ satisfies 
$\bigwedge_m u_m \red u_m^\rho$.
\end{subdefinition}

In practice, one can solve the equality and disequality part of the constraint 
and then check whether the solution $\theta$ satisfies the reduction
formulas.
This is trivial when $\theta$ only instantiates the $x'_i$, since it
can be extended by setting $\theta(x_i) = \theta(x'_i)$. Unfortunately, when 
$\theta$ also instantiates the $x_i$, we get the undecidable problem of 
reachability: given two 
ground terms $t$ and $t'$, can $t$ be transformed into $t'$ by repeated 
application of a given set of rewriting rules?

So here again, we can either test satisfiability of the formula of
cumulated constraints, or unsatisfiability.  
As satisfiability is in general more difficult to show than in the 
innermost case, we only present here inference rules checking unsatisfiability.

\OUT{
{\bf Est-ce que ca peut servir commme dans le cas innermost, ou on
decide de la supprimer?}
Je propose de supprimer.

To Definition~\ref{def:constraint-problem} corresponds the following
definition, specific to the outermost case.

\begin{definition}
Let $t,u_1,\ldots,u_m \in \TXF$ and $\Sigma$ an SCF. The  constraint 
$(t>u_1,\ldots,u_m) \sur \Sigma$ is said to be satisfiable iff 
there exists an $\F$-stable ordering $\succ$ on $\TF$ having the subterm property
such that $\theta t \succ \theta u_i, i \in [1..m]$,
for every ground substitution 
$\theta$ satisfying $\Sigma$, and whose domain contains the variables of $t$ and of 
the $u_i, i \in [1..m]$.
Such an ordering $\succ$ is said to satisfy $(t>u_1,\ldots,u_m) \sur \Sigma$.
\end{definition}

The satisfiability of  $(t>u_1,\ldots,u_m) \sur \Sigma$ is undecidable.
But here again, a  sufficient condition for an
ordering $\succ$ to satisfy this constraint is that $\succ$ is stable under
substitution (the induction ordering is then a simplification
ordering) and $t \succ u_1,\ldots,u_m $.
Remark that consequently, $(t>u_1,\ldots,u_m) \sur \Sigma$ may be proved 
satisfiable, even if $\Sigma$ is not.
}

\subsection{Inference rules for the outermost case}

The inference rules \rname{Abstract}, \rname{Narrow} and \rname{Stop}
instantiate respectively the proof steps \emph{abstract}, \emph{narrow}, and 
\emph{stop}.

They  work as follows: 

\begin{itemize}
\item
The narrowing step is expressed by a rule \rname{Narrow} applying on 
$(\{t\}, A, C)$:
the variables of $t$ are renamed as specified in Definition~\ref{def:renaming-red}.
Then $t^\rho$ is outermost narrowed in all possible ways in one step, with all 
possible rewrite rules of the rewrite system $\R$, 
into terms $u$. For any possible $u$, we generate the state
$(\{u\}, R(t) \wedge A \wedge \sigma, C)$
where $\sigma$ is the constrained substitution allowing outermost narrowing 
of $t^\rho$ into $u$.

\item
The rule \rname{Abstract} works as in the innermost case, except that the 
abstraction positions are such that the abstracted term is not narrowable 
at prefix positions of the abstraction positions.

 \item
The rule \rname{Stop} also works as in the innermost case.
\end{itemize}

\begin{table*}
\centering
\caption{Inference rules for the outermost strategy} 
\label{outermost-inference}
\framebox[\textwidth]{
\begin{minipage}{\textwidth}
\setlength{\parindent}{0pt}
\setlength{\parskip}{2ex}


\dubblewcinfr{~Abstract}
{ \{t\},~~A,~~C }
{ \{u\},~A \wedge  t|_{i_1} \con X_{i_1} \ldots  \wedge t|_{i_p} \con X_{i_p},
~C \wedge  H_C(t|_{i_1}) \ldots  \wedge H_C(t|_{i_p})}
{{\rm ~where~} t {\rm ~is~abstracted~into~} 
u {\rm~at~positions~} i_1, \ldots ,i_p \neq \epsilon}
{{\rm ~if~} C \wedge  H_C(t|_{i_1}) \ldots  \wedge H_C(|t_{i_p}) {\rm
~is~satisfiable}} 
{{\rm ~and~} u {\rm ~is~not~narrowable~at~prefix~positions~of~} i_1, \ldots ,i_p }

\cinfr{~Narrow}
{ \{t\},~~A,~~C }
{ \{u\},~~R(t) \wedge A \wedge \sigma , ~~ C}
{t^\rho \surred_{R}^{Out,\sigma} u }

\cinfr{~Stop}
{ \{t\},~~A,~~C }
{ \emptyset,~~A \wedge H_A(t),~~C \wedge H_C(t)}
{C \wedge H_C(t) {\rm ~is~satisfiable~or~} {A} {\rm  ~is~unsatisfiable}}

${\rm ~and~}  H_A(t) = 
\left\{
\begin{array}{ll}
\mathit{\top}           & {\rm ~if~any~ground~instance~of~} t \\
& {\rm ~is~in~normal~form}\\
t \con X     & {\rm ~otherwise.}
\end{array}
\right.
$
\\ 
$ ~~~~~~~~H_C(t) = 
\left\{
\begin{array}{ll}
\mathit{\top}           & {\rm ~if~} \mathit{TERMIN(Outermost,t)}\\
t_{\mathit{ref}}>t      & {\rm ~otherwise.}
\end{array}
\right.
$

\end{minipage}}
\end{table*}

To prove outermost termination of $\R$
on every term $t \in \TF$, for each defined symbol
$g \in \Def$, we apply the rules on the initial state
$(\{t_{\mathit{ref}}=g(x_1, \ldots , x_m)\}, \top, \top)$, with the strategy:
\begin{center}
$\mathit{Strat{-}Rules(Outermost) = repeat*(try{-}skip(\rname{Abstract}); try{-}skip(\rname{Narrow});}$ 
$\mathit{try{-}skip(\rname{Stop}))}$.
\end{center}

There are three cases for the behavior of the strategy: either
there is a branch in the proof tree with  infinite applications of
\rname{Abstract} and \rname{Narrow}, in which case we cannot say anything 
about termination, or the procedure stops on each branch with the rule 
\rname{Stop}. Then,
outermost termination is established, if all proof trees are finite.

According to the remark following Definition~\ref{sat-red-formula},
the reduction formulas in $A$ may often be reduced to simple variable renamings.
In this case, $A$ only contains variable renamings and constrained 
substitutions, that can be used to show that the ordering constraint
needed to apply \rname{Abstract} or \rname{Stop} is satisfiable 
(see Examples B.1 and B.4 in ~\cite{FGK-out-interne-2002}).
The following lemma can also be used, if satisfiability of $C$ 
is considered with Definition~\ref{def:constraint-problem}
(see Examples B.2, B.3 and B.4 in ~\cite{FGK-out-interne-2002}).

\begin{sublemma}
\label{lemme:var}
Let $(\{t_i\}, A_i, C_i)$ be the $i^{th}$ state of any
branch of the derivation tree obtained by applying
the strategy $S$ on $(\{t_{\mathit{ref}}\}, \top, \top)$,
and $\succ$ an $\F$-stable ordering having the subterm property.
If every reduction formula in $A_i$ can be reduced to a formula
$\bigwedge_j x_j = x'_j$, then
we have:
\begin{center}
for all variable $x$ of $t_i$ in $\X$:
$(t_{\mathit{ref}} > x)/A_i$ is satisfiable by $\succ$.
\end{center}
\end{sublemma}

\subsection{Examples}

\begin{subexample}

Consider the previous example $\R= \{f(g(a)) \ra a,  f(f(x)) \ra b ,
g(x)$ $\ra f(g(x))\}$, that is outermost
terminating, but not terminating for the standard rewriting relation.
We prove that $\R$ is outermost terminating on $\TF$ where $\F =\{f:1, g:1, a:0, b:0\}$.

The defined symbols of $\F$ for $\R$ are $f$ and $g$.
Applying the rules on $f(x_1)$, we get:

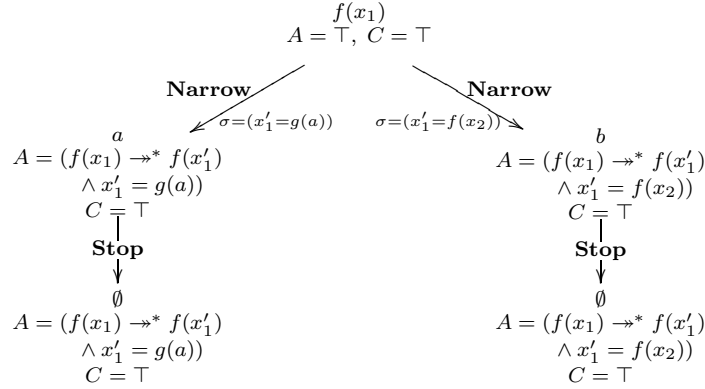
\footnotesize{
$$
\xymatrix{
& *{\txt{$f(x_1)$\\$A = \top, \; C=\top$\\$$}} \ar_{\rname{Narrow}}[dl]^{\sigma = (x'_1 = g(a))} \ar^{\rname{Narrow}}[dr]_{\sigma = (x'_1 = f(x_2))} &\\
*{\txt{$a$\\$A = (f(x_1) \red f(x'_1)$\\$~~~~~~ \wedge x'_1 = g(a))$\\$C=\top$}} \ar|-{\rname{Stop}}[d]
&& *{\txt{$b$\\$A = (f(x_1) \red f(x'_1)$\\$~~~~~~ \wedge x'_1 = f(x_2))$\\$C=\top$}} \ar|-{\rname{Stop}}[d]\\
*{\txt{$ $\\$\emptyset$\\$A = (f(x_1) \red f(x'_1)$\\$~~~~~~ \wedge x'_1 = g(a))$\\$C=\top$}} &&
*{\txt{$ $\\$\emptyset$\\$A = (f(x_1) \red f(x'_1)$\\$~~~~~~ \wedge x'_1 = f(x_2))$\\$C=\top$}}\\
}
$$
}
\normalsize

The first \rname{Stop} is applied because $a$ is in normal form,
the second \rname{Stop} because  $b$ is in normal form. 
Applying the rules on $g(x_1)$, we get:

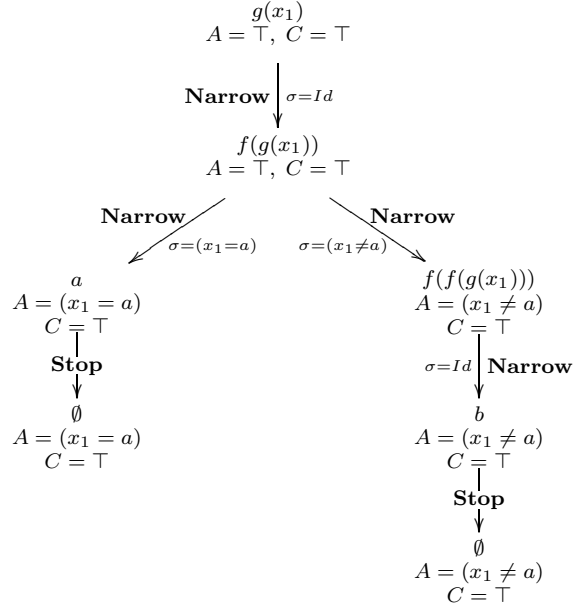
\gtree{
\[
\begin{array}{lll}
\epsilon & ~~~g(x_1)   & ~~~A = \top & ~~~C=\top \\

\rname{Narrow}  & &\\
1 & ~~~f(g(x'_1))       & ~~~A = \top (g(x_1) \red g(x'_1))\\
&       & ~~~(\sigma = (x' = x'_1)) \\

\rname{Narrow}  & &\\
1.1 & ~~~a      & ~~~A = (x_1 \red x'_1)[g(x_1)] \wedge x'_1 \red x'_2)[f(g(x'_1))] \wedge x'_2 = a)\\
& & ~~~(\sigma = (x'_2 = a))\\
1.2 & ~~~f(f(g(x'_2)))  & ~~~A = (x_1 \red x'_1)[g(x_1)] \wedge x'_1 \red x'_2)[f(g(x'_1))] \wedge x'_2 \neq a)\\
&& ~~~(\sigma = (x'' = x'_2) \mbox{ satisfies } (x'_2 \neq a)) \\

\rname{Abstract}        & &\\
1.1.1 & ~~~\emptyset    & ~~~A =\\$A = \top$ (x_1 \red x'_1)[g(x_1)] \wedge x'_1 \red x'_2)[f(g(x'_1))] \wedge x'_2 = a)\\
1.2 & ~~~f(f(g(x_3)))   & ~~~A = (x_1 \red x_2)[g(x_1)] \wedge x_2 \red x_3)[f(g(x_2))] \wedge x_3 \neq a)\\

\rname{Narrow}  & &\\
1.1.1 & ~~~\emptyset    & ~~~A = (x_1 \red x_2)[g(x_1)] \wedge x_2 \red x_3)[f(g(x_2))] \wedge x_3 = a)\\
1.2.1 & ~~~b            & ~~~A = (x_1 \red x_2)[g(x_1)] \wedge x_2 \red x_3)[f(g(x_2))] \wedge x_3 \red x_4)[f(f(g(x_3)))] \wedge x_3 \neq a)\\
&& ~~~(\sigma = (x''' = g(x_4))\\

\rname{Abstract}        & &\\
1.1.1 & ~~~\emptyset    & ~~~A = (x_1 \red x_2)[g(x_1)] \wedge x_2 \red x_3)[f(g(x_2))] \wedge x_3 = a)\\
1.2.1.1 &~~~\emptyset   & ~~~A = (x_1 \red x_2)[g(x_1)] \wedge x_2 \red x_3)[f(g(x_2))] \wedge x_3 \red x_4)[f(f(g(x_3)))] \wedge x_3 \neq a)\\

\end{array}
\]
}

\footnotesize{
\begin{center}
{
$$
\xymatrix{
& *{\txt{$ $ \\$g(x_1)$\\$A = \top, \; C = \top$\\$ $}} 
\ar_{\rname{Narrow}}[d]^{\sigma = Id}& \\
& *{\txt{$ $\\$f(g(x_1))$\\$A = \top, \; C = \top$\\$ $}} 
\ar_{\rname{Narrow}}[dl]^{\sigma = (x_1 = a)} 
\ar^{\rname{Narrow}}[dr]_{\sigma = (x_1 \neq a)} 
&\\
*{\txt{$ $ \\$a$\\$A = (x_1 = a)$\\$C = \top$}} 
\ar|-{\rname{Stop}}[d] 
&& *{\txt{$ $ \\$f(f(g(x_1)))$\\$A = (x_1 \neq a)$\\$C = \top$}} 
\ar^{\rname{Narrow}}[d]_{\sigma = Id}\\
*{\txt{$ $\\$\emptyset$\\$A = (x_1 = a)$\\$C = \top$}} 
&& *{\txt{$ $\\$b$\\$A = (x_1 \neq a)$\\$C = \top$}} \ar|-{\rname{Stop}}[d]\\
&& *{\txt{$ $\\$\emptyset$\\$A = (x_1 \neq a)$\\$C = \top$}}
}
$$
}
\end{center}
}
\normalsize

There is no reduction renaming before the \rname{Narrow} steps,
since $g(x_1)$, $f(g(x_1))$ and $f(f(g(x_1)))$ are reducible at prefix positions 
of the position of $x_1$. 

When narrowing $f(g(x_1))$, we first try the top position, 
and find a possible unification with the first rule (the left
branch).
One also must consider 
the third rule if $x_1$ is such that $x_1 \neq a$ (second branch).
\rname{Stop} is applied on $a$ and $b$ 
as previously.

\end{subexample}

\begin{subexample}

Let $\R$ be the  rewrite system cited in the introduction, built 
on $\F = \{cons:2, inf:1, big:0\}$ :
\[
\begin{array}{ll}
cons(x,cons(y,z))       & \ra big\\
inf(x)                  & \ra cons(x, inf(s(x)))
\end{array}
\]

Applying the inference rules on $inf(x_1)$, we get :

\footnotesize{
$$
\xymatrix{
*{\txt{$inf(x_1)$\\$A = \top, \; C = \top$\\$ $}} 
\ar[d]^{\rname{Narrow}}_{\sigma = Id}\\
*{\txt{$ $\\$cons(x_1, inf(s(x_1)))$\\$A = \top, \;C = \top$\\$ $}} 
\ar[d]^{\rname{Narrow}}_{\sigma = Id}\\
*{\txt{$ $\\$cons(x'_1,cons(s(x_1),inf(s(s(x_1)))))$\\
$A = (cons(x_1,inf(s(x_1))) \red cons(x'_1,inf(s(x_1))))$\\$C = \top$\\$ $}} 
\ar[d]^{\rname{Narrow}}_{\sigma = Id}\\
*{\txt{$ $\\$big$\\$A = (cons(x_1,inf(s(x_1))) \red cons(x'_1,inf(s(x_1))))$\\$C = \top$\\$ $}} 
\ar|-{\rname{Stop}}[d]\\
*{\txt{$ $\\$\emptyset$\\
$A = (cons(x_1,inf(s(x_1))) \red cons(x'_1,inf(s(x_1))))$\\
$C = \top$\\$ $}}\\
}
$$
}
\normalsize

Applying the inference rules on $cons(x_1,x_2)$, we get~:

\footnotesize{
$$
\xymatrix{
*{\txt{$cons(x_1,x_2)$\\$A = \top, \; C = \top$\\$ $}} 
\ar^{\rname{Narrow}}[d]_{\sigma = (x'_2 = cons(x_3,x_4))}\\
*{\txt{$ $\\$big$\\$A = (cons(x_1,x_2) \red cons(x'_1,x'_2))$\\
$ \wedge~ x'_2 = cons(x_3, x_4))$\\$C = \top$}} 
\ar|-{\rname{Stop}}[d]\\
*{\txt{$ $\\$\emptyset$\\$A = (cons(x_1,x_2) \red cons(x'_1,x'_2))$\\
$ \wedge~ x'_2 = cons(x_3, x_4))$\\$C = \top$}}
}
$$
}
\normalsize

\end{subexample}

Other examples can be found in~\cite{FGK-out-interne-2002}.

\section{Local strategies on operators}
\label{sec:local-strat}


We now address the termination problem for rewriting with local
strategies on operators.

\subsection{Abstraction and narrowing}



The information that variables are abstraction
variables can be very important to conclude the proofs here: 
if the current term is an
abstraction variable, its strategy is set to $[]$ in the \emph{Narrow}
step, and then the \emph{Stop} step applies.  This information can be
easily deduced when new variables are introduced: the abstracting
process directly introduces abstraction variables, by definition.
But the resulting term may still have variables of $\cal X$ since
the abstracted subterms of a term may not cover all variables of
the term.

Moreover, narrowing is performed on terms of $\TFXXA$. 
Indeed, there is no  variable renaming before the narrowing steps, that
could transform all variables into abstraction variables.
In addition, even if the variables of a narrowed term are all in $\XA$,
the range of the narrowing substitution can introduce variables of
$\X$, according to the LS-strategies, if these variables do not appear at 
LS-positions.

However some variable occurrences can be particularized into variables 
of $\XA$ in the narrowing process:
the narrowing substitution $\sigma$, whose range only contains new
variables of ${\cal X}$, can be transformed into a new substitution
$\sigma_A$ by replacing some of these variables by abstraction
variables.  Let us consider an equality of the form $X=u$, introduced
by the narrowing substitution $\sigma$, where $X \in \XA$,
and $u \in \TXF$.  
As $X$ is an abstraction variable, every
ground instance of $u$ must be in normal form.  So the variables in
$u$ that occur at an LS{-}position can be replaced by abstraction 
variables.
Let now $\mu$ be the
substitution $(x_i = X_i)$, for all $x_i \in Var(u)$ such that  
$X=u$ is an equality of $\sigma$ with $X \in \XA, \; u \in \TFXXA$,
and $x_i$ occurs at an LS{-}position  in  $u$. 
Then $\sigma_A = \mu
\sigma$.\\

Combining abstraction and narrowing is achieved here in the following
way. The abstraction positions are chosen so that the abstraction
mechanism captures the greatest possible number of rewriting steps: we
try to abstract the immediate subterms of the current term. If the
abstraction is possible, then a narrowing step is applied, only at the
top position, which limits the number of narrowing steps, more
complicated here than for the other strategies, since, as we will see
later, they involve complementary branches.

If \rname{Abstract} cannot be applied at all LS-positions of the term,
the process is stopped, and nothing can be concluded about termination.

\subsection{The termination proof procedure for local strategies}
\label{sec:rules}

The inference rules \rname{Abstract}, \rname{Narrow} and \rname{Stop}
instantiate respectively the proof steps \emph{abstract}, \emph{narrow}, and 
\emph{stop}.
They work in the following way on a state $(\{t^{[p_1,\ldots,p_n]}\},A,C)$,
where $top(t)=f$ and $LS(f)=[p_1,\ldots,p_n]$.

\begin{itemize}

\item The rule \rname{Abstract} processes the abstracting step. It
can apply:

\begin{itemize}
\item
when there exists $k \in
[2..n]$, $p_j \neq 0$ for $ 1 \leq j \leq k-1$ and $p_k=0$. 
The term $t$ is abstracted at positions 
$p_j \neq 0$ for $ 1 \leq j< k$ if
there exists an $\F$-stable ordering having the subterm property and 
such that $C \wedge (t_{\mathit{ref}}> t|_{p_j}, 1 \leq j< k)$
is satisfiable. Indeed, by induction hypothesis,
all ground instances of $t|_{p_j}, 1 \leq j< k$ LS-terminate. 
We  can instead have $\mathit{TERMIN(Local{-}Strat, t|_{p_j})}$ for some of the previous $t|_{p_j}$.
The list of positions then becomes $[0, p_{k+1}, \ldots ,p_n]$.

\item 
when there is no position $0$ in the strategy of the current term. Any ground
instance of the term obtained after abstraction is irreducible, by
definition of the LS-strategy, which ends the proof on the 
current derivation chain. The set containing the current term is then
replaced by the empty set.

\item when $p_1=0$. The rule applies but does not change the
state on which the {\emph narrow} step can be applied.

\end{itemize}


\item The rule \rname{Narrow} works as follows:
\begin{itemize}
\item
if the current term $t$ is narrowable at position 0, 
$t$ is  narrowed in all possible ways in one step, with all 
possible rewrite rules of the rewrite system $R$, and all possible
substitutions $\sigma_i$, into $u_i, i \in [1..l]$.
Then from the state $(\{t^{[0,p_1, \ldots ,p_n]}\}, A, C)$ we generate the states
$(\{u_i^{LS(top(u_i))}\}, A \wedge \sigma_i , C),$ $i \in [1..l]$, 
where the $\sigma_i$ are all  most general unifiers
allowing  narrowing of $t$ into terms $u_i$, such that $A \wedge \sigma_i$ is satisfiable.
This narrowing step means that $\sigma_1 t,  \ldots , \sigma_l t$ are all 
most general instances of $t$ that are reducible at the top position.
As a consequence, if 
$\Phi = \overline{\sigma_1} \wedge  \ldots  \wedge \overline{\sigma_l}$ 
is satisfiable, for each {\bf instantiation} $\mu$ satisfying 
$\Phi$, $\mu t$ is not reducible at the top position.
Then, as these $\mu t$ have to be reduced at positions
$[p_1, \ldots ,p_n]$, we also generate the complementary state
$(\{t^{[p_1, \ldots ,p_n]}\}, A \wedge \bigwedge_{i=1}^l \overline{\sigma_i}, C)$.

Let us also notice that if $u_i$ is a variable $x \in {\cal X}$, we
cannot conclude anything about termination of ground instances of
$x$. 
Setting $LS(x)$ to $[0]$ or $[]$ would wrongly lead to conclude, with the rule
\rname{Narrow}, that ground instances of $x$ are terminating.
So we force the proof process to stop in setting $LS(x)$ to
a particular symbol $\sharp$. 
However, if $u_i = X \in {\XA}$,
$LS(X)$ is set to $[]$, which is coherent with the fact that any
ground instance of $X$ is in LS- normal form.

\item if $t$ is not narrowable at position $0$ or is narrowable with a 
substitution {\bf that is not compatible with} the current constraint 
formula $A$, then no narrowing
is applied and the current term is evaluated at positions following 
the top position in the strategy. The list of positions then becomes 
$[p_1, \ldots ,p_n]$.
\end{itemize}

\item 
We also can check for the current term whether there
exists an ordering having the subterm property such that
$C \wedge t_{\mathit{ref}}>t$ is satisfiable. Then, by induction
hypothesis, any ground instance of $t$  terminates for the LS-strategy, 
which ends the proof on the current derivation chain. 
The \rname{Stop} rule then replaces the set containing the current term by
the empty set.

The rule \rname{Stop} also allows to stop the inference process
when the list of positions is empty.

\end{itemize}

The set of inference rules is given in Table \ref{tab:inference-strat-lo}.
In the conditions of these rules, satisfiability of $A$ is checked. 
Working with unsatisfiability of $A$ would be more technical to
handle here than in the innermost case, because of the complementary
branches generated by the \rname{Narrow} rule.

\begin{table*}
\centering
\caption{Inference rules for $t_{\mathit{ref}}$ LS-termination}
\label{tab:inference-strat-lo}
\framebox[\textwidth]{
\begin{minipage}{\textwidth}
\setlength{\parskip}{6ex}

\sixcinfr{Abstract}
{\{t^{[p_1, \ldots ,p_n]}\},~~A,~~C}
{\{u^S\},~~A \wedge \bigwedge_{j
\in \{i_1, \ldots ,i_p\}} (t|_j \da = X_j),~~C \wedge \bigwedge_{j \in \{i_1, \ldots ,i_p\}} H_C(t|_j)}
{{\rm ~where~} t {\rm ~is~abstracted~into~}
u {\rm ~at~the~positions~} 
i_1,\ldots,i_p \in \mathit{POS}}
{{\rm ~if~} A \wedge \bigwedge_{j \in \{i_1, \ldots ,i_p\}} (t|_j \da = X_j), \; C
\wedge \bigwedge_{j \in \{i_1, \ldots ,i_p\}}   H_C(t|_j){\rm
~are~satisfiable~and~}}
{\mathit{~POS} = \{p_1,\ldots,p_{k-1}\},S=[0,p_{k+1}, \ldots ,p_n] 
{\rm ~if~} \exists k \in [2..n] : p_1,\ldots,p_{k-1} \neq 0}  
{{\rm ~~~~~~~~~~~~and~} p_k = 0}
{\mathit{~POS} = \{p_1,\ldots,p_n\}, S=[]
{\rm ~if~} p_1,\ldots,p_n \neq 0 {\rm ~or~} [p_1, \ldots ,p_n]=[]}
{\mathit{~POS} = \emptyset, S=[p_1, \ldots ,p_n]
{\rm ~if~} p_1=0}

\fivecinfr{Narrow}
{\{t^{[0,p_1, \ldots ,p_n]}\},~~A,~~C}
{\{u^S\}, ~~A',~~C }
{{\rm where~} u=u_i,  S=LS(top(u_i)),  A'= A \wedge \sigma_i {\rm ~if~} 
t \leadsto^{\epsilon, \sigma_i}_\R u_i {\rm ~and~} A \wedge \sigma_i {\rm ~is~satisfiable~}}
{{\rm ~~or~} u^S = t^{[p_1, \ldots ,p_n]}, 
A'=A \wedge (\bigwedge_{i=1}^l \overline{\sigma_i}),
{\rm ~and~} \sigma_i, i \in [1..l] {\rm ~are~all~nar.~subst.~as~above~}}
{{\rm ~~or~} u^S = t^{[p_1, \ldots ,p_n]}, A'=A }
{{\rm ~~~~~~~~~if~} t {\rm ~is~not~narrowable~at~the~top~position}}
{{\rm ~~~~~~~~~or~} \forall \sigma {\rm ~nar.~subst.~of~} t 
{\rm ~at~the~top~position}, ~~A \wedge \sigma {\rm ~is~not~satisfiable}}

\cinfr{Stop}
{\{t^{[p_1, \ldots ,p_n]}\},~~A,~~C}
{\emptyset,~~A \wedge H_A(t),~~C \wedge H_C(t)}
{A \wedge  H_A(t), \; C \wedge H_C(t) {\rm ~are~satisfiable}
}

${\rm ~and~} H_A(t) =
\left\{
\begin{array}{ll}
\mathit{\top}           & {\rm ~if~} [p_1, \ldots ,p_n]=[] \\
& {\rm ~or~any~ground~instance~of~} t \\
& {\rm ~is~in~normal~form}\\

t \da = X       & {\rm ~otherwise.}
\end{array}
\right.
$
\\
$~~~~~~~~H_C(t) =
\left\{
\begin{array}{ll}
\mathit{\top}           & {\rm ~if~} [p_1, \ldots ,p_n]=[] \\
& {\rm ~or~} \mathit{TERMIN(Local{-}Strat,t)}\\
t_{\mathit{ref}}>t      & {\rm ~otherwise.}
\end{array}
\right.
$

\end{minipage}}
\end{table*}



The strategy for applying these rules is:


\begin{center}
$\mathit{repeat*
(try{-}stop(\rname{Abstract}); try{-}stop(\rname{Narrow}); try{-}skip(\rname{Stop}))}$.
\end{center}


There are here also three cases for
the behavior of the proof process. It can diverge as previously, or
stop and the states in the leaves have then to be considered.
The good case is when
the process stops and all final states of all proof trees are of the
form $(\emptyset,A,C)$. \\



\subsection{Examples}

\begin{subexample}
Let us recall the rules of the  example given in the introduction.

{\footnotesize
\[
\begin{array}{ll}
f(i(x))         & \ra ite(zero(x),g(x),f(h(x)))\\
zero(0)         & \ra true\\
zero(s(x))      & \ra false\\
ite(true,x,y)   & \ra x\\
ite(false,x,y)  & \ra y\\
h(0)            & \ra i(0)\\
h(x)            & \ra s(i(x))
\end{array}
\]
}
The LS-strategy is the following :

{\footnotesize
\begin{itemize}
\item $LS(ite) = [1 ; 0]$,
\item $LS(f) = LS(zero) = LS(h) = [1 ; 0]$ and
\item $LS(g) = LS(i) = [1]$.
\end{itemize}
}

Let us prove the termination of this system on the signature
$\F = \{f:1, zero:1, ite:3, h:1, s:1, i:1, g:1, 0:0\}$.

Applying the inference rules on $f(x_1)$, we get :

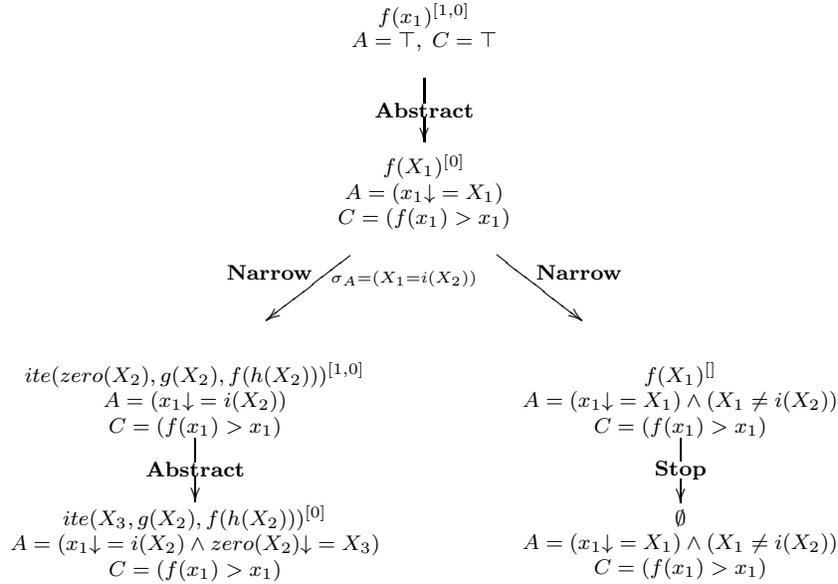
\footnotesize{
$$
\xymatrix{
& \txt{$f(x_1)^{[1,0]}$\\$A = \top, \; C = \top$\\$ $} 
\ar|-{\rname{Abstract}}[d] & \\
& \txt{$ $\\$f(X_1)^{[0]}$\\$A = (x_1 \con X_1)$\\$C = (f(x_1)>x_1)$\\$ $}
\ar_-{\rname{Narrow}}^{\sigma_A = (X_1=i(X_2))}[dl]
\ar^-{\rname{Narrow}}[dr] &\\
\txt{}
&& \txt{}
}
$$
$$
\xymatrix{
*{\txt{$ $\\$ite(zero(X_2),g(X_2),f(h(X_2)))^{[1,0]}$\\$A = (x_1 \con i(X_2))$\\$C = (f(x_1)>x_1)$}} 
\ar|-{\rname{Abstract}}[d] 
&& *{\txt{$ $\\$f(X_1)^{[]}$\\$A = (x_1 \con X_1) \wedge (X_1\neq i(X_2))$\\$C = (f(x_1)>x_1)$}} 
\ar|-{\rname{Stop}}[d]\\
*{\txt{$ $\\$ite(X_3,g(X_2),f(h(X_2)))^{[0]}$\\$A = (x_1 \con i(X_2) \wedge zero(X_2) \con X_3)$\\$C = (f(x_1)>x_1)$}} 
&& *{\txt{$ $\\$\emptyset$\\$A = (x_1 \con X_1) \wedge (X_1\neq i(X_2))$\\$C = (f(x_1)>x_1)$}} 
}
$$
}

\normalsize

\rname{Abstract} applies on $f(x_1)$, since $C$ is satisfiable by any 
ordering having the subterm property.
$A$ is satisfiable with any instantiation $\theta$ such that
$\theta x_1 = \theta X_1 = 0$.

\rname{Narrow} expresses the fact that $\sigma f(X_1)$ is reducible
if $\sigma$ is such that $\sigma X_1 = i(X_2)$, and that
the other instances ($\sigma' f(X_1)$ with $\sigma' X_1 \neq i(X_2)$) 
cannot be reduced.

The renaming of $x_2$ into $X_2$ in $\sigma_A$ comes from the fact that
$x_2$ occurs in $i(x_2)$ at an LS-position in $\sigma = (X_1 = i(x_2))$.

Then, the constraint formula $A$ on the left branch is satisfiable by any
instantiation $\theta$ such that $\theta X_2 = 0$ and $\theta x_1 = i(0)$.
The constraint formula on the complementary branch
is satisfied by any instantiation
$\theta$ such that $\theta x_1 = \theta X_1 = \theta X_2 = 0$.

\rname{Abstract} applies here on the first branch, since $zero(X_2)$ can be 
abstracted, thanks to a version of Proposition~\ref{prop:TC} adapted to local
strategies~\cite{FGK-local-strat-ENTCS-2001}.
Indeed, $\U(zero(X_2)) = \{zero(0) \ra true, zero(s(x)) \ra false\}$,
and both rules can be oriented by a LPO $\succ$
with the precedence $zero \succ_{\F} true$ and $zero \succ_{\F} false$.
Then we have $\mathit{TERMIN}(Local{-}strat, zero(X_2))$.

The next constraint formula $A$ is satisfiable with any instantiation $\theta$
such that $\theta X_2 = 0$, $\theta X_3 = true$ and 
$\theta x_1 = i(0)$.

Then, \rname{Narrow} applies on the left branch:

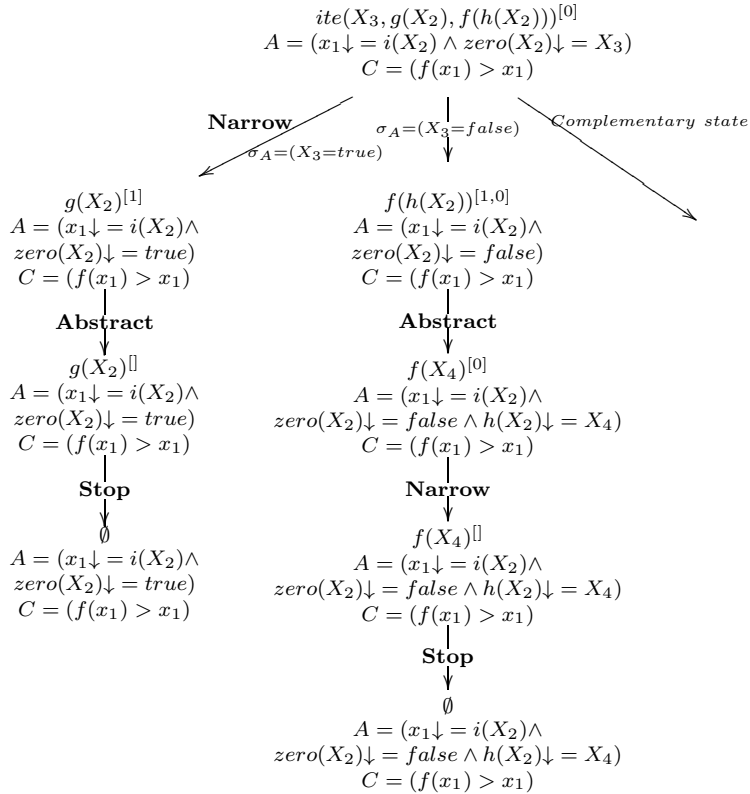
\footnotesize{
{
$$
\xymatrix{
& *{\txt{$ $\\$ite(X_3,g(X_2),f(h(X_2)))^{[0]}$\\$A = (x_1 \con i(X_2) \wedge zero(X_2) \con X_3)$\\$C = (f(x_1)>x_1)$\\$ $}} 
\ar_-{\rname{Narrow}}^{\sigma_A = (X_3 = true)}[dl] 
\ar|-{\sigma_A = (X_3 = false)}[d] 
\ar^{Complementary~state}[dr] & \\
*{\txt{$ $\\ $ $\\ $g(X_2)^{[1]}$\\$A = (x_1 \con i(X_2) \wedge$ \\$zero(X_2) \con true)$\\$C = (f(x_1)>x_1)$}} 
\ar|-{\rname{Abstract}}[d]
& *{\txt{$ $\\ $ $ \\ $f(h(X_2))^{[1,0]}$\\$A = (x_1 \con i(X_2) \wedge$ \\ $zero(X_2) \con false)$\\$C = (f(x_1)>x_1)$}} 
\ar|-{\rname{Abstract}}[d] 
&& *{\txt{$ $\\ $\bullet$}}\\
*{\txt{$g(X_2)^{[]}$\\$A = (x_1 \con i(X_2) \wedge$ \\$zero(X_2) \con true)$\\$C = (f(x_1)>x_1)$}} 
\ar|-{\rname{Stop}}[d] 
& *{\txt{$ $\\ $f(X_4)^{[0]}$\\$A = (x_1 \con i(X_2) \wedge$ \\ $zero(X_2) \con false \wedge h(X_2) \con X_4)$\\$C = (f(x_1)>x_1)$}} 
\ar|-{\rname{Narrow}}[d]\\
*{\txt{$\emptyset$\\$A = (x_1 \con i(X_2) \wedge$ \\$zero(X_2) \con true)$\\$C = (f(x_1)>x_1)$}} 
& *{\txt{$ $\\ $f(X_4)^{[]}$\\$A = (x_1 \con i(X_2) \wedge$ \\ $zero(X_2) \con false \wedge h(X_2) \con X_4)$\\$C = (f(x_1)>x_1)$}}  
\ar|-{\rname{Stop}}[d] \\
& *{\txt{$ $\\ $\emptyset$\\$A = (x_1 \con i(X_2) \wedge$ \\ $zero(X_2) \con false \wedge h(X_2) \con X_4)$\\$C = (f(x_1)>x_1)$}} 
}
$$
}
}
\normalsize

The first constraint formula $A$ is satisfiable by any 
instantiation $\theta$ such that
$\theta X_2 = 0$ and $\theta x_1 = i(0)$.
The second one is satisfiable by any instantiation $\theta$ such that
$\theta X_2 = s(0)$ and $\theta x_1 = i(s(0))$.
The third one (see below) is satisfiable by any instantiation $\theta$ such that
$\theta X_3 = zero(i(0))$, $\theta X_2 = i(0)$ and $\theta x_1 = i(i(0))$.

\rname{Abstract} trivially applies on $g(X_2)$: since $X_2$ is an abstraction 
variable, there is no need to abstract it.

The second \rname{Abstract} applies on $f(h(X_2))$, thanks to
the previous adaptation of Proposition~\ref{prop:TC} to local strategies.
Indeed, $\U(h(X_2)) = \{h(0) \ra i(0), h(x) \ra s(i(x))\}$, and both rules
can be oriented by the same LPO as previously with the additional precedence
$h \succ_{\F} i$ and $h \succ_{\F} s$.
Then we have $\mathit{TERMIN}(Local{-}strat,$ $ h(X_2))$.

The constraint formula associated to $f(X_4)^{[0]}$
is satisfiable by any instantiation $\theta$ such that
$\theta X_4 = s(i(s(0)))$, $\theta X_2 = s(0)$ and $\theta x_1 = i(s(0))$.

One could have tried to narrow $f(X_4)$, by using the first rule
and the narrowing substitution
$\sigma_A = (X_4=i(X_5))$.
But then $A \wedge \sigma_A$ would lead to $(x_1 \con i(X_2) 
\wedge zero(X_2) \con false \wedge h(X_2) \con i(X_5))$.
For any $\theta$ satisfying $A \wedge \sigma_A$, $\theta$ must be such that 
$\theta h(X_2) \con h(\theta X_2 \da) \con i(\theta X_5)$.
If $\theta X_2 \da \neq 0$, then, according to $\R$, 
$h(\theta X_2 \da) \ra s(i(\theta X_2 \da))$, where $s$ is a constructor.
Then we cannot have $h(\theta X_2 \da) \con i(\theta X_5)$, so
$\theta$ must be such that $\theta X_2 \con 0$.
But then $\theta zero(X_2) \con true$, which makes $A \wedge \sigma_A$ 
unsatisfied.
Therefore there is no narrowing.

For the third branch, we have:
\footnotesize{
{
$$
\xymatrix{
& *{\txt{$ $\\ $ $ \\ $\bullet$ \\$ite(X_3,g(X_2),f(h(X_2)))^{[]}$\\$A = (x_1 \con i(X_2) \wedge$ \\ $zero(X_2) \con X_3$ \\$\wedge X_3 \neq true \wedge X_3 \neq false)$\\$C = (f(x_1)>x_1)$\\$ $}} 
\ar|-{\rname{Stop}}[d] \\
& *{\txt{$ $\\ $\emptyset$\\$A = (x_1 \con i(X_2) \wedge$ \\ $zero(X_2) \con X_3 \wedge $ \\$X_3 \neq true \wedge X_3 \neq false)$\\$C = (f(x_1)>x_1)$\\$ $}} \\
}
$$
}
}

\normalsize

Like for the defined symbols $ite, zero, h$, the inference rules apply
successfully through one \rname{Abstract}, \rname{Narrow},
\rname{Abstract} with no abstraction position, \rname{Narrow} and
\rname{Stop} application.
Therefore $\R$ is LS-terminating.

\end{subexample}

Let us now give an example that cannot be handled with the
context-sensitive approach.

\begin{subexample}

Let \(\R\) be the following rewrite system
{\footnotesize
\[
\begin{array}{lll}

f(a,g(x))        \ra  f(a,h(x)) \\
h(x) \ra g(x)
\end{array}
\]
}
with the LS-strategy : 
{\footnotesize
$LS(f) = [0 ; 1 ; 2], \; LS(h) = [0]$ and
$LS(g) = [1]$.
}

The context-sensitive strategy would allow to permute the reducible
arguments of $f$, so that we also could evaluate terms with $LS(f) =
[1; 2; 0]$.
We let the user check that, with this strategy, \(\R\) does not terminate.

Applying the rules on $f(x_1,x_2)$, we get:
\newpage

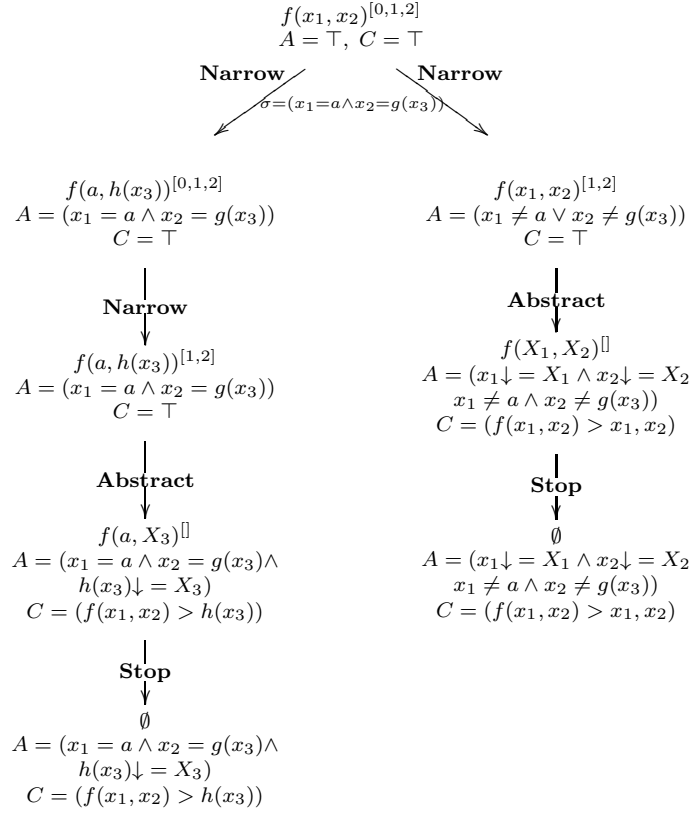
\footnotesize{
{
$$
\xymatrix{
& *{\txt{$f(x_1,x_2)^{[0,1,2]}$\\$A = \top, \; C = \top$\\$ $}} 
\ar_{\rname{Narrow}}^{\sigma = (x_1 = a \wedge x_2 = g(x_3))}[dl]
\ar^{\rname{Narrow}}[dr] &\\
\txt{}
&&\txt{}
}
$$
$$
\xymatrix{
*{\txt{$ $\\$f(a,h(x_3))^{[0,1,2]}$\\
$A = (x_1 = a \wedge x_2 = g(x_3))$\\$C = \top$\\$ $}} 
\ar|-{\rname{Narrow}}[d] 
&& *{\txt{$ $\\$f(x_1,x_2)^{[1,2]}$\\$A = (x_1 \neq a \vee x_2 \neq g(x_3))$\\$C = \top$\\$ $}} 
\ar|-{\rname{Abstract}}[d] \\
*{\txt{$ $\\$f(a,h(x_3))^{[1,2]}$\\
$A = (x_1 = a \wedge x_2 = g(x_3))$\\$C = \top$\\$ $}} 
\ar|-{\rname{Abstract}}[d]
&& *{\txt{$ $\\$f(X_1,X_2)^{[]}$\\$A = (x_1 \con X_1 \wedge x_2 \con X_2 $\\ $x_1 \neq a \wedge x_2 \neq g(x_3))$\\$C = (f(x_1,x_2) > x_1,x_2)$\\$ $}}  
\ar|-{\rname{Stop}}[d]  \\
*{\txt{$ $\\$f(a,X_3)^{[]}$\\
$A = (x_1 = a \wedge x_2 = g(x_3) \wedge$ \\$h(x_3) \con X_3)$\\$C = (f(x_1,x_2) > h(x_3))$\\$ $}}  
\ar|-{\rname{Stop}}[d] 
&& *{\txt{$ $\\$\emptyset$\\
$A = (x_1 \con X_1 \wedge x_2 \con X_2 $\\ $x_1 \neq a \wedge x_2 \neq g(x_3))$\\$C = (f(x_1,x_2) > x_1,x_2)$\\$ $}} \\
*{\txt{$ $\\$\emptyset$\\
$A = (x_1 = a \wedge x_2 = g(x_3) \wedge$ \\$h(x_3) \con X_3)$\\$C = (f(x_1,x_2) > h(x_3))$\\$ $}} 
&& 
}
$$
}
}

\normalsize

Applying the rules on $h(x_1)$, we get:

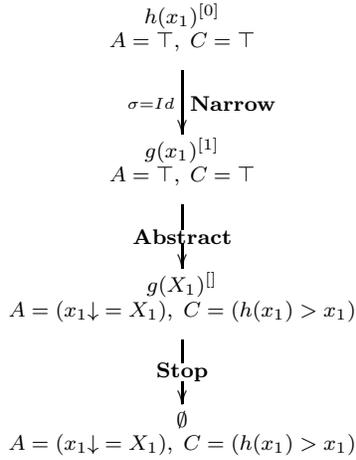
\footnotesize{
\begin{center}
{
$$
\xymatrix{
& *{\txt{$h(x_1)^{[0]}$\\$A = \top, \; C = \top$\\$ $}} 
\ar^{\rname{Narrow}}_{\sigma = Id}[d] &\\
& *{\txt{$ $\\$g(x_1)^{[1]}$\\ $A = \top, \; C = \top$\\$ $}} 
\ar|-{\rname{Abstract}}[d] &\\
& *{\txt{$ $\\ $g(X_1)^{[]}$\\ $A = (x_1 \con X_1), \; C = (h(x_1) > x_1)$\\$ $}} 
\ar|-{\rname{Stop}}[d] &\\
& *{\txt{$ $\\ $\emptyset$\\ $A = (x_1 \con X_1), \; C = (h(x_1) > x_1)$\\$ $}} &
}
$$
}
\end{center}
}

\normalsize

\end{subexample}

\OUT{
{\bf Choisir ce qu'on garde et developpe:}
Let us finally give another example, that cannot be be handled with the
context-sensitive approach:the rewrite system 
$\{f(b)  \ra c, g(x) \ra h(x), h(c) \ra g(f(a)), a \ra b\}$
with the LS-strategy $LS(f) =[0 ; 1], LS(g) = LS(h) = [1 ; 0], LS(a) = [0]$, and two 
examples of innermost rewriting: the well-known Toyamas'example
$\{f(0,1,x) \ra f(x, x, x), g(x,y) \ra x, g(x,y) \ra y\}$
with the LS-strategy $LS(f) = [1 ; 2 ; 3 ; 0], LS(g) = [1 ; 2 ; 0], 
LS(0) = LS(1) = [0]$, and $\{f(f(x)) \ra f(f(x)), f(a) \ra a\}$ 
with the LS-strategy $LS(f) = [1 ; 0], LS(a) = [0]$, that is
innermost terminating on $\TF$, but is not on $\TXF$. The complete development of these 
examples can be found in the \cite{FGK-local-strat-2001-interne}.
}

\section{Conclusion}

The generic termination proof method presented in this paper is based
on the simple ideas of schematizing and observing the derivation trees 
of ground terms and of using an induction ordering to stop
derivations as soon as termination is ensured by induction.
The method  makes clear the schematization  power of narrowing,
abstraction and constraints. Constraints are heavily used on one hand 
to gather conditions that the induction ordering must satisfy, on the
other hand to represent the set of ground instances of generic terms.

Our technique is implemented in a system named 
CARIBOO~\cite{FGK-PPDP-2002,Fissore-These-2003,CARIBOO-APP-2004},
providing a termination proof tool for the innermost, the outermost, 
and the local strategies~\footnote{Available at http://protheo.loria.fr/softwares/cariboo/}.
CARIBOO consists of two main parts~:
\begin{enumerate}

\item The proof procedure, written in ELAN, 
which is a direct translation of the inference rules. It generates
the proof trees, dealing with the ordering and the abstraction constraints.
It is worth emphasizing the reflexive aspect of this proof procedure,
written in a rule-based language, to allow termination of rule-based programs.

\item A graphical user interface (GUI), written in Java.
It provides an edition tool to define specifications of rewrite systems
which are then transformed into an ELAN specification 
used by the proof procedure.
It also displays the detailed results of the proof process~: 
which defined symbols have already been treated and, for
each of them, the proof tree together with the detail of each state.
Trace files can be generated in different formats (HTML, ps, pdf...)

\end{enumerate}

To deal with the generated constraints, 
the proof process of CARIBOO can use integrated features, 
like the computation of usable rules, 
the use of the subterm ordering or the Lexicographic Path ordering 
to satisfy ordering constraints, and the test of sufficient conditions of 
Section~\ref{subsec:cumulating-constraints} for detecting unsatisfiability of 
$A$.

It can also delegate features, as solving the ordering constraints or
orienting the usable rules when the LPO fails, proving termination of a term,
or testing satisfiability of $A$. Delegation is either proposed to the user, 
or automatically ensured by the ordering constraint solver C{\it i}me2.

CARIBOO provides several automation modes for dealing with constraints.
Dealing with unsatisfiability of $A$ allows a complete automatic mode, 
providing a termination proof for a large class of examples 
(a library is available with the distribution of CARIBOO).

It is interesting to note that thanks to the power of induction, 
and to the help of usable rules,
the generated ordering constraints are often simple, and are easily 
satisfied by the subterm ordering or an LPO.

Finally, the techniques presented here have also been applied to weak 
termination in~\cite{FGK-ter-faible-ICTAC-2004}.

As our proof process is very closed to the rewriting mechanism, it could easily 
be extended to conditional, equational and typed rewriting, 
by simply adapting the narrowing definition.
Our approach is also promising to tackle inductive proofs of other term 
properties like confluence or ground reducibility.

\appendix
\section*{APPENDIX}

\section{The lifting lemma}

The lifting lemma for standard narrowing~\cite{MiddeldorpH-AAECC94}
can be locally adapted to $S${-}rewriting with non-normalized substitutions
provided they fulfill some constraints on the positions of rewriting.
To do so, we need the following
two propositions (the first one is obvious).

\begin{proposition}\label{prop:obvious}
Let $t \in \TXF$ and $\sigma$ a substitution of $\TXF$. 
Then $Var(\sigma t) = (Var(t) - Dom(\sigma)) \cup Ran(\sigma_{Var(t)})$.
\end{proposition}

\begin{proposition}\label{prop:congruence}
Suppose we have substitutions $\sigma, \mu, \nu$ and sets $A,B$ of variables 
such that
$(B - Dom(\sigma)) \cup Ran(\sigma) \subseteq A$.
If $\mu = \nu [A]$ then $\mu \sigma = \nu \sigma [B]$.
\end{proposition}

\begin{proof}
Let us consider $(\mu \sigma)_B$, which can be divided as follows:
$(\mu \sigma)_B = (\mu \sigma)_{B \cap Dom(\sigma)} \cup 
(\mu \sigma)_{B - Dom(\sigma)}$.\\
For $x \in B \cap Dom(\sigma)$, we have $\Var(\sigma x) 
\subseteq Ran(\sigma)$, and then
$(\mu \sigma) x = \mu(\sigma x) = \mu_{Ran(\sigma)}(\sigma x) = 
(\mu_{Ran(\sigma)} \sigma) x$.
Therefore $(\mu \sigma)_{B \cap Dom(\sigma)} =$\\ 
$(\mu_{Ran(\sigma)} \sigma)_{B \cap Dom(\sigma)}$.\\
For $x \in B - Dom(\sigma)$, we have $\sigma x = x$, and then 
$(\mu \sigma) x = \mu(\sigma x) = \mu x$.
Therefore we have $(\mu \sigma)_{B-Dom(\sigma)} = \mu_{B - Dom(\sigma)}$.
Henceforth we get 
$(\mu \sigma)_B = (\mu_{Ran(\sigma)} \sigma)_{B \cap Dom(\sigma)}$ 
$\cup \mu_{B - Dom(\sigma)}$.\\
By a similar reasoning, we get 
$(\nu \sigma)_B = (\nu_{Ran(\sigma)} \sigma)_{B \cap Dom(\sigma)} 
\cup \nu_{B - Dom(\sigma)}$.\\
By hypothesis, we have $Ran(\sigma) \subseteq A$ and $\mu = \nu [A]$. Then 
$\mu_{Ran(\sigma)} = \nu_{Ran(\sigma)}$. Likewise, since $B - Dom(\sigma) 
\subseteq A$, we
have $\mu_{B - Dom(\sigma)} = \nu_{B - Dom(\sigma)}$.\\
Then we have $(\mu \sigma)_B = (\mu_{Ran(\sigma)} \sigma)_{B \cap Dom(\sigma)} 
\cup \mu_{B - Dom(\sigma)}=$\\
$({\nu}_{Ran(\sigma)} \sigma)_{B \cap Dom(\sigma)} 
\cup {\nu}_{B - Dom(\sigma)} = (\nu \sigma)_B$.
Therefore $(\mu \sigma)= (\nu \sigma) [B]$.
\end{proof}

\begin{lemma}[\ref{lemma:surred-S}\ ($S${-}lifting Lemma)]

Let $\R$ be a rewrite system.
Let $s \in \TXF$, $\alpha$ a ground
substitution  such that $\alpha s$ is $S${-}reducible 
at a non variable position $p$ of $s$, and ${\cal Y} \subseteq {\cal X}$
a set of variables 
such that $Var(s) \cup Dom(\alpha) \subseteq {\cal Y}$.
If $\alpha s \ra^{S}_{p, l \ra r} t'$, then there exist a term $s' \in
\TXF$ and substitutions
$\beta, \sigma = \sigma_0 \wedge \bigwedge_{j \in [1..k]} \overline{\sigma_j}$ 
such that:

\[
  \begin{array}{ll}

        1.~s \surred^{S}_{p, l \ra r, \sigma} s', \\
        2.~\beta s' = t', \\
        3.~\beta \sigma_0 = \alpha [{\cal Y}]\\
        4.~\beta {\rm ~satisfies~} \bigwedge_{j \in [1..k]} \overline{\sigma_j}.

  \end{array}
\]

where $\sigma_0$ is the most general unifier of $s|_p$ and $l$ and 
$\sigma_j, j \in [1..k]$ are
all most general unifiers of $\sigma_0 s|_{p'}$ and a left-hand side
$l'$ of a rule of $\R$, for all position $p'$  which are  $S$-better
positions than $p$ in $s$.
\end{lemma}

\begin{proof}
In the following, we assume that ${\cal Y} \cap \Var(l) = \emptyset$ for every
$l \ra r \in \R$.\\
If $\alpha s \ra^{S}_{p, l \ra r} t'$, then there exists a substitution $\tau$ 
such that $Dom(\tau) \subseteq \Var(l)$ and $(\alpha s)|_p = \tau l$.
Moreover, since $p$ is a non variable position of $s$, we have $(\alpha s)|_p 
= \alpha(s|_p)$.
Denoting $\mu = \alpha \tau$, we have: 
\\
\begin{tabular}{lll}
$\mu(s|_p)$ & $= \alpha(s|_p)$ & for $Dom(\tau) \subseteq \Var(l)$ and 
$\Var(l) \cap \Var(s) = \emptyset$\\
& $= \tau l$ & by definition of $\tau$\\
& $= \mu l$ & for $Dom(\alpha) \subseteq {\cal Y}$ and ${\cal Y} \cap 
\Var(l) = \emptyset$,
\end{tabular}
\\
and therefore $s|_p$ and $l$ are unifiable.
Let us note $\sigma_0$ the most general unifier of $s|_p$ and $l$, and
$s' = \sigma_0(s[r]_p)$.

Since $\sigma_0$ is more general than $\mu$, there exists a substitution 
$\rho$ such that $\rho \sigma_0 = \mu$.
Let ${\cal Y}_1 = ({\cal Y} - Dom(\sigma_0)) \cup Ran(\sigma_0)$.
We define $\beta = \rho_{{\cal Y}_1}$.
Clearly $Dom(\beta) \subseteq {\cal Y}_1$.\\
We now show that $\Var(s') \subseteq {\cal Y}_1$, by the following reasoning:
\begin{itemize}
        \item since $s' = \sigma_0(s[r]_p)$, we have 
$\Var(s') = \Var(\sigma_0(s[r]_p))$;
        \item the rule $l \ra r$ is such that $\Var(r) \subseteq \Var(l)$, 
therefore we have
$\Var(\sigma_0(s[r]_p)) \subseteq \Var(\sigma_0(s[l]_p))$, and then, 
thanks to the previous point,
$\Var(s') \subseteq \Var(\sigma_0(s[l]_p))$;
        \item since $\sigma_0(s[l]_p) = \sigma_0 s [\sigma_0 l]_p$ and
since $\sigma_0$ unifies $l$ and $s|_p$, we get 
$\sigma_0(s[l]_p) = (\sigma_0 s) [\sigma_0(s|_p)]_p = \sigma_0 s[s|_p]_p = 
\sigma_0 s$ and, thanks to the
previous point: $\Var(s') \subseteq \Var(\sigma_0 s)$;
        \item according to Proposition \ref{prop:obvious}, we have
$\Var(\sigma_0(s)) = (\Var(s)$ $ - Dom(\sigma_0)) \cup Ran(\sigma_{0\Var(s)})$;
by hypothesis, $\Var(s) \subseteq {\cal Y}$. Moreover, since
$Ran(\sigma_{0\Var(s)}) \subseteq Ran(\sigma_0)$, we have \\
$\Var(\sigma_0(s)) \subseteq ({\cal Y} - Dom(\sigma_0)) \cup Ran(\sigma_0)$, 
that is
$\Var(\sigma_0 s) \subseteq {\cal Y}_1$.
Therefore, with the previous point, we get  $Var(s') \subseteq {\cal Y}_1$.
\end{itemize}

From $Dom(\beta) \subseteq {\cal Y}_1$ and 
$Var(s') \subseteq {\cal Y}_1$, we infer
$Dom(\beta) \cup Var(s') \subseteq {\cal Y}_1$.

Let us now prove that $\beta s' = t'$.\\
Since $\beta = \rho_{{\cal Y}_1}$, we have $\beta = \rho [{\cal Y}_1]$.
Since $Var(s') \subseteq {\cal Y}_1$, we get
$\beta s' = \rho s'$.
Since $s' = \sigma_0(s[r]_p)$, we have
$\rho s' = \rho \sigma_0(s[r]_p) = \mu(s[r]_p) = \mu s[\mu r]_p$.
Then $\beta s' = \mu s[\mu r]_p$.\\
We have $Dom(\tau) \subseteq \Var(l)$ and ${\cal Y} \cap \Var(l) = \emptyset$,
then we have ${\cal Y} \cap Dom(\tau) = \emptyset$. Therefore,
from $\mu = \alpha \tau$, we get $\mu = \alpha [{\cal Y}]$. 
Since $\Var(s) \subseteq {\cal Y}$, we get $\mu s = \alpha s$.\\
Likewise, by hypothesis we have $Dom(\alpha) \subseteq {\cal Y}$,
$\Var(r) \subseteq \Var(l)$ and ${\cal Y} \cap \Var(l) = \emptyset$, then we get
$Var(r) \cap Dom(\alpha) = \emptyset$, and then we have $\mu = \tau [Var(r)]$,
and therefore $\mu r = \tau r$.\\
From $\mu s = \alpha s$ and $\mu r = \tau r$ we get $\mu s [\mu r]_p = 
\alpha s [\tau r]_p$.
Since, by hypothesis, $\alpha s \ra^p t'$, with $\tau l = (\alpha s)|_p$, 
then $\alpha s [\tau r]_p = t'$.
Finally, as $\beta s' = \mu s[\mu r]_p$, we get $\beta s' = t'$ (2).

Next let us prove that $\beta \sigma_0 = \alpha [{\cal Y}]$.
Reminding that ${\cal Y}_1 = ({\cal Y} - Dom(\sigma_0)) \cup Ran(\sigma_0)$,
Proposition \ref{prop:congruence} (with the notations $A$ for ${\cal Y}_1$,
$B$ for $\cal Y$, $\mu$ for $\beta$, $\nu$ for $\rho$ and $\sigma$ for $\sigma_0$)
yields $\beta \sigma_0 = \rho \sigma_0 [{\cal Y}]$.
We already noticed that $\mu = \alpha [{\cal Y}]$.
Linking these two equalities via the equation $\rho \sigma_0 = \mu$ yields 
$\beta \sigma_0 = \alpha [{\cal Y}]$ (3).

Let us now suppose that there exist a rule $l' \ra r' \in \R$, a
position $p'$ $S${-}better than $p$ and a substitution $\sigma_i$ such
that $\sigma_i(\sigma_0(s|_{p'})) = \sigma_i l'$. 

Let us now suppose that $\beta$ does not satisfy 
$\bigwedge_{j \in [1..k]} \overline{\sigma_j}$.
There exists $i \in [1..k]$ such that $\beta$ satisfies $\sigma_i =
\bigwedge _{i_l \in [1..n]} (x_{i_l} = u_{i_l})$. So $\beta$ is such that 
$\bigwedge _{i_l \in [1..n]} (\beta x_{i_l} = \beta u_{i_l})$.

Thus, on $Dom(\beta) \cap Dom(\sigma_i) \subseteq \{x_{i_l}, i_l
\in[1..n]\}$, we have $(\beta x_{i_l} = \beta u_{i_l})$, so $\beta
\sigma_i = \beta$.
Moreover, as $\beta$ is a ground substitution, $\sigma_i \beta = \beta$. Thus,
$\beta \sigma_i = \sigma_i \beta$.

On $Dom(\beta) \cup Dom(\sigma_i) -(Dom(\beta) \cap Dom(\sigma_i))$,
either $\beta = Id$, or $\sigma_i = Id$, so $\beta \sigma_i = \sigma_i
\beta$.

As a consequence,
$\alpha(s) = \sigma_i \alpha(s) = \sigma_i \beta \sigma_0(s) = \beta
\sigma_i \sigma_0 (s)$ is reducible at position $p'$ with the rule
$l'$, which is impossible by definition of S-reducibility of
$\alpha(s)$ at position $p$.
So the ground substitution $\beta$ satisfies 
$\bigwedge_{i \in [1..k]} \overline{\sigma_i}$
for all most general unifiers $\sigma_i$ of $\sigma_0 s$ and a left-hand side of 
rule of $\R$ at $S${-}better positions of $p$ (4).

Therefore, denoting $\sigma = \sigma_0 \wedge \bigwedge_{i \in [1..k]} 
\overline{\sigma_i}$,
from the beginning of the proof, 
we get $s \surred^{S}_{[p, l \ra r, \sigma]} s'$, and 
then the point (1) of the current lemma holds.
\end{proof}

\section{Proof of the generic termination result}

Let us remind that $\mathit{SUCCESS(g,\succ)}$ means that 
the application of $\mathit{Strat{-}Rules(S)}$
on $(\{g(x_1, \ldots , x_m)\},$ $\top,\top)$  
gives a finite  proof tree, whose sets $C$ of ordering constraints
are satisfied by a same ordering $\succ$, and
whose leaves
are either states of the form $(\emptyset,A,C)$  or states
whose set of constraints $A$ is unsatisfiable.

\begin{subtheorem}
[\ref{theo:termin}]

Let $R$ be a rewrite system on a set $\F$ of symbols containing at least a
constructor constant.  If there exists an $\F$-stable ordering $\succ$
having the subterm property, such that for each symbol $g \in \Def$,
we have $\mathit{SUCCESS(g,\succ)}$, then every term of $\TF$ terminates with
respect to the strategy $S$.

\end{subtheorem}

\begin{proof} 
We use an emptyness lemma, an abstraction lemma, a narrowing lemma, 
and a stopping lemma, which are given after this main proof.

We prove by induction on $\TF$ that any ground instance    
$\theta f(x_1, \ldots,$ $x_m)$ of   
any term $f(x_1, \ldots ,x_m) \in \TXF$ S-terminates. 
The induction ordering is constrained along the proof. At the 
beginning, it has at least to be $\F$-stable and to have the  
subterm property, which ensures its noetherianity. 
Such an ordering always exists on $\TF$ (for instance the embedding  
relation). Let us denote it $\succ$.\\ 


If $f$ is a constructor, then $\theta f(x_1, \ldots ,x_m) \da = 
f(\theta x_1,  \ldots ,\theta x_m) \da =  
[f(\theta x_1, \ldots,\theta x_m)$ 
$[\theta x_{i_1} \da]_{i_1}\ldots[\theta x_{i_p} \da]_{i_p}]\da$, 
where $\{i_1, \ldots, i_p\} \in [1..m]$  
are the highest positions in $f(\theta x_1,$ $ \ldots,\theta x_m)$, where
subterms can be normalized, according to the strategy $S$. 
(More specifically, $\{i_1, \ldots, i_p\} = [1..m]$ if $\mathit{S = Innermost}$
or $\mathit{S = Outermost}$, $\{i_1, \ldots,
i_p\}$ $ = \{j|$ $j \in \{p_1, \ldots p_n\}, j \neq 0\}$ where $[p_1,
\ldots, p_n]=LS(f)$ if $\mathit{S = Local{-}Strat}$.)

By subterm property of $\succ$, we have $\theta f(x_1, \ldots ,x_m) =
f(\theta x_1, \ldots ,\theta x_m)$ $ \succ \theta x_{i_1},$ $ \ldots ,\theta
x_{i_p}$.  Then, by induction hypothesis, we suppose that $\theta
x_{i_1},$ $\ldots ,\theta x_{i_p}$ S-terminate, and so their respective
normal forms $\theta x_{i_1} \da, \ldots,$ $\theta x_{i_p} \da$ exist
and $f(\theta x_1, \ldots,\theta x_m)$ $[\theta x_{i_1}
\da]_{i_1}\ldots[\theta x_{i_p} \da]_{i_p}$ is in normal form.  We may
thus restrict our attention to terms headed by a defined symbol.\\


If $f$ is not a constructor, let us denote it $g$ and prove that
$g(\theta x_1 , \ldots,$ $\theta x_m)$ S-terminates for any $\theta$
satisfying $A$ $= \top$ if we have $\mathit{SUCCESS{-}S(h,\succ)}$ for every
defined symbol $h$.  Let us denote $g(x_1 , \ldots ,x_m)$ by
$t_{\mathit{ref}}$ in the sequel of the proof.

To each state $s$ of the proof tree of $g$, characterized by a current
term $t$ and the set of constraints $A$, we associate the set of
ground terms $G = \{\alpha t ~|~ \alpha {\rm ~satisfies~} A\}$, that
is the set of ground instances represented by $s$.

Inference rule \rname{Abstract} (resp. \rname{Narrow}) transforms
$(\{t\},A)$ into $(\{t'\},A')$ to which is associated $G' = \{\beta t'
~|~ \beta {\rm ~satisfies~} A'\}$ (resp. into $(\{t'_i\},A'_i), i
\in [1..l]$ to which are associated $G' = \{\beta_i t'_i ~|~
\beta_i {\rm ~satisfies}$ $A'_i\}$).

By abstraction (resp. narrowing) Lemma, applying \rname{Abstract}
(resp. \rname{Narrow}), for each $\alpha t$ in
$G$, there exists a 
$\beta t'$ (resp. $\beta_i t'_i$) in $G'$ and such that
S-termination of $\beta t'$ (resp. of the $\beta_i t'_i$) implies
S-termination of $\alpha t$.

When the inference rule \rname{Stop} applies on  $(\{t\},A,C)$:
\begin{itemize}
\item either $A$ is satisfiable, in which case, 
by stopping lemma, every term of 
$G = \{\alpha t ~|~ \alpha {\rm ~satisfies~} A\}$
is S-terminating,

\item {or $A$ is unsatisfiable.} In this case, $G$ is empty.
By emptyness lemma, all previous states on the branch correspond 
to empty sets $G_i$, until an ancestor state $(\{t_p\},A_p,C_p)$, 
where $A_p$ is satisfiable.
Then every term $\alpha t$ of $G_p$ is irreducible, otherwise, by 
Abstraction and Narrowing lemmas, $G_{p+1}$ would not be empty.
\end{itemize}

Therefore, S-termination is ensured for all terms in all sets $G$
of the proof tree.\\

As the process is initialized with $\{t_{\mathit{ref}}\}$ and a
constraint problem satisfiable by any ground substitution, we get that
$g(\theta x_1, \ldots , \theta x_m)$ is S-terminating, for any
$t_{\mathit{ref}}=g(x_1, \ldots , x_m)$, and any ground instance
$\theta$.
\end{proof}

\vspace{3mm}

\begin{lemma}
[(Emptyness lemma)]
Let $(\{t\}, A,C)$ be a state of any proof tree, giving 
$(\{t'\}, A',C')$ by application of \rname{Abstract} or \rname{Narrow}.
If $A$ is unsatisfiable, then so is $A'$.
\end{lemma}

\begin{proof}
If \rname{Abstract} is applied, then if $A$ is unsatisfiable,
$A' = A \wedge t|_{i_1} \con X_{i_1} \ldots  \wedge t|_{i_p} \con
X_{i_p}$ is also unsatisfiable.

If \rname{Narrow} is applied, then 
if $A$ is unsatisfiable (which does not occur for local strategies),
$A'=A \wedge \sigma$ in the innermost case, and $A'=R(t) \wedge A \wedge \sigma$
in the outermost case are also unsatisfiable.
\end{proof}
\vspace{3mm}

\begin{lemma}
[(Abstraction lemma)]
Let $(\{t\}, A,C)$ be a state of any proof tree, giving 
the state
$(\{t'= t[X_j]_{j \in \{i_1,\ldots,i_p\} }\},$ $ A',C')$ 
by application of \rname{Abstract}.

For any ground substitution $\alpha$ satisfying $A$, 
if $\alpha t$ is reducible, there exists $\beta$ such that S-termination of 
$\beta t'$ implies S-termination of $\alpha t$. 
Moreover, $\beta$ satisfies $A'$.
\end{lemma}

\begin{proof}

We prove that $\alpha t \ra^{* S} \beta
t'$, where $\beta = \alpha \cup \bigcup_{j \in \{i_1,\ldots,i_p\}} X_j =
\alpha  t|_j \da$.

First, whatever the strategy $S$, the abstraction positions in $t$ are
chosen so that the $\alpha t|_j$ can be supposed terminating w.r.t. $S$.
Indeed, each term $t|_{i_j}$ is such that:
\begin{itemize}

\item
either $\mathit{TERMIN(S,t|_j)}$ is true, and then by definition of the predicate
$\mathit{TERMIN}$, $\alpha t|_j$ S-terminates;

\item
or $t_{ref} > t|_j$ is satisfiable by $\succ$, and then, by induction
hypothesis, $\alpha t|_j$ S-terminates.

\end{itemize}
So the $\alpha t|_j \da exist.$

Then, let us consider the different choices of abstraction positions w.r.t 
the strategy S:
\begin{itemize}
\item either $\mathit{S=Innermost}$, and whatever the positions $i_1,\ldots,i_p$ in
the term $t$, we have $\alpha t \ra^{* Inn}$ $  \alpha t[\alpha t|_{i_1} \da]_{i_1}\ldots$
$[\alpha t|_{i_p} \da]_{i_p} = \beta t'$;

\item either $\mathit{S=Outermost}$ and $t$ is abstracted 
at positions $i_1,\ldots,i_p$ if $t[X_j]_{j\in \{i_1,\ldots,i_p\} }$
is not outermost narrowable at prefix positions of $i_1,\ldots,i_p$, 
which warrants that the only redex positions of
$\alpha t$ are suffixes of the $j$, and then that
$\alpha t \ra^{* Outermost}
\alpha t[\alpha t|_{i_1} \da]_{i_1}\ldots$
$[\alpha t|_{i_p} \da]_{i_p} = \beta t'$;

\item or $\mathit{S=Local{-}Strat}$ and $top(t)=f$ with $LS(f)=[p_1,\ldots,p_n]$. 
The term $t$ is abstracted 
at positions $i_1,\ldots,i_p \in \{p_1,\ldots,p_{k-1}\}$, 
if  $\exists k \in [2..n] : p_1,\ldots,p_{k-1} \neq 0, 
p_k = 0$, or at positions $i_1,\ldots,i_p \in \{p_1,\ldots,p_n\}$ 
if $p_1,\ldots,p_n \neq 0$.
According to the definition of local strategies, $\alpha t \ra^{* Local{-}Strat}
\alpha t[\alpha t|_{i_1} \da]_{i_1}\ldots$
$[\alpha t|_{i_p} \da]_{i_p} = \beta t'$.

If $LS(f)=[]$ or $LS(f)=[0,p_2,\ldots,p_n]$, then $t=t'$ and $A=A'$,
so $\alpha t = \beta t'$.\\

So $\alpha t \ra^{* S} \beta t'$ for any normal form $\alpha  t|_j \da$ of 
$\alpha  t|_j$, for $j \in \{i_1,\ldots,i_p]\}$.
Then, S-termination of 
$\beta t'$ implies S-termination of $\alpha t$.

Clearly in all cases, $\beta$ satisfies 
$A' =  A \wedge t|_{i_1} \con X_{i_1} \ldots  \wedge t|_{i_p} \con
X_{i_p}$,
provided the $X_i$ are not in $Dom(\alpha)$, which is true
since the $X_i$ are fresh variables not appearing in $A$.

\end{itemize}
\end{proof}
\vspace{3mm}

\begin{lemma}
[(narrowing lemma)]
Let $(\{t\}, A,C)$ be a state of any proof tree, giving 
the states
$(\{v_i\}, $ $A'_i, C'_i), i \in [1..l]$,
by application of \rname{Narrow}.
For any ground substitution $\alpha$ satisfying $A$,
if $\alpha t$ is reducible, then, for each $i \in [1..l]$, 
there exist  $\beta_i$
such that S-termination of the $\beta_i v_i, i \in [1..l]$,
implies S-termination of $\alpha t$.
Moreover, $\beta_i$ satisfies $A'_i$ for each $i \in [1..l]$.
\end{lemma}

\begin{proof}
\vspace{1mm}

We reason by case on the different strategies.

\begin{itemize}
\item 
Either $\mathit{S = Innermost}$, and
By lifting lemma, there is a
term $v$ and substitutions $\beta$ and 
$\sigma = \sigma_0 \wedge \bigwedge_{j \in [1..k]} \overline{\sigma_j}$, 
corresponding to each rewriting step 
$\alpha f(u_1,\ldots,u_m)$ $
\ra^{Inn}_{p, l \ra r} t'$, such that:

\[
  \begin{array}{ll}
        1.~t=f(u_1,\ldots,u_m) \surred^{Inn}_{p, l \ra r,\sigma} v,\\
        2.~\beta v = t', \\
        3.~\beta \sigma_0 = \alpha [{\cal Y}]\\
        4.~\beta {\rm ~satisfies~} \bigwedge_{j \in [1..k]} \overline{\sigma_j}.
  \end{array}
\]
where $\sigma_0$ is the most general unifier of $t|_p$ and $l$ and 
$\sigma_j, j \in [1..k]$ are
all most general unifiers of $\sigma_0 t|_{p'}$ and a left-hand side
$l'$ of a rule of $\R$, for all position $p'$  which are  
suffix positions of $p$ in $t$.

These narrowing steps are effectively produced by the rule
\rname{Narrow}, applied in all possible ways on $f(u_1, \ldots ,u_m)$.
So a term $\beta v$ is 
produced for every innermost rewriting branch starting from 
$\alpha t$.
Then innermost termination of the $\beta v$ implies
innermost termination of $\alpha t$.

Let us prove that $\beta$ satisfies 
$A'= A \wedge \sigma_0 \wedge \bigwedge_{j \in [1..k]}
\overline{\sigma_j}$.

By lifting lemma, we have $\alpha = \beta \sigma_0$ on ${\cal Y}$.
As we can take ${\cal Y} \supseteq Var(A)$, we have  $\alpha = \beta
\sigma_0$ on $Var(A)$.

More precisely, on $Ran(\sigma_0)$, $\beta$ is such that  $\beta \sigma_0 = \alpha$ 
and on $Var(A) \setminus Ran(\sigma_0)$, $\beta = \alpha$.
As $Ran(\sigma_0)$ only contains fresh variables, we have  $Var(A)
\cap Ran(\sigma_0) = \emptyset$, so $Var(A) \setminus Ran(\sigma_0) =
Var(A)$.
So $\beta = \alpha$ on $Var(A)$ and then, $\beta$ satisfies $A$.

Moreover, as $\beta \sigma_0 = \alpha$ on $Dom(\sigma_0)$,
$\beta$ satisfies $\sigma_0$.

So $\beta$ satisfies $A \wedge \sigma_0$.
Finally, with the point 4. of the lifting lemma, we conclude that 
$\beta$ satisfies $A'=  A \wedge \sigma_0 \wedge \bigwedge_{j \in [1..k]}
\overline{\sigma_j}$.

\vspace{3mm}
\item Either $\mathit{S = Local{-}Strat}$, and
\rname{Narrow} is applied on $\{t=f(u_1, \ldots, $ $u_m)\}$ with $l =
[0,p_1, \ldots ,p_n]$.
For any $\alpha$ satisfying $A$, 

\begin{itemize}

        \item either 
$\alpha f(u_1, \ldots ,u_m)$  is irreducible at the top position, but may
be reduced at the positions $p_1, \ldots ,p_n$. 
In this case, either $f(u_1, \ldots ,u_m)$
is not narrowable at the top position, 
either $ f(u_1, \ldots ,u_m)$ $ \leadsto_{\epsilon, \sigma_i} v_i$ 
for $i \in [1..l]$ and 
$A \wedge \sigma_i$ is unsatisfiable for each $i$,
or there exists $i \in [1..l]$ such that 
$ f(u_1, \ldots ,u_m) \leadsto_{\epsilon, \sigma_i} v_i$ and 
$A \wedge \sigma_i$ is satisfiable.

In the first two cases, \rname{Narrow} produces the state 
$(\{t^{[p_1, \ldots ,p_n]}\},$ $A,C)$, and setting $\beta = \alpha$, 
we obtain that termination of $\beta t^{[p_1, \ldots ,p_n]}$ implies termination of 
$\alpha t^{[0,p_1, \ldots ,p_n]}$, 
and that $\beta$ satisfies $A'=A$.

In the third case, \rname{Narrow} produces the state 
$(\{t^{[p_1, \ldots ,p_n]}\}, 
A \wedge (\bigwedge_{i=1}^l \overline{\sigma_i}),C)$,
and setting $\beta = \alpha$, 
we have termination of $\beta t^{[p_1, \ldots ,p_n]}$ implies termination of 
$\alpha t^{[0,p_1, \ldots ,p_n]}$.
Moreover, as $\alpha t$ is not reducible at the top position, 
$\alpha = \beta$ satisfies $(\bigwedge_{i=1}^l \overline{\sigma_i})$.
Thus, as $\alpha$ satisfies $A$, $\beta$ satisfies $A'=A 
\wedge (\bigwedge_{i=1}^l \overline{\sigma_i})$.

       \item or $\alpha f(u_1, \ldots ,u_m)$ is reducible at the top
position,
and by lifting lemma, there is a
term $v$ and substitutions $\beta$ and 
$\sigma_0$ corresponding to each rewriting step $\alpha f(u_1,\ldots,u_m)
\ra_{\epsilon, l \ra r} t'$, such that:

\[
  \begin{array}{ll}
        1.~t=f(u_1,\ldots,u_m) \surred_{\epsilon, l \ra r,\sigma_0} v,\\
        2.~\beta v = t', \\
        3.~\beta \sigma_0 = \alpha [{\cal Y}].
  \end{array}
\]

where $\sigma_0$ is the most general unifier of $t$ and $l$.

These narrowing steps are effectively produced by
\rname{Narrow}, which is applied in all possible ways on
$f(u_1, \ldots ,u_m)$ at the top position.
So a term $\beta v$ is 
produced for every LS-rewriting step applying on
$\alpha t$ at the top position.
Then termination of the $\beta v$ implies
termination of $\alpha t$ for the given LS-strategy.


We prove that $\beta$ satisfies $A \wedge \sigma_0$
like in the innermost case, except that there is no negation of 
substitution here.

\end{itemize}
\vspace{3mm}
 



\vspace{3mm}

\item Or $\mathit{S= Outermost}$, and 
in this case, $t=f(u_1, \ldots ,u_n)$ is renamed into $t_0=f(u_1,\ldots,u_n)^{\rho}$. 
$A$ then becomes 
$A_0 = A \cup R(f(u_1,\ldots,$ $u_n))$ where
$\rho = (x_1 \red x'_1) \ldots (x_k \red x'_k)$.

We first show that if every $\beta_0 t_0$ outermost
terminates, for $\beta_0 {\rm ~satisfying~} A_0$,
then every $\alpha t$ outermost terminates.

If $A$ is satisfiable, then $A_0$ is satisfiable.
Indeed, $A_0 = A \cup 
f(u_1, \ldots$ $ ,u_m) \red 
f(u_1, \ldots$ $ ,u_m)^{\rho}$, with $\rho = (x_1 \red x'_1) 
\ldots (x_k \red x'_k)$.
In addition, the $x_i$ are the 
variables of $f(u_1,\ldots,u_n)$. 

If $A = \top$, then  $A_0 = f(u_1, \ldots$ $ ,u_m) \red 
f(u_1, \ldots$ $ ,u_m)^{\rho}$,
which is always satisfiable.
If $A \neq \top$, since they are the variables of $f(u_1,\ldots,u_n)$,
the $x_i$ can appear in $A$, either in abstracted subterms, 
either as new abstraction variables,
either in the right hand-sides of equalities and disequalities defining 
the substitution of the previous narrowing step,  
or as new variables introduced by the previous reduction renaming step. 
In any case, the formula in which they appear is compatible with 
$f(u_1, \ldots$ $ ,u_m) \red f(u_1, \ldots, u_m)^{\rho}$.
More precisely, for the $\theta x_i$ such that $\theta$ satisfies $A$, 
$\theta$ can be extended on the variables $x'_i$, in such a way that 
$A_0$ is satisfiable.
Then $A_0 = A \cup f(u_1, \ldots, u_m) \red 
f(u_1, \ldots$ $ ,u_m)^{\rho}$ is satisfiable.\\

By definition of $A_0$, the $\beta_0$ are the $\alpha$ verifying
the reduction formula $f(u_1, \ldots, u_m)$ $\red 
f(u_1, \ldots, u_m)^{\rho}$, with $\rho = (x_1 \red x'_1) 
\ldots (x_k \red x'_k)$.
We have $Dom(\alpha) = Var(A) \cup
\{x_1,\ldots,x_k\}$.
The domain of $\beta_0$ is $Dom(\alpha) \cup \{x'_1,\ldots,x'_k\}$.
Then $\beta_0 = \alpha \; [Dom(\alpha)]$ and by definition of the
reduction formula, the $\beta_0 x'_i$ are  such that
$t [\beta_0 x'_1]_{p_1} \ldots [\beta_0 x'_k]_{p_k}$ is
the first reduced form of $\alpha f(u_1, \ldots ,u_n)$
in any outermost rewriting chain starting from 
$\alpha f(u_1, \ldots ,u_n)$, having an outermost rewriting position
at a non variable position of $f(u_1, \ldots ,u_n)$.

Then, by definition of the outermost strategy, the $\beta_0 t_0$
represent any possible outermost reduced form of $\alpha t$
just before the reduction occurs at a non variable occurence of 
$f(u_1,\ldots ,u_n)$. Thus, outermost termination of the $\beta_0 t_0$
implies outermost termination of the $\alpha t$.\\

Then $t_0$ is narrowed in all possible ways 
into terms $v_i$ at positions $p_i$
with substitutions $\sigma_i$, provided $p_i$ and 
$\sigma_i$ satisfy the outermost
narrowing requirements, as defined in Definition \ref{def:narrowing}.
We now show that if $\beta_0 t_0$ is reducible,
then there exist $\beta_i$ satisfying $A'$ such that 
outermost termination of the $\beta_i v_i$ implies 
outermost termination of $\beta_0 t_0$.

We have $\beta_0 t_0 \ra^{Out}_{p, l \ra r}
t'$ and $p \in \OS(t_0)$ since $t_0 = t^{\rho}$. 

By lifting lemma, there is a
term $v$ and substitutions $\beta$ and 
$\sigma = \sigma_0 \wedge \bigwedge_{j \in [1..k]} \overline{\sigma_j}$, 
corresponding to each rewriting step $\alpha t_0
\ra^{Out}_{p, l \ra r} t'$, such that:

\[
  \begin{array}{ll}
        1.~t_0 \surred^{Out}_{p, l \ra r,\sigma} v,\\
        2.~\beta v = t', \\
        3.~\beta \sigma_0 = \beta_0 [{\cal Y}]\\
        4.~\beta {\rm ~satisfies~} \bigwedge_{j \in [1..k]} \overline{\sigma_j}.
  \end{array}
\]

where $\sigma_0$ is the most general unifier of $t_0|_p$ and $l$ and 
$\sigma_j, j \in [1..k]$ are
all most general unifiers of $\sigma_0 t_0|_{p'}$ and a left-hand side
$l'$ of a rule of $\R$, for all position $p'$  which are  
prefix positions of $p$ in $t_0$.

These narrowing steps are effectively produced by the rule
\rname{Narrow}, applied in all possible ways.
So a term $\beta v$ is 
produced for every outermost rewriting branch starting from 
$\beta_0 t_0$. 
Then outermost termination of the $\beta v$ implies
outermost termination of $\beta_0 t_0$.

We prove that $\beta$ satisfies 
$A' = A_0 \wedge \sigma_0  \bigwedge_{j \in [1..k]} \overline{\sigma_j}$
like in the innermost case.
\end{itemize}
\end{proof}
\vspace{3mm}

 \begin{lemma}[(Stopping lemma)]
Let $(\{t\}, A,C)$ be a state of any proof tree, {\bf with $A$ satisfiable,}
and giving  the state $(\emptyset, A',C')$  by application of an inference rule.
Then for any ground substitution $\alpha$ satisfying $A$,
$\alpha t$ S-terminates. 
\end{lemma}

\begin{proof} 
The only rule giving the state $(\emptyset, A',C')$  
is \rname{Stop}.
When \rname{Stop} is applied, then 

\begin{itemize}
\item either $\mathit{TERMIN(S,t)}$ and then $\alpha t$ S-terminates for any
ground substitution $\alpha$,

\item or $(t_{\mathit{ref}} > t)$ is satisfiable.  
Then, for any ground substitution $\alpha$ satisfying $A$, 
$\alpha t_{\mathit{ref}} \succ \alpha t$. 
By induction hypothesis, $\alpha t$ S-terminates. 
\end{itemize}
\end{proof}

\section{The usable rules} 

To prove Lemma~\ref{lemma:U-correction}, we  need the next three
lemmas. 
The first two ones are pretty obvious from the definition of the
usable rules.

\begin{lemma}\label{lemma:U-def}
Let $\R$ be a rewrite system on a set $\F$ of symbols and $t \in \TFXXA$.
Then, every  symbol $f \in \F$ occuring in $t$ is such that 
$Rls(f) \subseteq \U(t)$.
\end{lemma}

\begin{proof}
We proceed by structural induction on $t$.

\begin{itemize}
        \item If $t \in \X \cup \XA$, the property is trivially true;
        \item if $t$ is a constant $a$, 
$\U(t = a) = Rls(a) \cup_{l \ra r \in Rls(a)} \U(r)$; the only symbol
of $t$ is $a$, and we have $Rls(a) \subseteq \U(t)$.
\end{itemize}

Let us consider a non-constant and non-variable term $t \in \TFXXA$, 
of the form $f(u_1, \ldots ,u_n)$.
Then, by definition of $\U(t)$, we have 
$\U(t) = Rls(f) \cup_{i=1}^n \U(u_i) \cup_{l \ra r \in Rls(f)} \U(r)$.
Then, whatever $g$ symbol of $t$, either $g = f$ and then
$Rls(g) \subseteq \U(t)$, or $g$ is a symbol occuring in some $u_i$ and,
by induction hypothesis on $u_i$,  $Rls(g) \subseteq \U(u_i)$,
with $\U(u_i) \subseteq \U(t)$.
\end{proof}

\begin{lemma}\label{lemma:U-rhs}
Let $\R$ be a rewrite system on a set $\F$ of symbols and $t \in \TFXXA$.
Then $l \ra r \in \U(t) \Rightarrow \U(r) \subseteq \U(t)$.
\end{lemma}

\begin{proof}
According to the definition of the usable rules, if a term $t$
is such that $\Var(t) \cap \X \neq \emptyset$, then $\U(t) = \R$,
and then the property is trivially true.
We will then suppose in the following that $t$ does not contain
any variable of $\X$.

Let $l \ra r \in \U(t)$. 
By definition of $\U(t)$, since $\Var(t) \cap \X = \emptyset$,
among  all recursive applications of the definition of $\U$ in
$\U(t)$, there is an application $\U(t')$ of $\U$ to some term $t'$
such that $\U(t') = Rls(g) \cup_i \U(t'|_i) \cup_{l' \ra r' \in Rls(g)} \U(r')$,
with $\U(t') \subseteq \U(t)$, and $l \ra r \in Rls(g)$, with $g = top(l)$.

Since $l \ra r \in Rls(g)$, by definition of $\U(t')$, we have 
$\U(r) \subseteq \cup_{l' \ra r' \in Rls(g)} \U(r')$, and then 
$\U(r) \subseteq \U(t') \subseteq \U(t)$.
\end{proof}

\begin{lemma}\label{lemma:U-reducible}
Let $\R$ be a rewrite system on a set $\F$ of symbols and $t \in \TFXXA$.
Whatever $\alpha$ ground normalized substitution and
$\alpha t \ra_{p_1,l_1 \ra r_1} t_1 \ra_{p_2,l_2 \ra r_2} t_2
\ra  \ldots  \ra_{p_n,l_n \ra r_n} t_n$ rewrite chain starting from $\alpha t$,
the defined symbol of $t_k, 1 \leq k \leq n$ at a redex position
of $t_k$ is either a symbol of $t$ or one of the $r_i, i \in [1..k]$.
\end{lemma}

\begin{proof}
We proceed by induction on the length of the derivation.
The property is obviously true
for an empty derivation i.e. on $\alpha t$.

Let us show the property for the first rewriting step 
$\alpha t \ra_{p_1, l_1 \ra r_1} t_1$.
By definition of rewriting, 
$\exists \sigma : \sigma l_1 = \alpha t|_{p_1}$ and 
$t_1 = \alpha t[\sigma r_1]_{p_1}$.
Let $f$ be the redex symbol of $t_1$ at a position $p$, and let us show 
that $f$ comes either from $t$ or from $r_1$.

Since $t_1 = \alpha t[\sigma r_1]_{p_1}$, either $p$ is a position 
of the context $\alpha t[]_{p_1}$, which does not change by rewriting, 
so we already have $f$ as redex symbol of $\alpha t$ at position $p$.
As $\alpha$ is normalized, $p$ is a position of $t$, so $f$ is a symbol of $t$.

Either $p$ corresponds in $t_1$ to a non variable position of $r_1$, 
so $f$ is a symbol of $r_1$.

Or $p$ corresponds in $t_1$ to a position $r$ in $\sigma x$,
for a variable $x \in \Var(r_1)$ at position $q$ in $r_1$: we have $p=p_1qr$.
In this case, since $\Var(r_1) \subseteq \Var(l_1)$, we have $x \in \Var(l_1)$,
so $\sigma x$ is also a subterm of $\alpha t$, and $f$ occurs in $\alpha t$ 
at position $p'=p_1q'r$, where $q'$ is a position of $x$ in $l_1$.

Moreover, as $p$ is a redex position in $t_1$, then by definition of the innermost 
strategy, there is no suffix redex position of $p$ in $t_1$.
As $t_1|_p = \alpha t|_{p'}$, then similarly $p'$ is a redex position in $\alpha t$.
As $\alpha$ is normalized, $p'$ is a position of $t$, so $f$ is a symbol of $t$.

Then, let us suppose the property true for any term of the rewrite chain
$\alpha t$ $\ra_{p_1,l_1 \ra r_1} t_1 \ra  \ldots  \ra_{p_k, l_k \ra r_k}
t_k$, i.e. any redex symbol $f$ of $t_k$ is also a symbol of $t$, or a symbol 
of one of the $r_i, i \in [1..k]$,
and let us consider
$t_k \ra_{p_{k+1}, l_{k+1} \ra r_{k+1}} t_{k+1}$.

By a similar reasoning than previously, we establish that any 
redex symbol $f$ of $t_{k+1}$ is also a symbol of $t_k$, or a symbol of $r_{k+1}$.
We then conclude with the previous induction hypothesis.
\end{proof}

We are now able to prove Lemma~\ref{lemma:U-correction}.
\vspace{2mm}

\begin{lemma}[\ref{lemma:U-correction}]

Let $\R$ be a rewrite system on a set $\F$ of symbols and $t \in \TFXXA$.
Whatever $\alpha t$ ground instance of $t$ and
$\alpha t \ra_{p_1,l_1 \ra r_1} t_1 \ra_{p_2,l_2 \ra r_2} t_2
\ra  \ldots  \ra_{p_n,l_n \ra r_n} t_n$ rewrite chain starting from $\alpha t$,
then $l_i \ra r_i \in \U(t), \;  \forall i \in [1..n]$.
\end{lemma}

\begin{proof}
If a variable $x \in \X$ occurs in $t$, then $\U(t) = \R$ and the property
is trivially true.
We  then consider in the following that $t \in \TFXA$, 
and then that $\alpha$ is a (ground) normalized substitution.\\
We proceed by induction on $\TFXA$ and on the length of the derivation.

The property is trivially true if $\alpha t$ is in normal form.
For any $\alpha t$ $ \ra_{p_1, l_1 \ra r_1} t_1$, 
since $\alpha$ is normalized, 
$p_1$ corresponds in $\alpha t$ to a non-variable position of $t$. 
Let $f$ be the symbol at
position $p_1$ in $t$.
Since $f$ is the symbol at the redex position $p_1$ of $\alpha t$
with the rule $l_1 \ra r_1$, then $l_1 \ra r_1 \in Rls(f)$.
Moreover, thanks to 
Lemma~\ref{lemma:U-def}, $Rls(f) \subseteq \U(t)$. Therefore,
$l_1 \ra r_1 \in \U(t)$.

Let us now suppose the property is true for any derivation chain 
starting from $\alpha t$ whose
length is less or equal to $k$, and consider the chain:
$\alpha t \ra_{p_1,l_1 \ra r_1} t_1 \ra_{p_2,l_2 \ra r_2} t_2
\ra  \ldots  \ra_{p_k,l_k \ra r_k} t_k \ra_{p_{k+1},l_{k+1} \ra r_{k+1}} t_{k+1}$.
Let $f$ be the symbol at position $p_{k+1}$ in $t_k$.
Since $p_{k+1}$ is a redex position of $t_k$ with the rule
$l_{k+1} \ra r_{k+1}$, then 
$l_{k+1} \ra r_{k+1} \in Rls(f)$.

By Lemma~\ref{lemma:U-reducible} with a derivation of length $k$, 
we have two cases:
\begin{itemize}
        \item either the symbol $f$ at position $p_{k+1}$ in $t_k$ is
a symbol of $t$; 
then, thanks to Lemma~\ref{lemma:U-def} on $t$, we get
$Rls(f) \subseteq \U(t)$;
henceforth $l_{k+1} \ra r_{k+1} \in
\U(t)$;

        \item or the symbol $f$ at position $p_{k+1}$ in $t_k$ is
a symbol of a $r_i, i \in [1..k]$;
then, thanks to Lemma~\ref{lemma:U-def} on
$r_i$, we get $Rls(f) \subseteq \U(r_i)$; 
henceforth $l_{k+1} \ra r_{k+1} \in \U(r_i)$;
by induction hypothesis we have $l_i \ra r_i \in \U(t)$ and, thanks to
Lemma~\ref{lemma:U-rhs}, we have $\U(r_i) \subseteq \U(t)$.
Henceforth $l_{k+1} \ra r_{k+1} \in \U(t)$.
\end{itemize}
\end{proof}

\begin{subproposition}[\ref{prop:TC}]
Let $\R$ be a rewrite system on a set $\F$ of symbols, and $t$ a term of $\TFXXA$.
If there exists a simplification ordering $\succ$ such that
$\forall l \ra r \in \U(t): l \succ r$, then any ground instance of $t$
is terminating.
\end{subproposition}

\begin{proof}
As $\succ$ orients the rules used in any reduction chain starting from 
$\alpha t$ for any ground substitution $\alpha$,
by properties of the simplification orderings, 
$\succ$ also orients the reduction chains, which are then finite.
\end{proof}

\section{A lemma specific to the outermost case} 

\begin{lemma}[\ref{lemme:var}]
Let $(\{t_i\}, A_i, C_i)$ be the
$i^{th}$ state of any branch of the derivation tree obtained by
applying the strategy $S$ on $(\{t_{\mathit{ref}}\}, \top, \top)$, and
$\succ$ an $\F$-stable ordering having the subterm property.  If every
reduction formula in $A_i$ can be reduced to a formula $\bigwedge_j
x_j = x'_j$, then we have:
\begin{center} 
for all variable $x$ of $t_i$ in $\X$: $(t_{\mathit{ref}} > x) \sur
A_i$ is satisfiable by $\succ$.
\end{center}
\end{lemma}

\begin{proof}
The proof is made by induction on the number $i$ of applications of
the inference rules from $(\{t_{\mathit{ref}}\}, \top, \top)$ to the
state $(\{t_i\}, A_i, C_i)$.

Let us prove that the property holds for $i=0$.  We have $t_0 =
t_{\mathit{ref}}$ and then $Var(t_0) = \Var(t_{\mathit{ref}})$.
Consequently, for every $x \in Var(t_0)$, whatever the ground
substitution $\alpha$ such that $\Var(t_{\mathit{ref}}) \subseteq
Dom(\alpha)$, $\alpha x$ is a subterm of $\alpha t_{\mathit{ref}}$.
The induction ordering $\succ$ satisfying the conditions of the rules
before the application of these rules can be any $\F$-stable ordering
having the subterm property.  We then have $\alpha t_{\mathit{ref}}
\succ \alpha x $.

We now prove that if the property holds for $i-1$, it also holds for
$i$.\\ 

If the rule used at the $i^{th}$ step is \rname{Stop}, then $Var(t_i)
= \emptyset$, and then, the property is trivially verified.\\

If the rule used at the $i^{th}$ step is \rname{Abstract}, 
as the rule \rname{Abstract} replaces subterms in $t_{i-1}$ by new 
variables of $\XA$, then $(Var(t_i) \cap \X) \subseteq (Var(t_{i-1}) \cap \X)$, 
so the property still holds.\\

If the rule used at the $i^{th}$ step is \rname{Narrow} then,
by hypothesis, the
reduction renaming applied to $t_{i-1}$ and giving a term $t'_{i-1}$
just consists in a mere
renaming of the variables of $t_{i-1}$.
Let $t_i$ be a term obtained
by narrowing $t'_{i-1}$ with the substitution $\sigma$.

Let $z \in \Var(t_i)$, and $\alpha$ a substitution satisfying $A_i$.
We show that $\alpha t_{\mathit{ref}} \succ \alpha z$.
We have two cases.

Either $z$ is a fresh variable introduced by the narrowing step.
Let $x' \in \Var(t'_{i-1})$ such that $z \in \Var(\sigma x')$,
and $x \in \Var(t_{i-1})$ such that $x'$ is a renaming of $x$.
By hypothesis, every reduction formula in $A_i$ can be reduced to a formula
$\bigwedge_j x_j = x'_j$. This is then the same for 
$A_{i-1}$. Moreover, since $\alpha$ satisfies $A_i$,
then it satisfies in particular $A_{i-1}$.
Then, by induction hypothesis, $\alpha t_{\mathit{ref}} \succ \alpha x$ and,
since $\alpha$ satisfies $x = x'$, we also have 
$\alpha t_{\mathit{ref}} \succ \alpha x'$.\\
By hypothesis, $\sigma$ contains the equality $x' = C[z]$, with $C[z]$ a 
(possibly empty) context of $z$.
Moreover, by definition of the rule \rname{Narrow}, 
$A_i = A_{i-1} \wedge R(t_{i-1}) \wedge \sigma$. 
So $A_i$ contains the equality $x' = C[z]$.\\
Then, as $\alpha$ satisfies $A_i$, $\alpha$ is such that
$\alpha x' = \alpha C[z]$.
Since $\alpha t_{\mathit{ref}} \succ \alpha x'$, we have 
$\alpha t_{\mathit{ref}} \succ \alpha C[z]$ and then, by subterm property,
$\alpha t_{\mathit{ref}} \succ \alpha z$.

Or $z \in \Var(t'_{i-1})$~; 
by the same reasoning as in the previous point
for $x'$, we have $\alpha t_{\mathit{ref}} \succ \alpha z$.
\end{proof}

\begin{acks}
We would like to thank Olivier Fissore for fruitful exchanges, we have had in 
previous works on the topic, the Protheo group for dynamically supporting our 
ideas, and Nachum Dershowitz for the interest he took in our approach, and for 
his advice on the manuscript of this paper.
\end{acks}

\bibliography{/local/protheo/bibtex/abbrev,/local/protheo/bibtex/protheo,/local/protheo/bibtex/gnaedig,/local/protheo/bibtex/hkirchne}


\end{document}